%% file: HRS_spectra_astroph.tex
\documentclass[traditabstract]{aa} 
\usepackage{graphicx}
\usepackage{url}
\usepackage{lscape}
\usepackage{longtable}
\usepackage{multirow}
\usepackage{txfonts}

\newcommand{\oii}{O{\sc ii}}

\newcommand{\hb}{H{\sc $\beta$}}
\newcommand{\oiii}{O{\sc iii}}

\newcommand{\nii}{N{\sc ii}}

\newcommand{\ha}{H{\sc $\alpha$}}
\newcommand{\sii}{S{\sc ii}}

\begin{document}
   \title{Integrated spectroscopy of the $Herschel$ Reference Survey.}

   \subtitle{The spectral line properties of a volume-limited, K-band selected
   sample of nearby galaxies\thanks{Based on observations collected at the Observatoire de Haute Provence (OHP) (France), operated by the CNRS}}

  \author{A. Boselli\inst{1}
  	  ,
	  T.M. Hughes\inst{2}
	  ,
	  L. Cortese\inst{3}
	  ,
	  G. Gavazzi\inst{4}
	  ,
	  V. Buat\inst{1}
         }

\institute{	
	Laboratoire d'Astrophysique de Marseille - LAM, Universit\'e d'Aix-Marseille \& CNRS, UMR7326, 38 rue F. Joliot-Curie, 13388 Marseille Cedex 13, France 
              \email{Alessandro.Boselli@oamp.fr, Veronique.Buat@oamp.fr}
	      \and
	      	Kavli Institute for Astronomy \& Astrophysics, Peking University, Beijing 100871, P.R. China
                \email{tmhughes@pku.edu.cn}
	       \and
	      	European Southern Observatory, Karl-Schwarzschild Str. 2, 85748 Garching bei Muenchen, Germany
                \email{lcortese@eso.org}
		\and
		Universita' di Milano-Bicocca, piazza della Scienza 3, 20126, Milano, Italy
             \email{giuseppe.gavazzi@mib.infn.it}
}

   \date{}

 
  \abstract
  {We present long-slit integrated spectroscopy of 238 late-type galaxies belonging to the $Herschel$ Reference Survey, 
  a volume limited sample representative of the nearby universe. This sample has a unique legacy value since ideally 
  defined for any statistical study of the multifrequency properties of galaxies spanning a large range in morphological type and
  luminosity.
  The spectroscopic observations cover the spectral range 3600-6900 \AA ~ at a resolution $R$ $\simeq$ 1000 and are thus suitable for 
  separating the underlying absorption from the emission of the H$\beta$ line as well as the two [\nii] \ lines from the \ha \ emission.
  We measure the fluxes and the equivalent widths of the strongest emission lines ([\oii]$\lambda$3727, \hb, [\oiii]$\lambda$4959 and [\oiii]$\lambda$5007,
  [\nii]$\lambda$6548, \ha, [\nii]$\lambda$6584, [\sii]$\lambda$6717 and [\sii]$\lambda$6731). The data are used to study the distribution of the equivalent width of all
  the emission lines,  
  of the Balmer decrement $C(H\beta)$ and of the observed underlying Balmer absorption under \hb \ (E.W.H$\beta_{abs}$) in this sample.
  Combining these new spectroscopic data with those available at other frequencies, we also study the dependence of $C(H\beta)$ and E.W.H$\beta_{abs}$
  on morphological type, stellar mass and stellar surface density, star formation rate, birthrate parameter and metallicity in galaxies belonging to different
  environments (fields vs. Virgo cluster). The distribution of the equivalent width of all
  the emission lines, of $C(H\beta)$ (or equivalently of $A(H\alpha)$) and  E.W.H$\beta_{abs}$ are systematically different in cluster and field galaxies. 
  The Balmer decrement increases with stellar mass, stellar surface density, metallicity and star formation rate of the observed galaxies, while it is unexpectedly almost
  independent from the column density of the atomic and molecular gas. The dependence of $C(H\beta)$ on stellar mass is steeper than that previously found in other works.
  The underlying Balmer absorption does not significantly change with any of these physical parameters. 
  }
   {}
   {}
   {}
   {}
   {}

   \keywords{Galaxies: spiral; Galaxies: ISM; ISM: dust, extinction
               }
	       
   \maketitle
%

\section{Introduction}

A complete understanding of the matter cycle in galaxies, i.e. of the process that transforms the primordial atomic gas into molecular clouds 
where stars are formed, and of the metal production and the formation of dust grains requires a 
multifrequency analysis. Indeed, the atomic gas can be directly observed using the 21 cm emission line, while the molecular component is generally traced 
by the emission of carbon monoxide emission lines. Star formation can be quantified under some assumptions through the observations of the ionised hydrogen 
or of the UV stellar continuum emitted by the youngest stellar populations. Dust, formed by the aggregation of metals produced in the final phases of stellar evolution
and injected into the interstellar medium by stellar winds and supernovae explosions, absorbs the stellar radiation and re-emits the acquired energy in the infrared domain (e.g. Boselli
2011).\\
With the aim of studying the matter cycle in galaxies of different morphological type and luminosity, we have defined a K-band selected, volume 
limited sample of nearby galaxies, the $Herschel$ Reference Survey (HRS; Boselli et al. 2010a), to be observed in guaranteed time with the 
SPIRE instrument (Griffin et al. 2010) on board of $Herschel$ (Pilbrat et al. 2010).
This sample, which includes just over three hundreds objects, is ideally defined for characterising the statistical properties of normal, nearby galaxies. Given the tight
relation between the near-infrared bands and the total stellar mass of galaxies (Gavazzi et al. 1996), the choice of the K 
band secures a stellar mass selection. The sample, which includes both isolated galaxies and objects in the Virgo cluster, 
is also appropriate for studying the effects of the environment on galaxy evolution (Boselli \& Gavazzi 2006).\\
Since the definition of the sample, which is extensively described in Boselli et al. (2010a), we are making
a huge effort at gathering multifrequency data spanning the whole range of the electromagnetic spectrum
to offer to the astronomical community a complete and homogeneous set of data suitable for any kind of statistical analysis.
Near infrared and optical images are already available thanks to the 2MASS (Jarrett et al. 2003; Skrutskie et al. 2006) and
SDSS (Adelman-McCarthy et al. 2008) surveys. The SPIRE imaging data at 250, 350 and 500 $\mu$m of the whole sample, collected during the 
first year of $Herschel$, have been recently published in a dedicated paper (Ciesla et al. 2012). 
A cycle 6 GALEX legacy survey has been completed (Cortese et al. 2012a). Combined with the data obtained during the Virgo cluster 
survey (GUViCS; Boselli et al. 2011), this provides FUV and NUV magnitudes for most of the HRS galaxies.
We are also undertaking an H$\alpha$+[\nii] imaging survey at San Pedro Martir, Mexico (Boselli et al., in prep.) and a $^{12}CO(1-0)$ survey
of the star forming spirals of the sample with the 12 m Kitt Peak radio telescope. The present paper 
is devoted to the publication of the long-slit, integrated spectroscopy of the late-type systems obtained with the 1.93 m 
telescope of the Observatoire de Haute Provence. \\
Given the complete nature of the $Herschel$ Reference Survey, 
we use these data to make a statistical analysis of the spectroscopic properties of a mass-selected sample of nearby galaxies in different environments.
We study the distribution of the equivalent width of the different emission lines. The equivalent widths are normalised indices and are thus ideally 
defined for a direct comparison of galaxies of different size and luminosity.
We focus our attention on the analysis of the Balmer decrement. This index traces the attenuation of the emission lines due to dust.
Its importance resides in the fact that, when far infrared data are missing, it is often used in cosmological surveys 
to quantify the amount of dust attenuation of the stellar emission with the Calzetti law (Calzetti 2001). 
Spectroscopic surveys are allowing the measurement of this quantity for galaxies at higher redshift (e.g. Caputi et al. 2008; Ly et al. 2012). We also analyse the statistical properties 
of the underlying Balmer absorption due to the absorption of the stellar continuum in the atmosphere of warm A-type stars (Poggianti \& Barbaro 1997; Thomas et al. 2004).
This index, crucial for an accurate measure of the Balmer decrement, is often used as an indicator of the mean age of the underlying stellar population.
In an accompanying paper (Hughes et al. 2012), we analyse the stellar mass - metallicity relation in the same complete sample of nearby galaxies and its dependence
on the environment. Several papers based on the exploitation of the SPIRE data of the HRS have been published in the $Herschel$ dedicated A\&A special issue
or in more recent communications.
Some of these are focused on the study of the statistical properties of the HRS sample: in Cortese et al. (2011a) and Cortese et al. (2012b) 
we analyse the HI gas and dust scaling relations of the whole sample. The far infrared colours and the spectral energy distributions are 
discussed in Boselli et al. (2010b) and Boselli et al. (2012), while the dust properties of the early-type systems in Smith et al. (2012a).
Some of these works are already taking advantage of the spectroscopic data presented here.\\
The spectroscopic data presented in this work, as well as those collected at other frequencies, will be made available to the community through the 
dedicated HeDaM database (http://hedam.oamp.fr/; Roehlly et al. 2012).

\section{The sample}

\begin{figure}
   \centering
   \includegraphics[width=8cm]{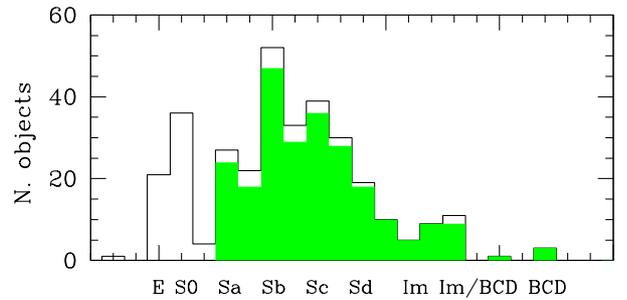}
   \caption{Distribution in morphological type of all the HRS galaxies (solid line) and of the late-type systems with available spectroscopic
   data (green). }
   \label{type}%
   \end{figure}

The $Herschel$ Reference Survey is a SPIRE guaranteed time key project aimed at observing with $Herschel$ a complete, 
K-band selected (K $\leq$ 8.7 mag for early-types, K $\leq$ 12 for type $\geq$ Sa), volume limited (15$\leq$ $D$ $\leq$ 25 Mpc)
sample of nearby galaxies at high galactic latitude. The sample, which is extensively presented in Boselli et al. (2010a),
is composed of 322 galaxies out of which 260 are late-type systems\footnote{With respect to the original sample given in Boselli et al. (2010a), we 
removed the galaxy HRS 228 whose new redshift indicates it as a background object.
We also revised the morphological type for three galaxies that moved from the early- to late-type class: 
NGC 5701, now classified as Sa, and NGC 4438 and NGC 4457 now Sb.}. 
Figure \ref{type} shows the distribution of the different morphological classes within the HRS.
The K-band selection has been chosen as a proxy for galaxy stellar mass (Gavazzi et al. 1996). The sample includes objects in environments of 
different density, from the core of the Virgo cluster, to loose groups and fairly isolated systems. As defined, the present
sample is ideal for any statistical study of the mean galaxy population of the nearby universe.\\
This paper is focused on the late-type galaxies 
of the sample. We present new spectroscopic observations of 134 galaxies\footnote{Four other galaxies with data available in the
literature have been also observed, bringing the total number of observed objects to 138.}, and we combine this new dataset with the one already
published by our team for Virgo cluster objects (Gavazzi et al. 2004) or available in the literature (Kennicutt 1992a, 1992b; Jansen et al. 2000; Moustakas \& Kennicutt 2006;
Moustakas et al. 2010) with the purpose of providing the most complete and
homogeneous list of spectroscopic parameters of these late-type systems. The resulting spectroscopic sample is fairly complete 
since it includes 238 out of the 260 late-type systems (see Figure \ref{type}).

\section{Observations}

Late-type HRS galaxies have been observed during 51 nights 
in the years 2004-2009 using the 1.93 m telescope of the Observatoire
d'Haute Provence (OHP; see Table \ref{tab:Tablogbook}). Data were acquired using a telescope configuration and sky conditions similar 
to those of Gavazzi et al. (2004). Observations were carried out using the CARELEC spectrograph and a grism with a dispersion of 133 \AA/mm
corresponding to 1.8 \AA/pixel. The spectrograph is equipped with a 2048 $\times$ 1024 EEV CCD with a pixel of 13.5 $\mu$m, and a
spatial scale of 0.58 arcsec/pixel. Galaxies where observed using a 5 arcmin slit of width of 2.5 arcsec, adapted for the typical
seeing conditions (2-3 arcsec). 
To obtain data representative for the whole galaxy, most targets have been observed in drift-scan
mode, i.e. with the slit, generally parallel to the galaxy major axis, drifting over the optical surface of the 
galaxy\footnote{No drift has been done on perfectly edge-on galaxies and on a few other objects.}. For those few
objects with angular size larger than the size of the slit, the drift was done in the direction perpendicular to the
major axis. Drifting was obtained by slewing manually several times the telescope between two extreme positions checked on one
offset star or on the galaxy itself. Not unexpectedly spectra obtained in this way have lower S/N ratio than traditional 
long-slit spectra of similar integration time, because a large fraction of the time is spent on low surface brightness regions.
Bright stars overposed on the disc of the galaxies were avoided, whenever possible, during the observations, or identified as high-surface brightness regions
in the spectra and removed during the data reduction.
Our spectra cover the wavelength range 3600-6900  (from [\oii] to [\sii]) with a resolution of  $R$ $\simeq$ 1000. 
Observations were taken either in photometric conditions or through thin cirrus\footnote{The 2006 run was totally lost due to bad weather conditions.}. 
Typical integration times are of 900-3600 sec,
depending on the surface brightness of the target. The spectra were calibrated using the
spectrophotometric standards Feige 34 and Hz 44 from the catalogue of Massey et al. (1988) observed twice on each night.

\begin{table}
\caption{Logbook of the observations}
\label{tab:Tablogbook}
{
\[
\begin{tabular}{cc}
\hline
\hline
\noalign{\smallskip}
\hline
Date		& N. of galaxies\\
\hline
17-25/4/2004	& 4	\\
4-10/3/2005	& 18	\\
11-22/4/2007	& 69	\\
3-9/3/2008 	& 23	\\
25-31/3/2009	& 24	\\
\noalign{\smallskip}
\hline
\end{tabular}
\]
}
\end{table}


The general properties of the target galaxies and the log-book of the observations are presented in Table \ref{tab:hrsgalaxies}. 
Each column contains:

\begin{itemize}
\item{Column 1: $Herschel$ Reference Sample (HRS) name, from Boselli et al. (2010a).}
\item Columns 2-6: Name as in the Catalogue of Galaxies and Clusters of Galaxies (CGCG; Zwicky et al. 1961-1968), Virgo Cluster Catalogue 
(VCC; Binggeli et al. 1985), Uppsala General Catalogue (UGC; Nilson 1973), New General Catalogue (NGC) or Index Catalogue (IC) (Dreyer 1888,
1908).
\item Column 7, 8: J2000 right ascension and declination, from NED.
\item Column 9: Distance in Mpc, determined assuming galaxies in Hubble flow outside the Virgo cluster (with $H_0$ = 70 km s$^{-1}$Mpc$^{-1}$)
or according to their membership to the different Virgo cluster substructures as indicated by Gavazzi et al. (1999) (17 Mpc for Virgo A and for 
all the other clouds with the exception of Virgo B, taken at 23 Mpc).
\item Column 10: Morphological classification, from NED. 
\item Column 11: Total 2MASS K-band magnitude (Jarrett et al. 2003; Skrutskie et al. 2006).
\item Column 12: Optical isophotal diameter $D_{25}$ (25 mag arcsec$^{-2}$), in arcmin, from NED.
\item Column 13: Heliocentric radial velocity, in km s$^{-1}$, from NED. 
\item Column 14: Cluster or cloud membership, from Gavazzi et al. (1999) for Virgo, and Tully (1988) or Nolthenius (1993) 
whenever available, or from our own estimate otherwise (Boselli et al. 2010a).
\item Column 15: Observing run.
\item Column 16: Photometric quality, classed as either photometric (P), transparent (T) or thin Cirrus (C). 
\item Column 17: Integration time (as number of exposures $\times$ individual exposure time). 
\end{itemize}

\section{Data reduction}

In order to form a homogeneous dataset with the existing observations of Gavazzi et al. (2004), we apply the same data reduction 
method to the new observations presented in this work. The reduction of the observations from the raw, two-dimensional images 
into calibrated, one-dimensional spectra is performed using standard tasks within the IRAF\footnote{IRAF is the Image Analysis and Reduction
Facility made available to the astronomical community by the National Optical Astronomy Observatories, which are operated by AURA, Inc., under
contract with the U.S. National Science Foundation. STSDAS is distributed by the Space Telescope Science Institute, which is operated by the
Association of Universities for Research in Astronomy (AURA), Inc., under NASA contract NAS 5-26555.} package. Bias subtraction is applied 
using the median of several bias frames and the median of exposures of quartz lamps is used for correcting the flat-field. 
Cosmic rays are removed by using \textsc{cosmicray}. Under visual inspection, remaining extended features are manually removed and 
any bad pixel is masked.

The wavelength calibration is carried out using \textsc{identify}, \textsc{reidentify} and \textsc{fitcoord} on combined 
exposures of helium and argon lamps, typically using 20-30 arc lines. The calibration solution is then applied to 
the science frames using \textsc{transform}. 
Using measurements of sky emission lines to test the accuracy of the calibration, we find the typical error in the 
solution being ~0.1 - 0.2 \AA. We manually correct the calibrations for systematic offsets, thus reducing the 
typical wavelength calibration errors to $\leq$ 0.5 \AA.

Subtraction of the sky background from the two-dimensional images is performed with 
the \textsc{background} task and then the one-dimensional spectra are extracted using \textsc{apsum} to define an aperture along 
which the signal is integrated.

The flux calibration to transform the measured intensities into flux densities is 
performed by the \textsc{standard}, \textsc{sensfunc} and \textsc{calibrate} tasks. The standard star frames, taken of Feige 34 and Hz 44 on 
each observing night, are used to determine the sensitivity function of the detector. The \textsc{standard} task integrates 
the standard star observations over calibration bandpasses and the measurements are corrected for atmospheric extinction using 
the reference extinction data in IRAF. The observational measurements from these bandpasses are then compared with 
the standard reference observations to determine the system sensitivity as a function of wavelength. The calibration 
factor used to convert from the measured intensities into flux densities at each wavelength is determined with \textsc{sensfunc}.


The \textsc{calibrate} task uses these sensitivity functions to convert the observed intensities at each wavelength in each spectra 
into flux densities, i.e. with units of erg cm$^{-2}$ s$^{-1}$ \AA$^{-1}$, corrected for atmospheric extinction. 
Finally, each spectra is transformed to the rest frame wavelength. Two template spectra are chosen to represent absorption-line 
objects or emission-line objects, and these templates are converted to their rest frame wavelength. The \textsc{fxcor} task cross-correlates 
the template spectrum to the remaining spectra of each object type, thus determining the relative shift for each spectra. 
This shift is then applied with \textsc{dopcor}. Due to the manual drifting of the telescope during the observations, however, 
we are in the impossibility of reconstructing the absolute flux within a given aperture of the observed galaxies.
For this reason we normalise all spectra to their mean intensity between $\lambda = 5400 - 5600$ \AA. 
The typical signal to noise of the reduced spectra, measured using the \textsc{DER\_SNR} package\footnote{\url{http://www.stecf.org/software/ASTROsoft/DER_SNR/}}, 
ranges in between S/N $\sim$ 15 and 70, but is significantly lower in the blue range ($\lambda$ $\leq$ 4000 \AA:
S/N $\sim$ 5-20) because of the low sensitivity of the EEV CCD.
The observed spectra of all HRS galaxies, including those with data available in the literature, 
are given in Fig. \ref{spectra} in order of increasing HRS name.

  \begin{figure*}
   \centering
   \includegraphics[width=0.3\textwidth]{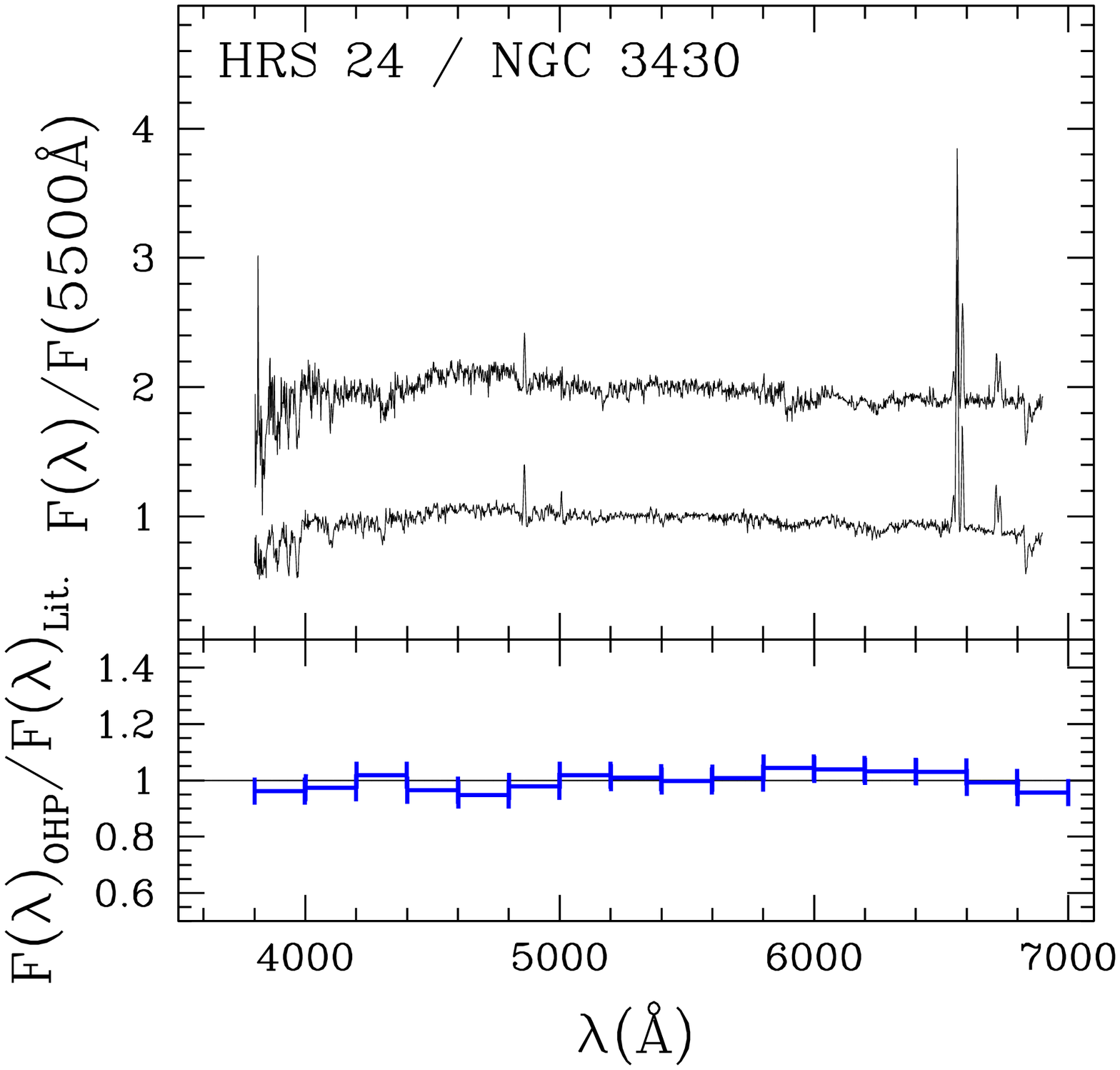}
   \includegraphics[width=0.3\textwidth]{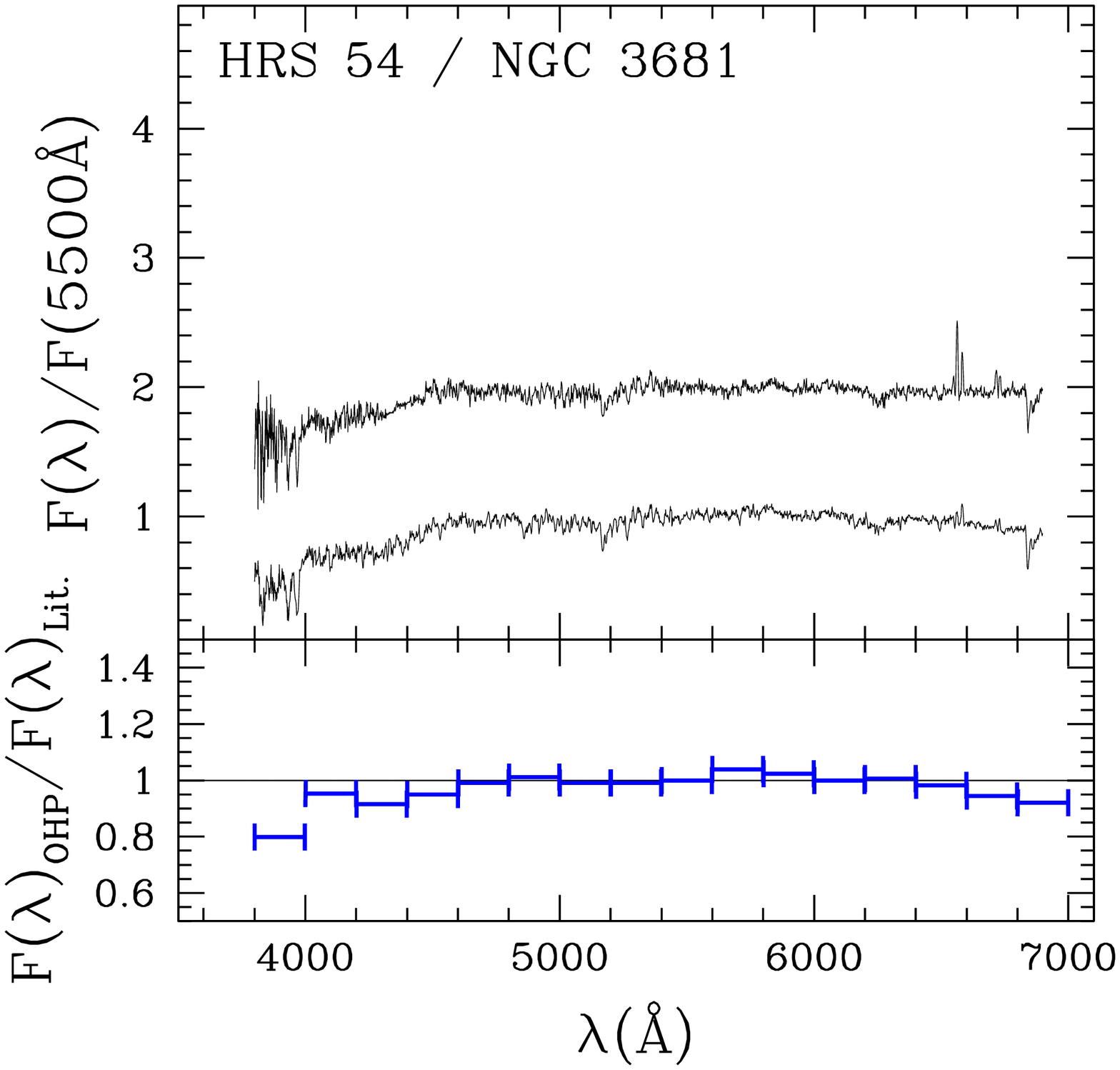}
   \includegraphics[width=0.3\textwidth]{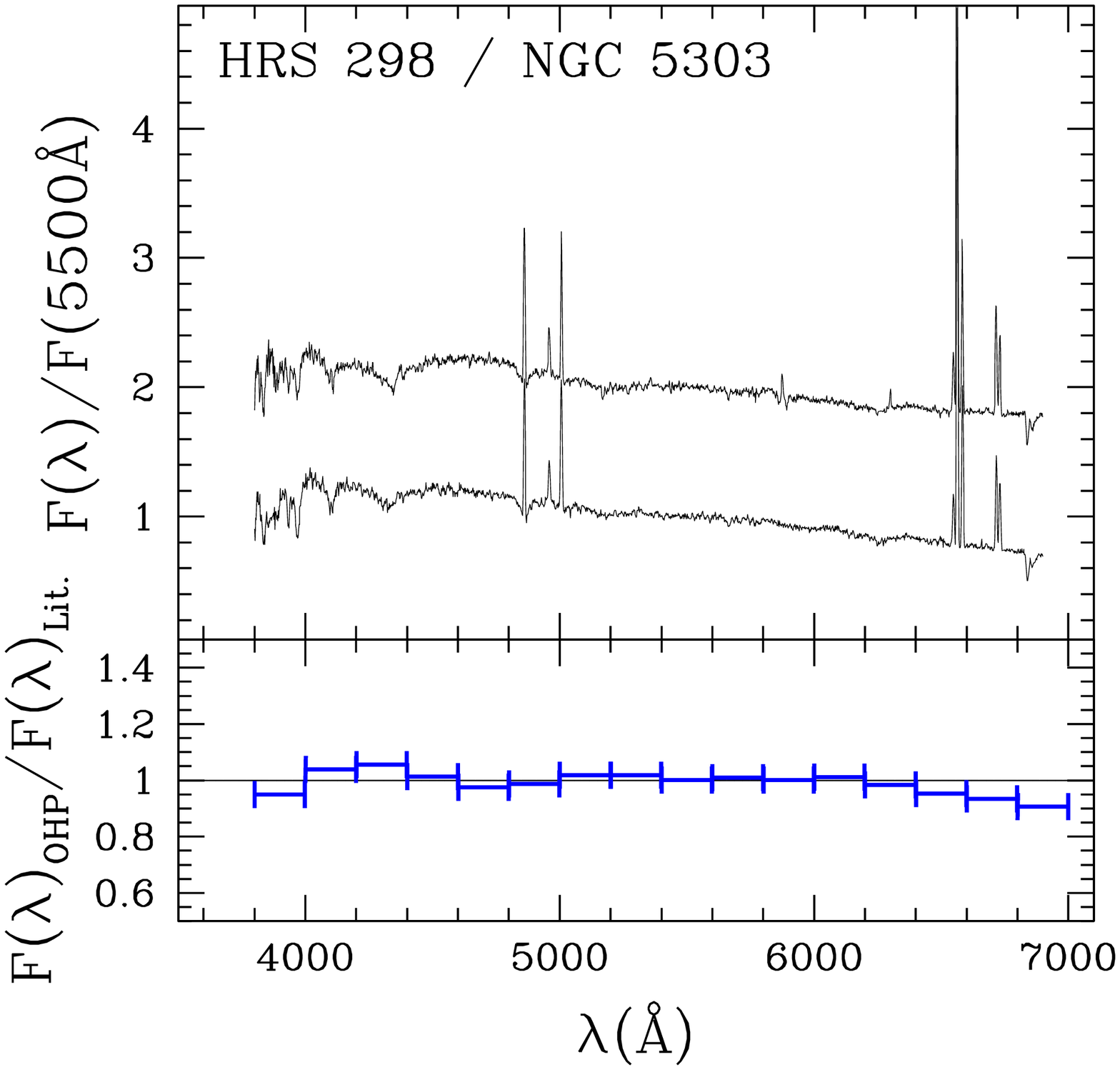}\\
   \caption{Comparison between the spectra of HRS 24, 54 and 298 observed multiple times between 
   2007 and 2009 (upper and lower spectra, respectively). The lower panel shows the ratio of the two
   measurements in 200 \AA ~ wide bins (blue line).}
   \label{compspecus}%
   \end{figure*}

\begin{figure*}
\centering
\includegraphics[width=0.3\textwidth]{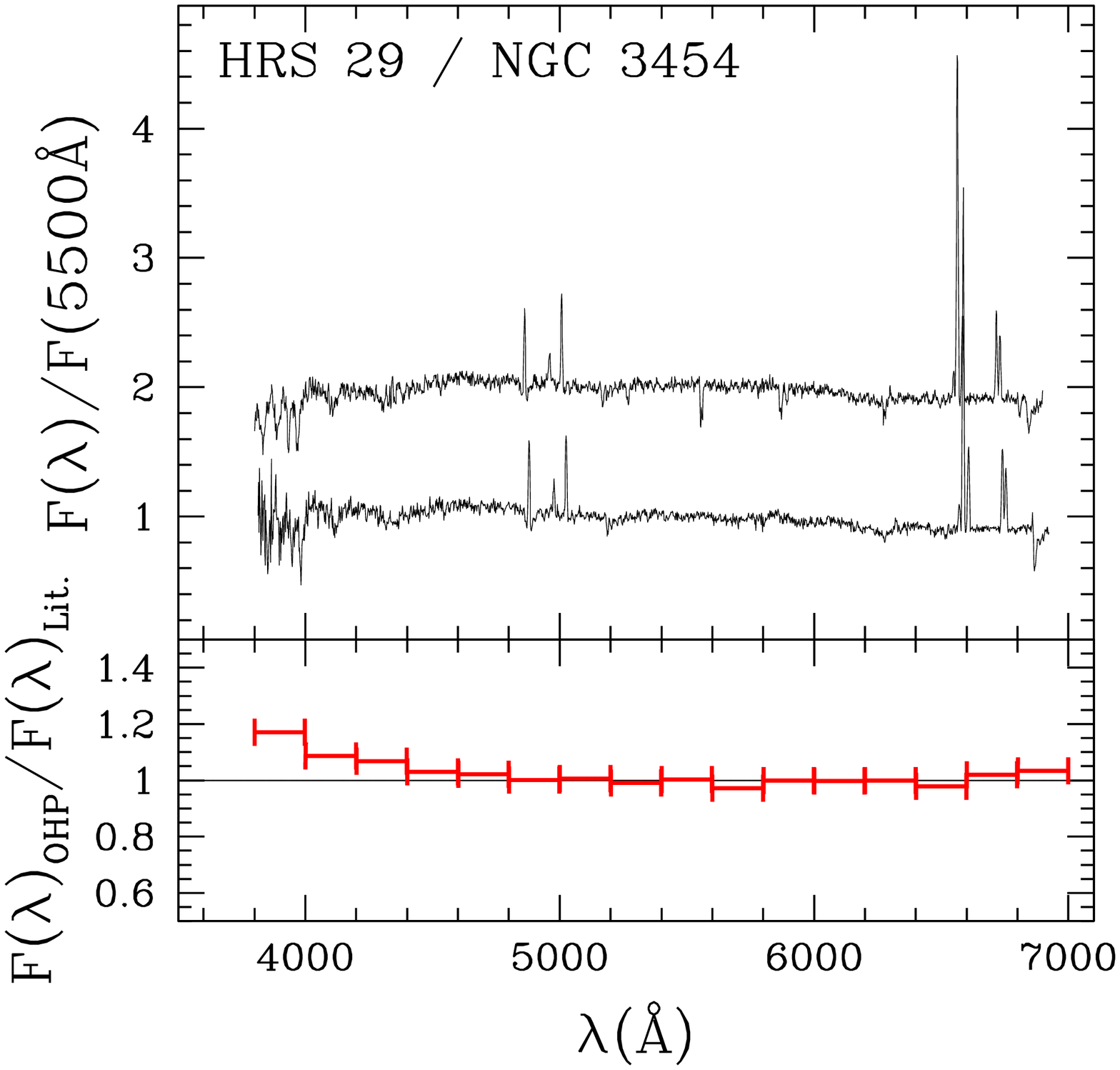}
\includegraphics[width=0.3\textwidth]{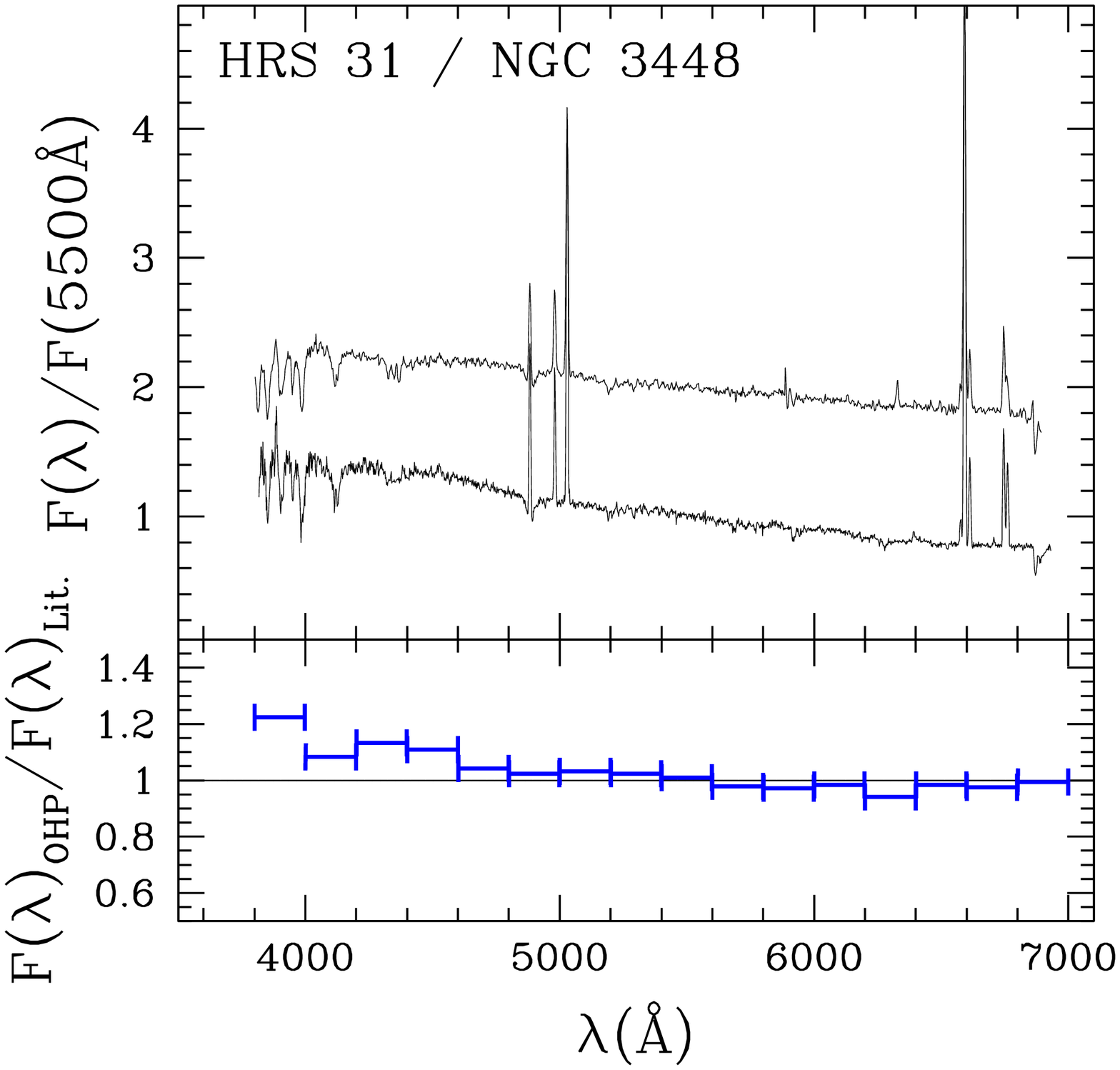}
\includegraphics[width=0.3\textwidth]{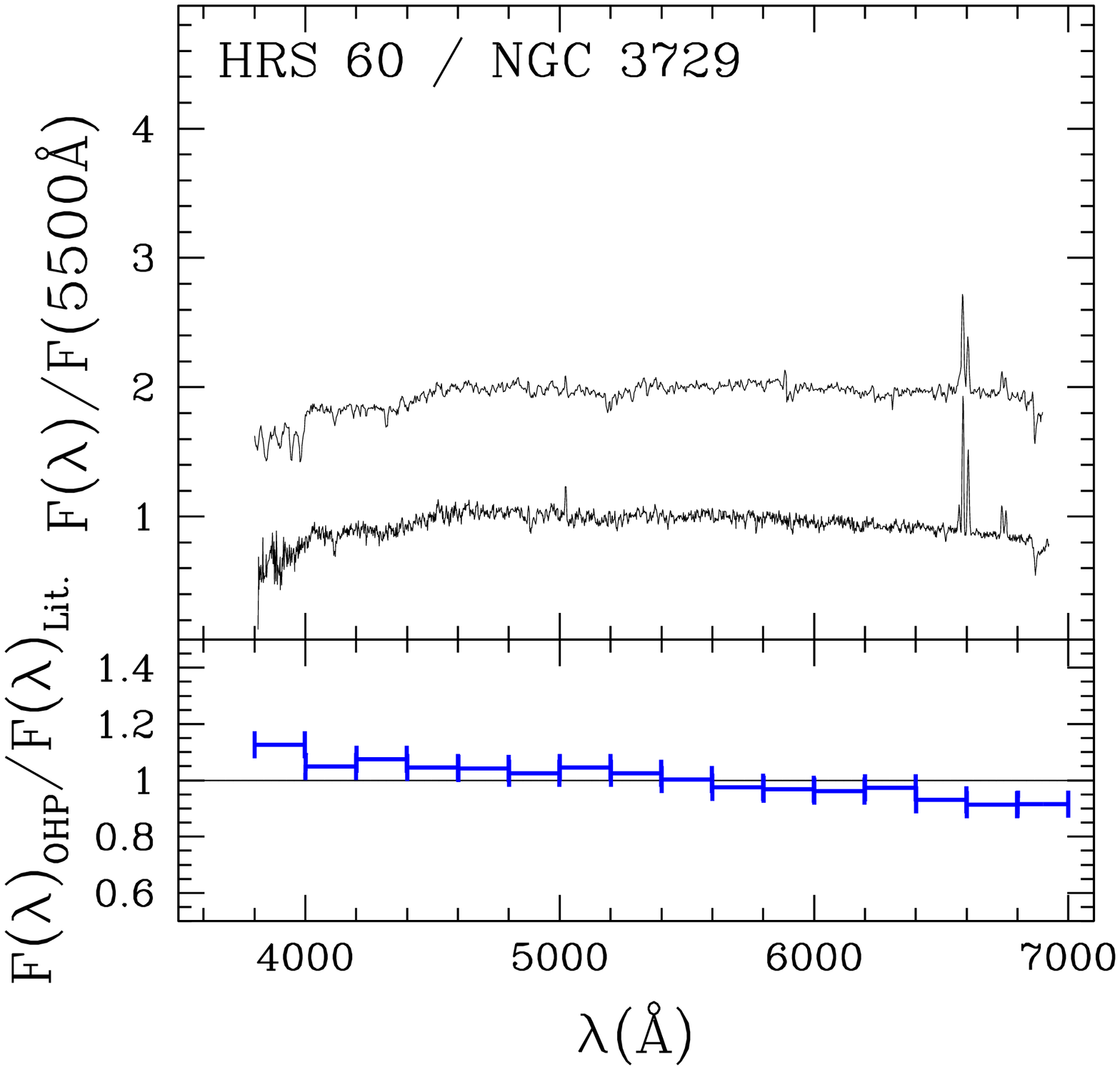}\\
\includegraphics[width=0.3\textwidth]{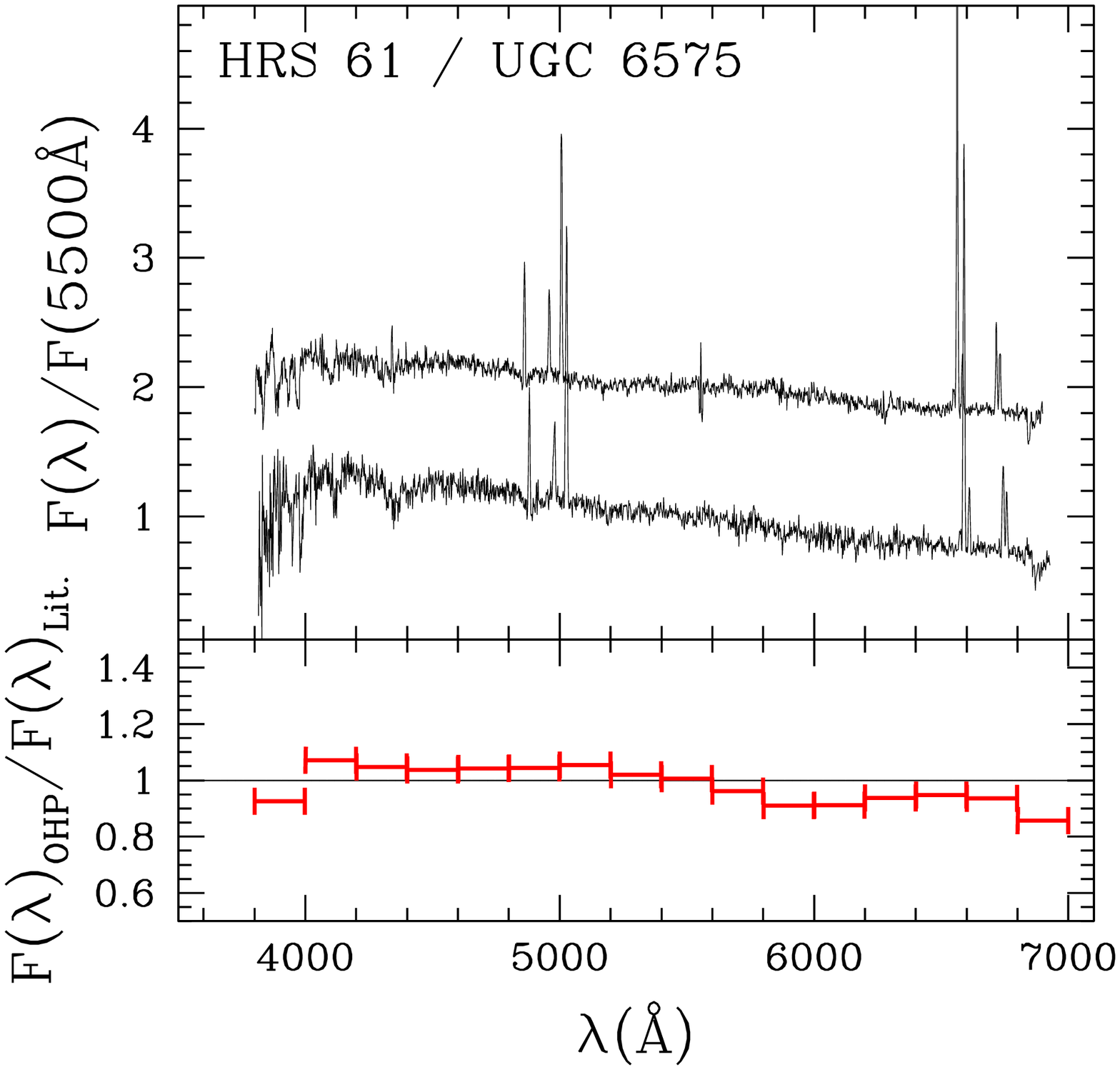}
\includegraphics[width=0.3\textwidth]{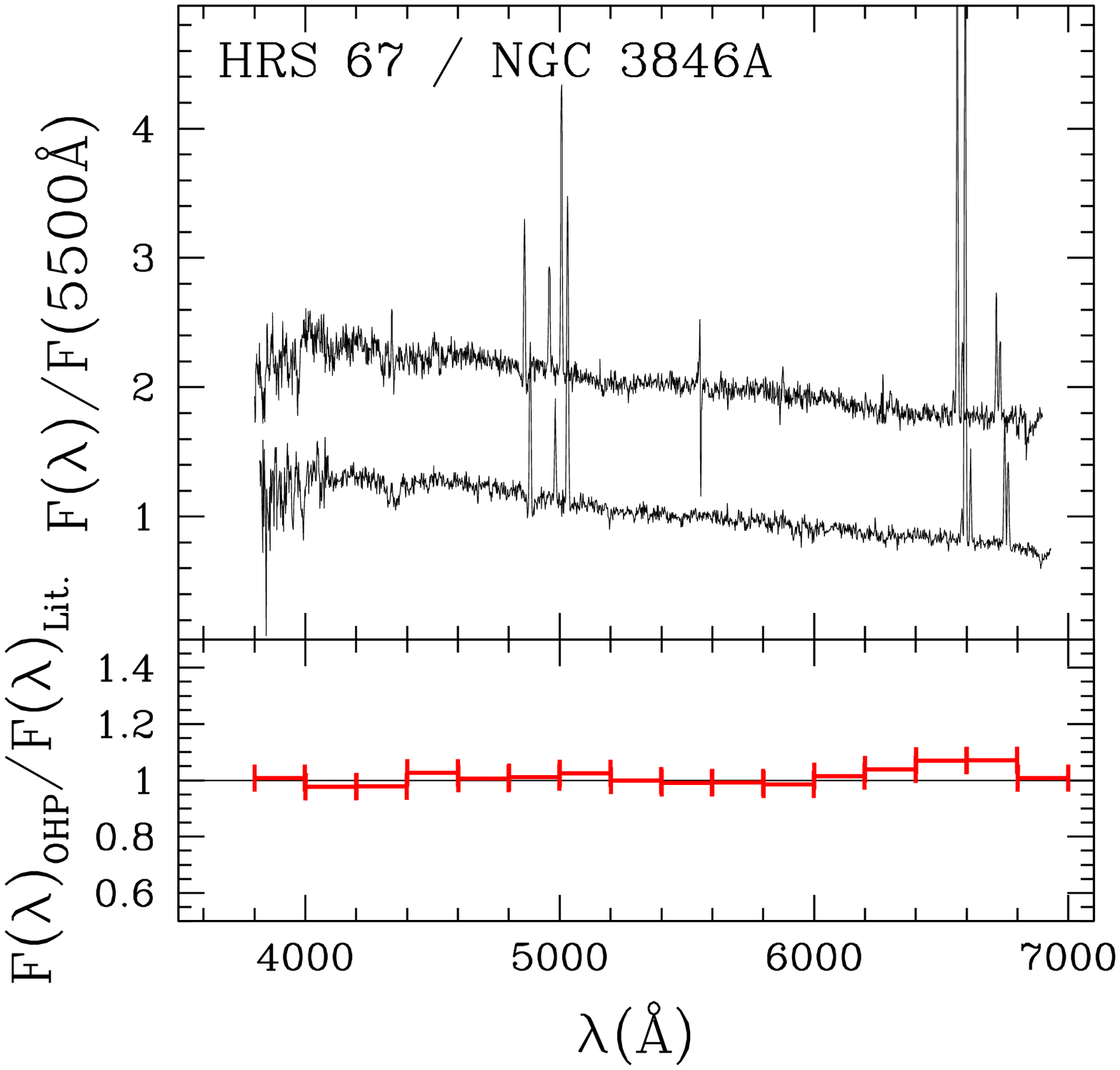}
\includegraphics[width=0.3\textwidth]{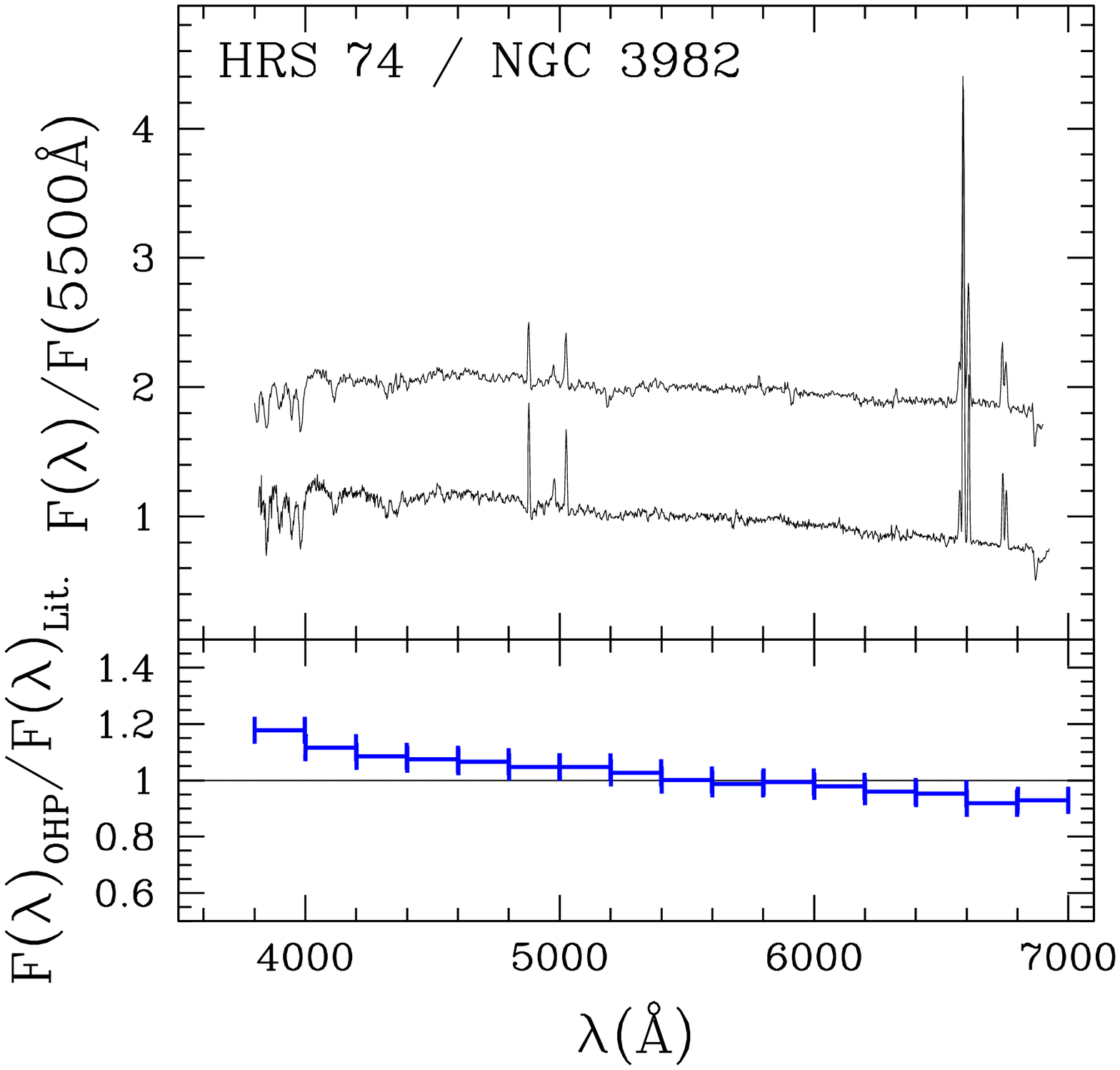}\\
\includegraphics[width=0.3\textwidth]{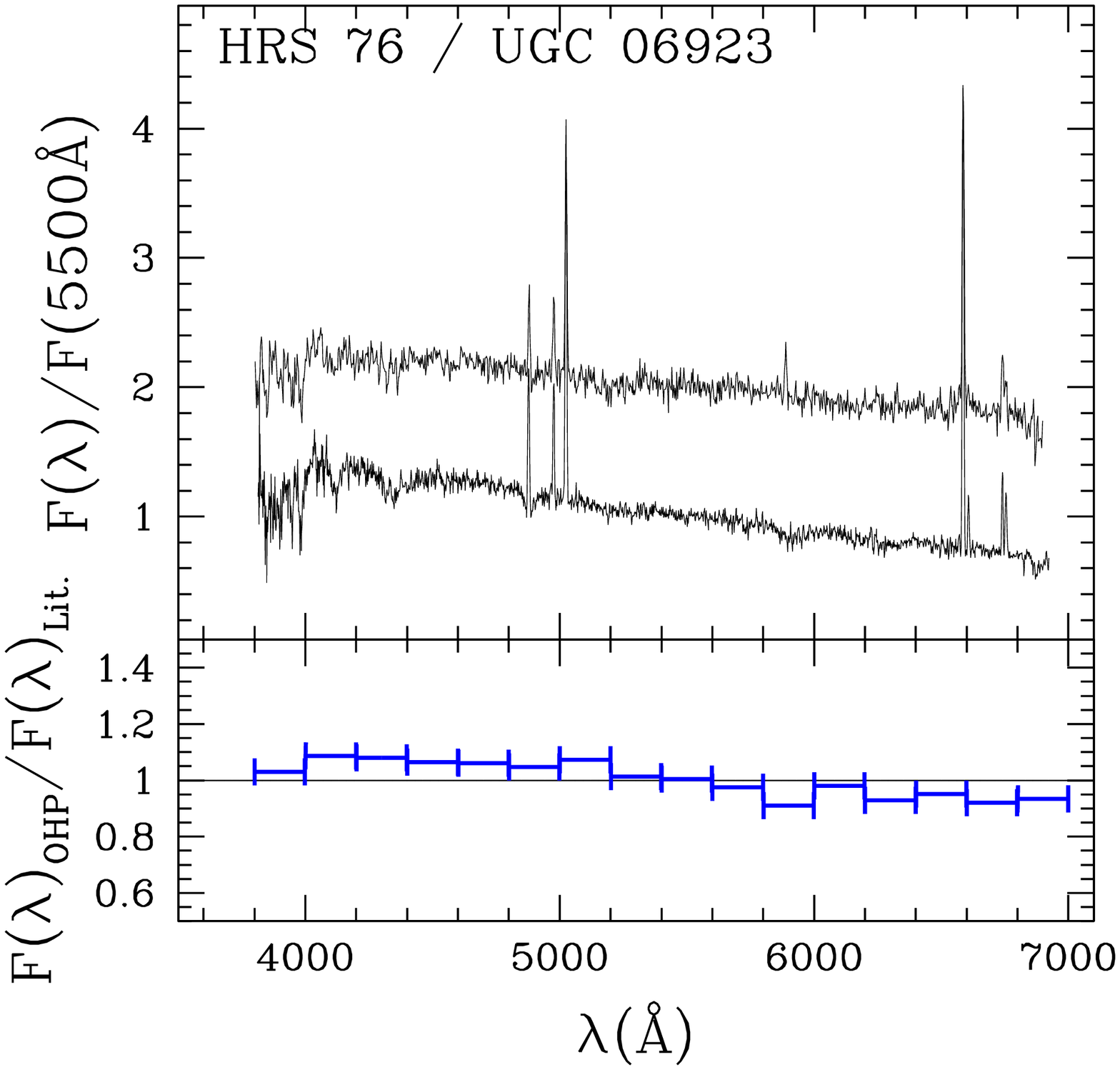}
\includegraphics[width=0.3\textwidth]{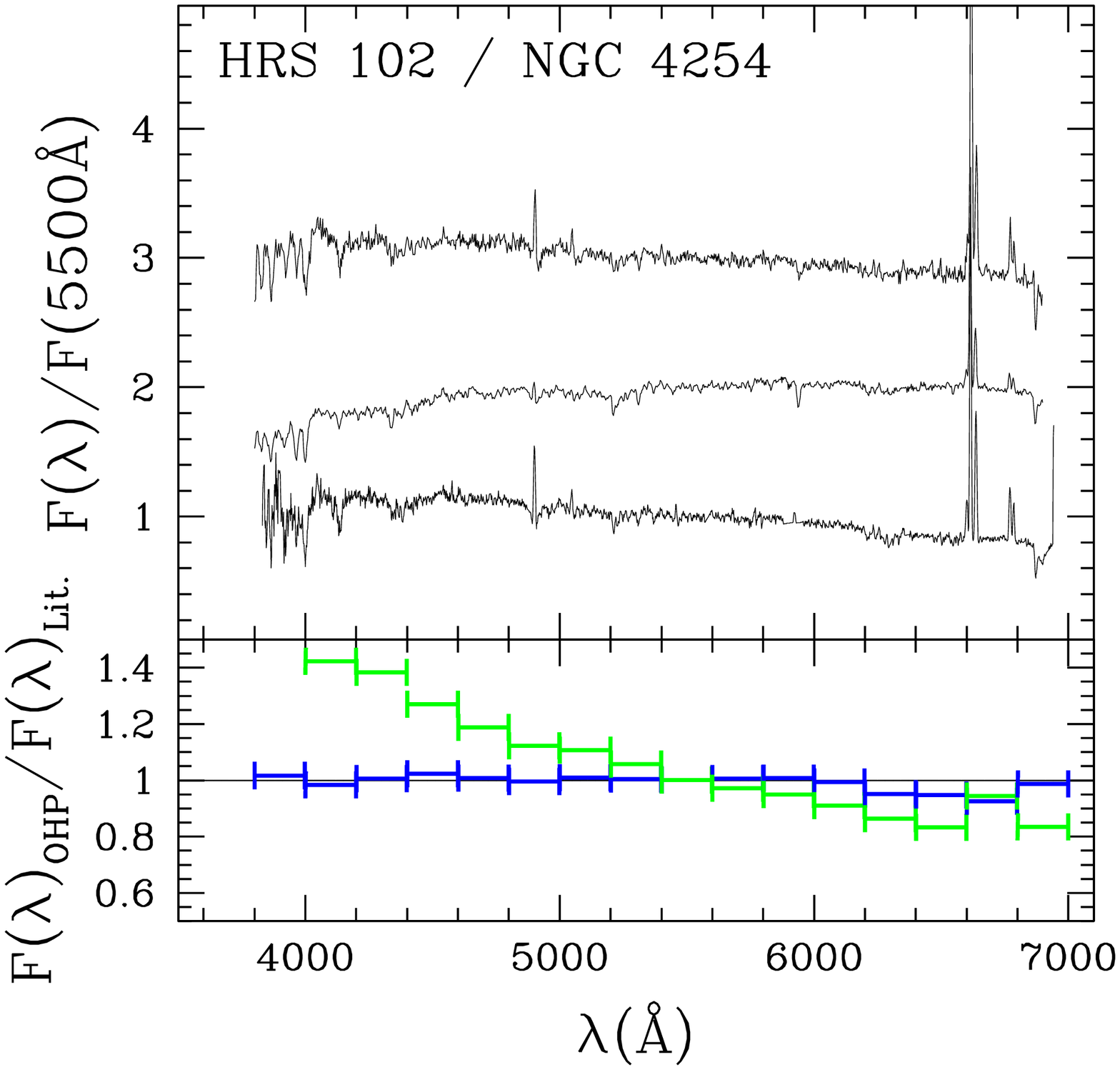}
\includegraphics[width=0.3\textwidth]{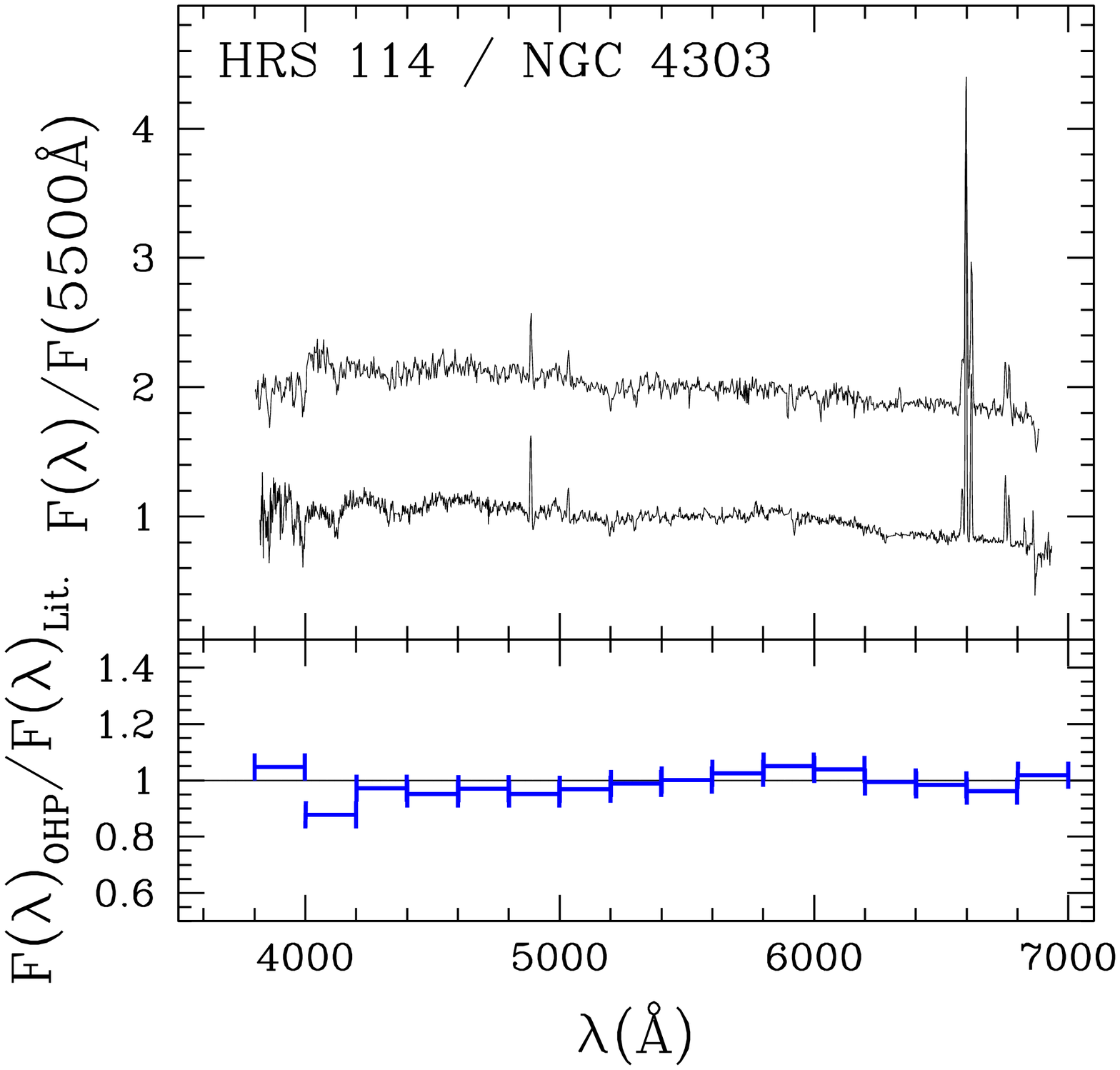}\\
\includegraphics[width=0.3\textwidth]{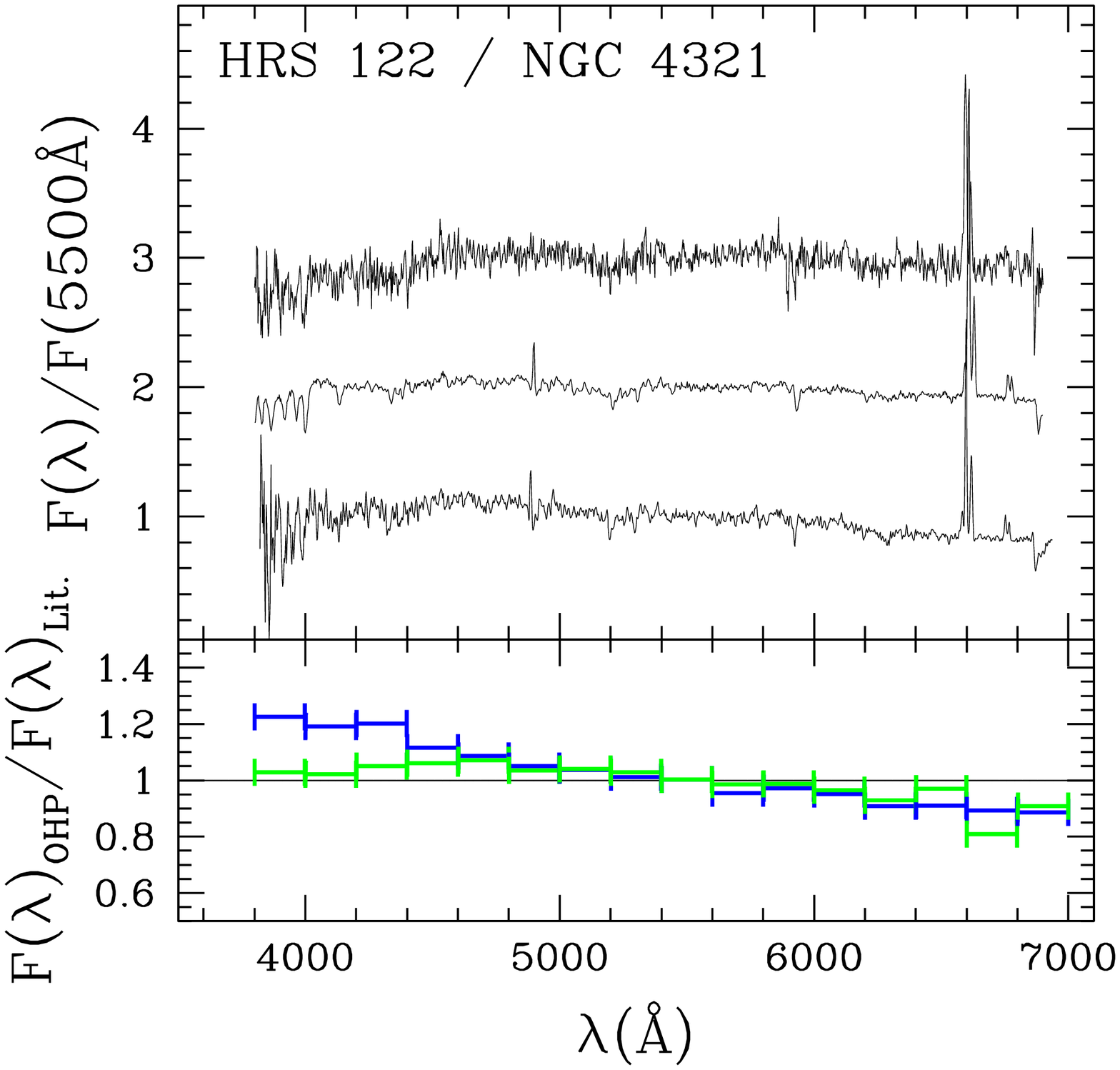}
\includegraphics[width=0.3\textwidth]{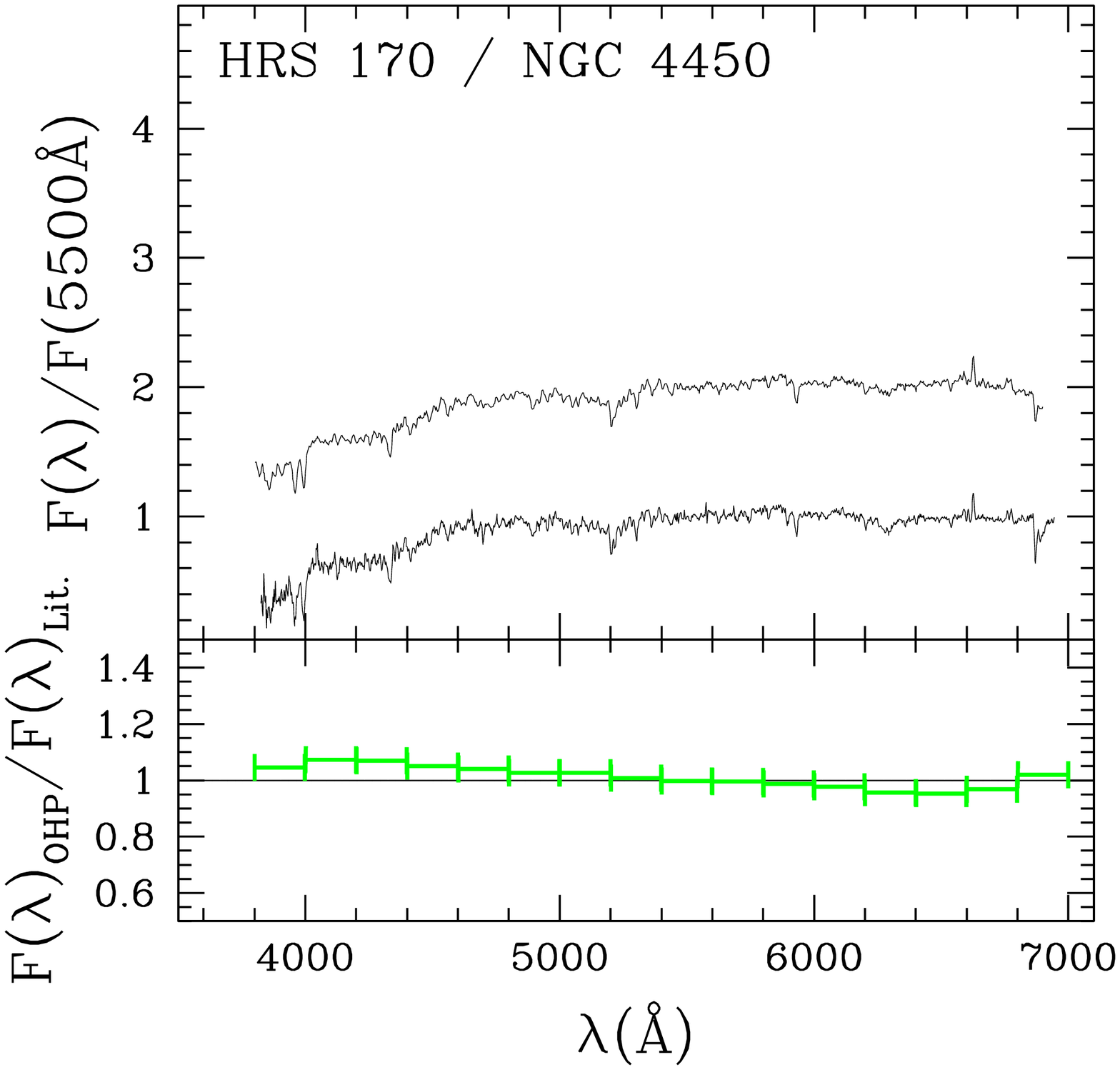}
\includegraphics[width=0.3\textwidth]{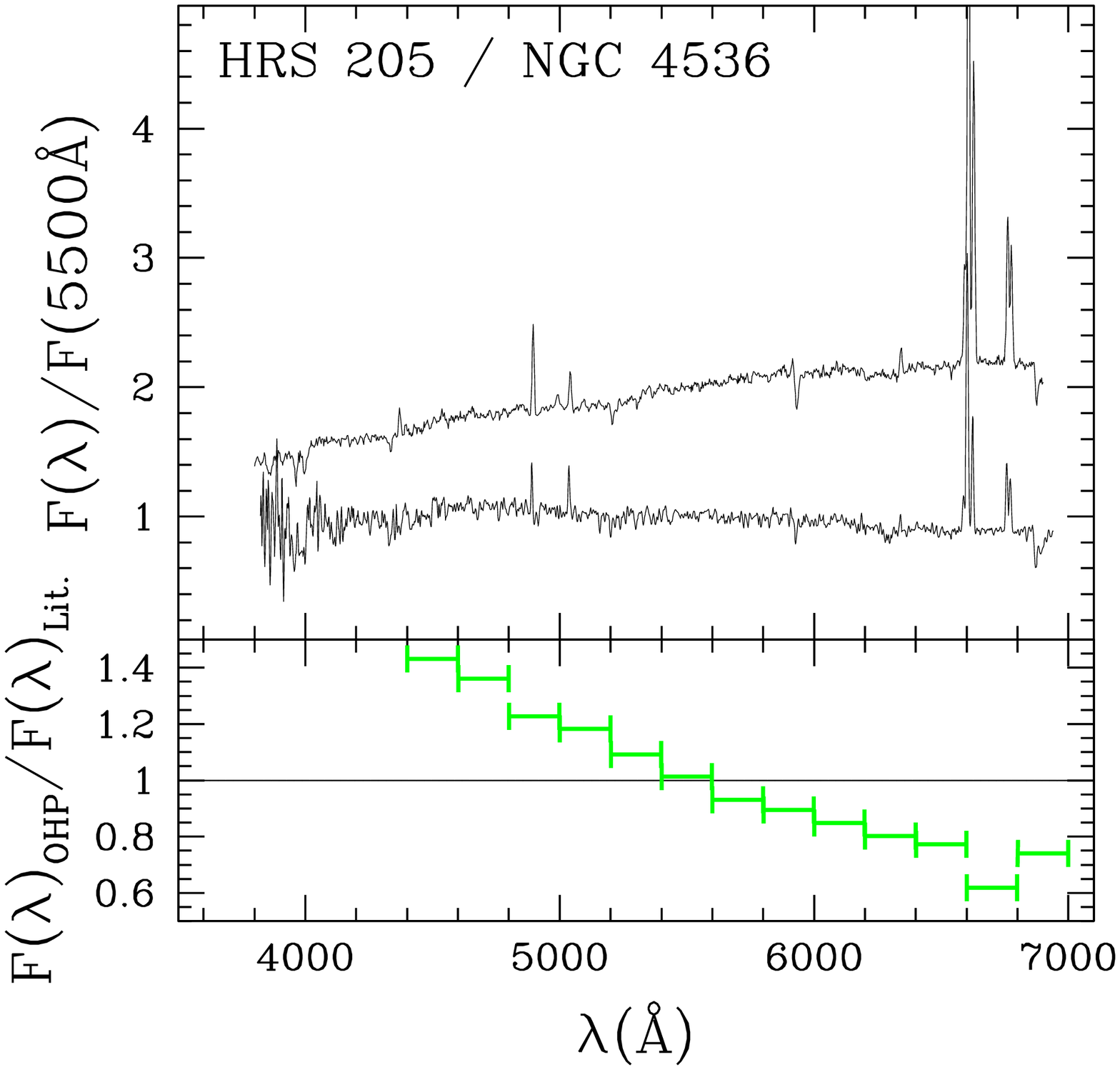}\\
\caption{Comparison between the spectra of galaxies observed in this work or in Gavazzi et al. (2004) (lower spectra) and 
with data available in the literature (upper spectra). The ratio of the different spectra is given in the lower panel, in bins 200 \AA ~ wide.
Different colour codes are used when data are from Moustakas et al. (2010) (green),
Moustakas \& Kennicutt (2006) (blue) and Jansen et al. (2000) (red). Whenever two independent 
spectra are available in the literature (HRS 102, 122 and 217), they are given from top to bottom 
following the order Moustakas \& Kennicutt (2006), Moustakas et al. (2010), this work or Gavazzi et al. (2004).
   }
   \label{spectracomplit}%
   \end{figure*}
\clearpage

 \addtocounter{figure}{-1}
   \begin{figure*}
   \centering   
\includegraphics[width=0.3\textwidth]{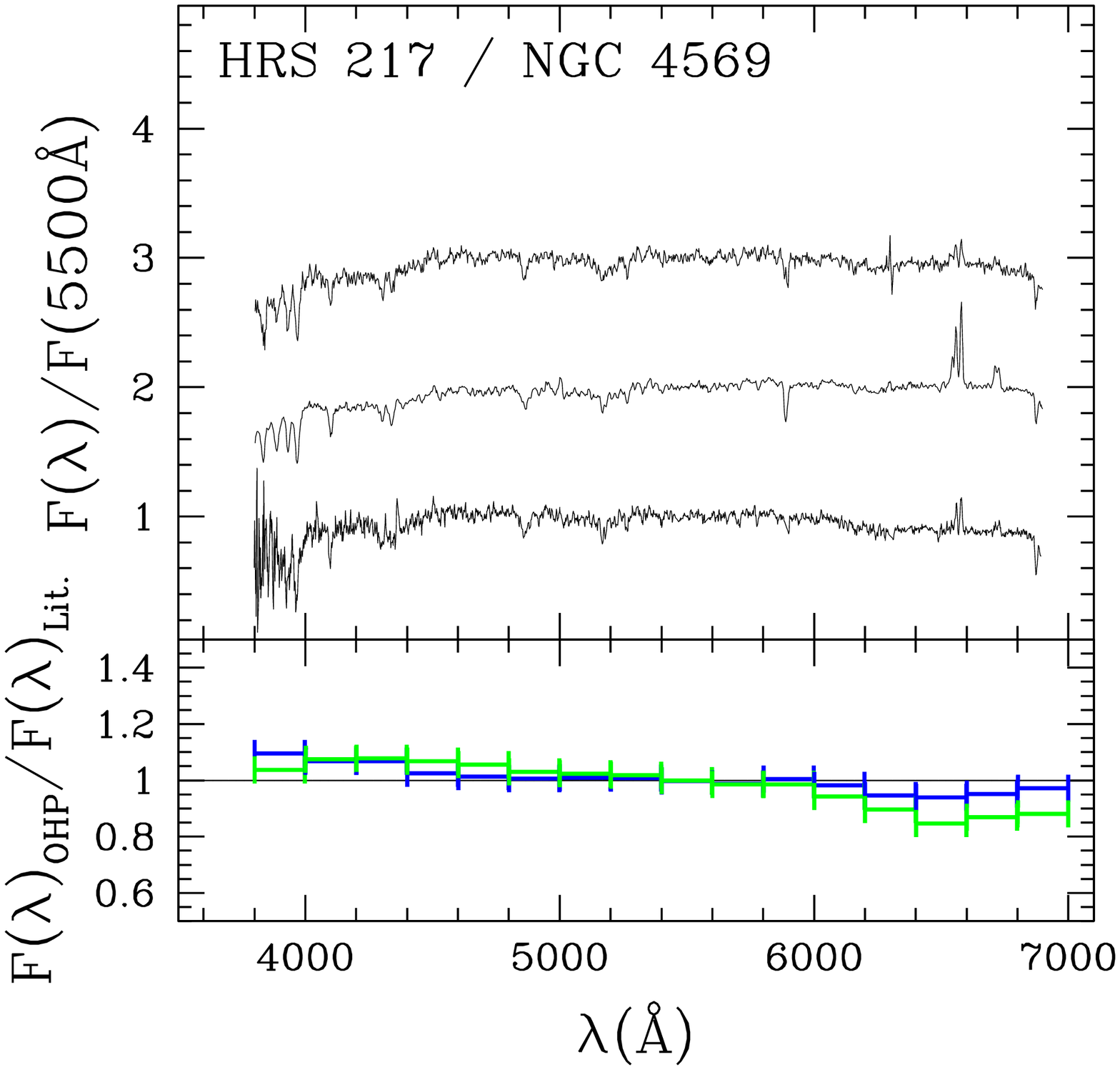}
\includegraphics[width=0.3\textwidth]{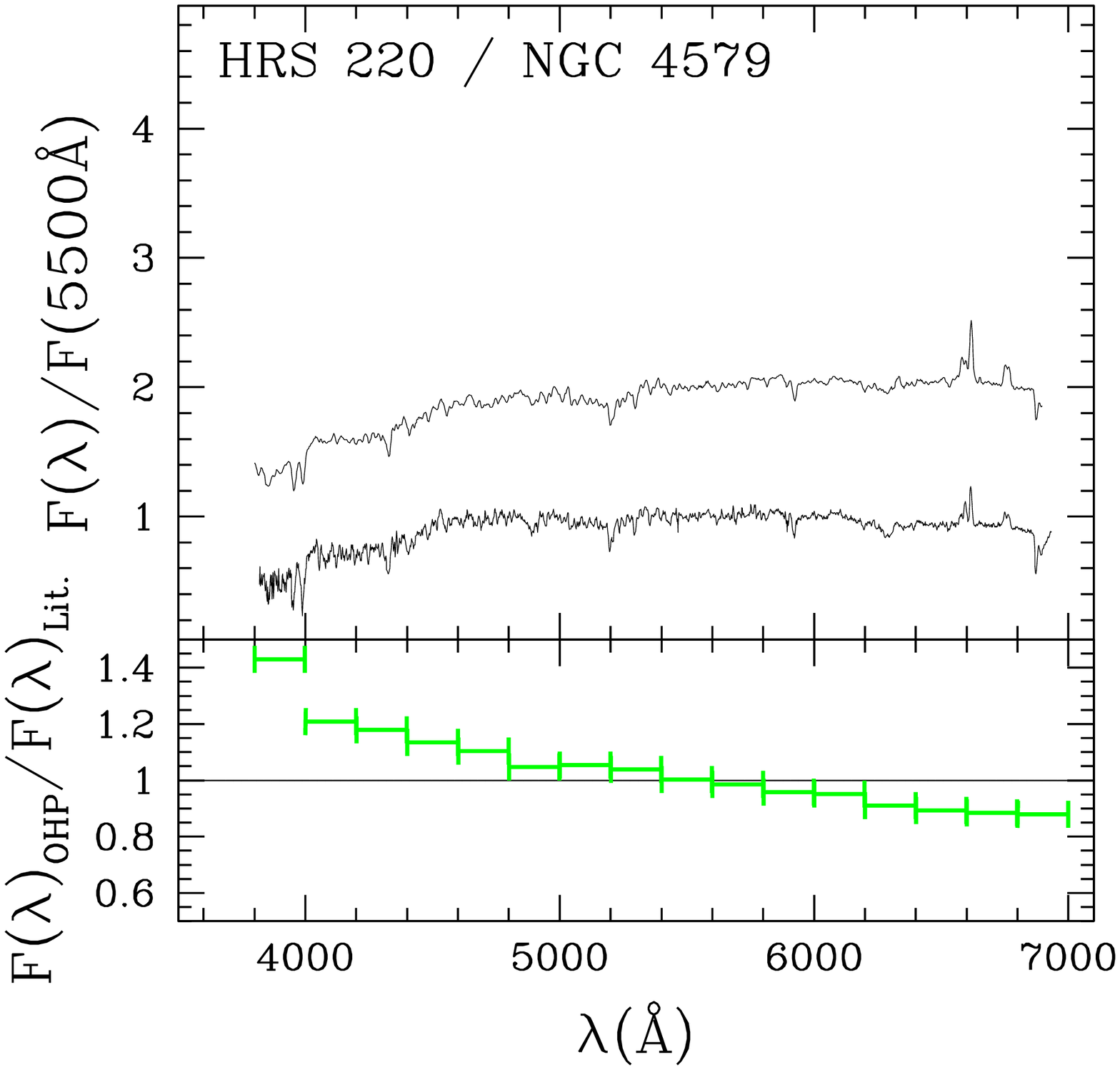}
\includegraphics[width=0.3\textwidth]{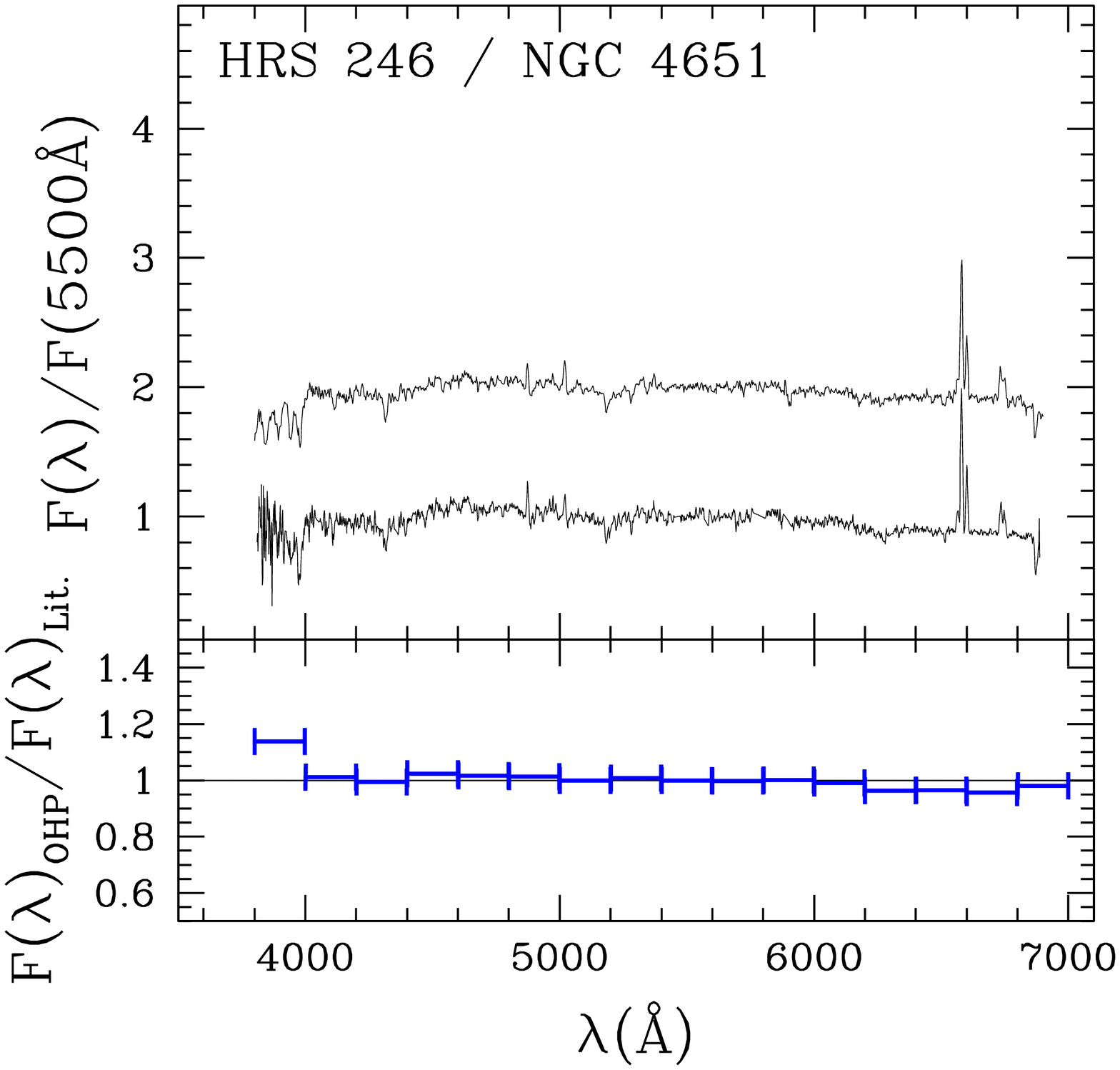}
\includegraphics[width=0.3\textwidth]{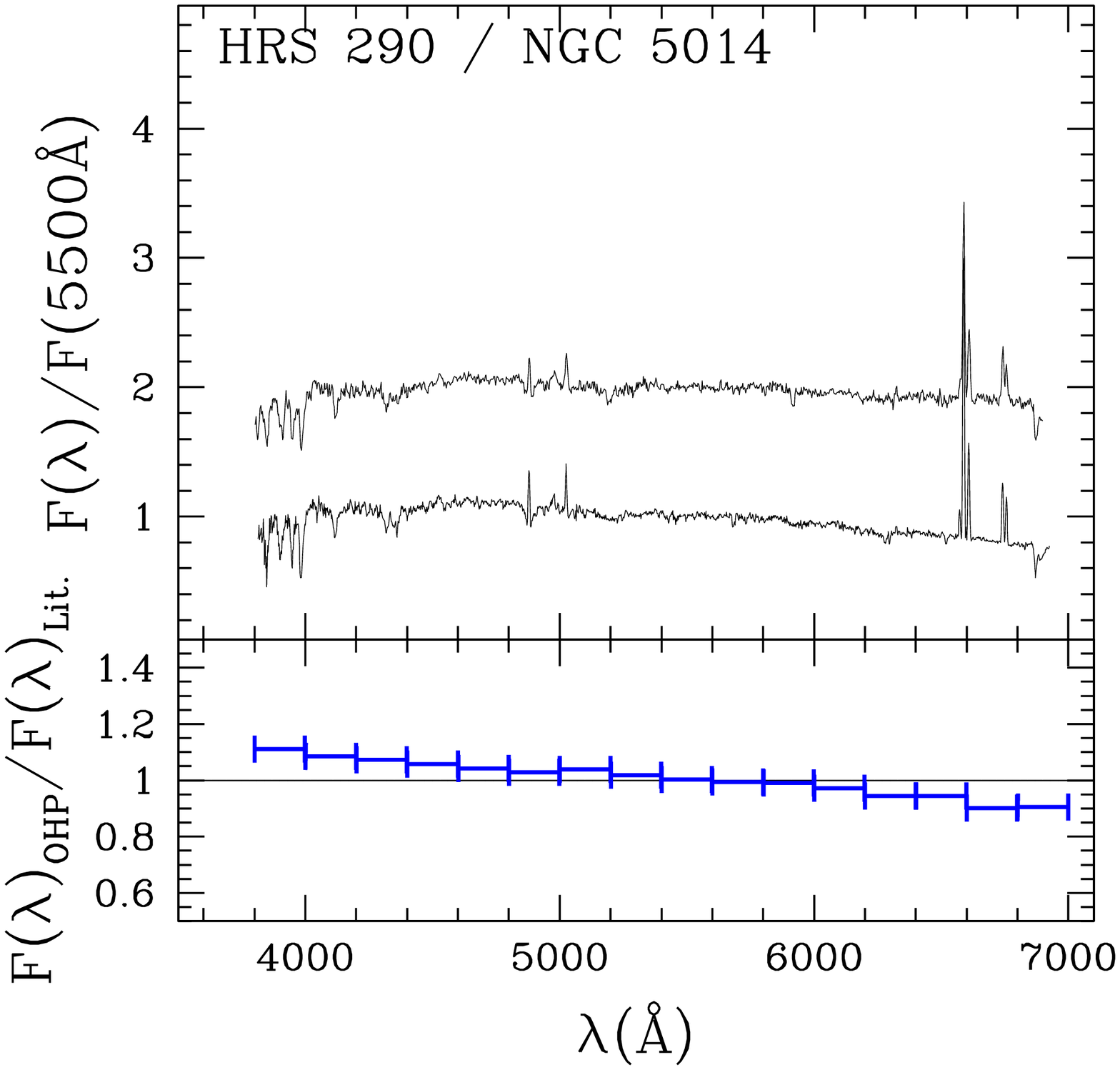}\\
\caption{Continued.
   }
   \end{figure*} \clearpage

  \begin{figure}
   \centering
   \includegraphics[width=8cm]{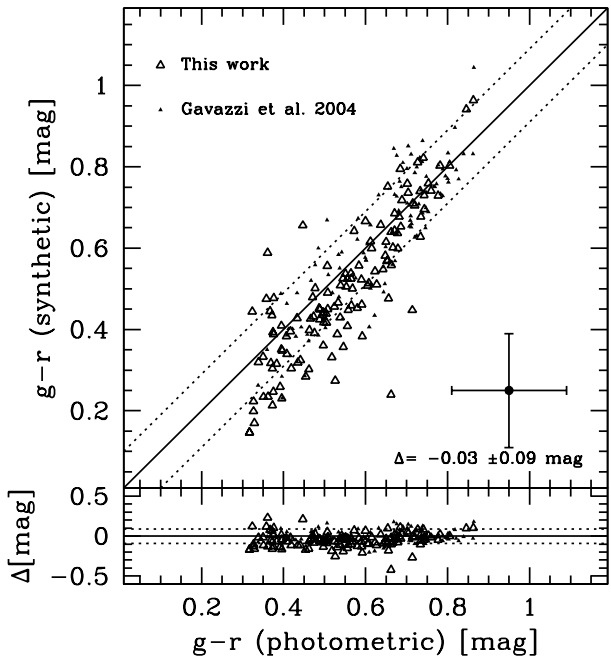}
   \caption{Comparison between the photometric and the
spectroscopic $g-r$ colours for galaxies observed in this work (empty triangles) and in Gavazzi et al. (2004) (filled triangles). 
The solid line gives the 1:1 relationship and the dotted lines the 1 sigma uncertainty. The cross indicates a typical error bar 
of 0.14 mag on the colour, corresponding to $\sim$ 10\% uncertainty in each band.}
   \label{synthetic}%
   \end{figure}

\subsection{Spectrophotometric accuracy}

We first test the spectrophotometric quality of our spectra
by comparing multiple observations of the same galaxy
both from our survey and the literature. Over the course of
our observing campaign, only three objects in our sample
have been observed two or more times using the drift-scan
technique and spectra were derived using the same reduction method.
We normalize each pair of spectra to the mean flux around 5500 \AA ~ and
calculate the flux ratio in 200 \AA ~ wide bins, in order to check
for any variation in continuum shape. In Fig. \ref{compspecus}, we compare the multiple observations of these
three galaxies. We find that the relative scatter in the
observations is 3.2\% with a slight wavelength dependence (see Table \ref{Tabspeclit}).
The relative scatter of the red portion ($\lambda$ $>$ 5500\AA) 
of the spectrum is around 3.1\%, and this increases to 3.6\% 
at the blue end ($\lambda$ $<$ 5500\AA). The largest discrepancy occurs at $\lambda$ $<$ 4000 \AA,
where the difference in one object reaches $\sim$ 20 \%. In the red ($\lambda$ $>$ 6800 \AA)
the differences are $\lesssim$ 15\%.
Although we are limited 
by the low number of objects for comparison, it appears that 
the largest discrepancies occurs in a galaxy observed in cirrus conditions (HRS 54). 

Ten galaxies observed in this survey and six in Gavazzi
et al. (2004)  have integrated spectroscopy data available
in the literature from Jansen et al. (2000), Moustakas \&
Kennicutt (2006) and Moustakas et al. (2010). We compare our
results in Fig. \ref{spectracomplit}. Despite various differences
in how these sets of spectra were obtained, we find an
average relative difference of 7.0\% with some wavelength
dependence present for some objects (see Table \ref{Tabspeclit}). The relative scatter 
of the red portion ($\lambda$ $>$ 5500\AA) 
of the spectrum is around 5.7\%, and it increases to 8.9\% 
at the blue end ($\lambda$ $<$ 5500\AA).

\begin{table*}
\caption{Comparison with spectra available in the literature or with multiple observations}
\label{Tabspeclit}
{
\[
\begin{tabular}{ccccccc}
\hline
\noalign{\smallskip}
\hline
Spectral~range	& All   &J00+MK06	& J00 	& MK06	& M10 	& Repeated\\
\hline
3700-7000 \AA	& 7.0\%	& 4.4\%		& 3.7\%	& 4.6\%	& 12.5\% & 3.2\%	\\
3700-5500 \AA	& 8.9\%	& 5.2\%		& 3.9\%	& 5.7\%	& 16.9\% & 3.1\%	\\
5500-7000 \AA	& 5.7\%	& 4.1\%		& 4.1\%	& 4.1\%	& 9.1\%  & 3.6\%	\\
\noalign{\smallskip}
\hline
\end{tabular}
\]
}
\end{table*}

\noindent
The Moustakas et al. (2010) spectra are typically redder than 
our spectra. The differences between our and the Moustakas et al. (2010) spectra 
reach $\sim$ 60 \% ~ at $\lambda$ $\sim$ 4000 \AA ~ and 35 \% ~ at $\lambda$ $\sim$ 7000 \AA ~ in HRS 205, 
while is generally $\lesssim$ 20 \% ~ and $\lesssim$ 15 \% ~ in the same spectral domain when our data are compared to Moustakas \& Kennicutt (2006)
and Jansen et al. (2000).
This large difference might result from the fact that the 
Moustakas et al. (2010) spectra are limited to the radial strip 
observed by Spitzer within the SINGS project. These strips cover the inner regions and are thus significantly
less extended than the optical discs of the observed galaxies. Because of the colour gradient, these inner spectra are redder, as
clearly evident in NGC 4254 (HRS 102). Here, our and the Moustakas \& Kennicutt (2006)
spectra, encompassing the whole galaxy disc, are consistent within 5\% at all wavelengths, 
and are both significantly bluer than the one of Moustakas et al. (2010) which is limited to the inner disc.
If we limit the comparison of our spectra with those published in Moustakas \& Kennicutt (2006) and Jansen et al. (2000),
the difference in the spectra reduces to $\sim$ 4.4\%, with 5.2\% in the blue and 4.1\% in the red. 

Contrary to what is found in the comparison of multiple observations of the same galaxies done
within this survey (see above), we do not see any strong systematic difference between galaxies observed
in photometric (HRS 29, 67, 102, 114, 122, 217; mean difference 3.8\%) or cirrus conditions (HRS 246; 2.3\%) 
also when the comparison is limited to $\lambda$ $<$ 4000 \AA ~ and $\lambda$ $>$ 6800 \AA \footnote{This comparison is done 
excluding the Moustakas et al. (2010) data.}.

We further test our spectrophotometric accuracy by comparing magnitudes 
synthesized from our spectra to the recently published broadband photometry from Cortese et al. (2012a).
Synthesized spectroscopic $g-r$ colours were obtained by deconvolving the continua with the profiles of the $g$ and $r$
SDSS filters. Figure \ref{synthetic} shows the comparison between the
photometric and the spectroscopic $g-r$ colours
synthesized for all the HRS galaxies with
available photometric data. The mean offset is 0.03 mag with
a rms scatter of 0.09 mag. Overall the agreement between the two sets of data 
is good, considering that the error on the photometric magnitudes is of the order of $\sim$ 10 \%.
A systematic difference is however present for the bluest galaxies, 
where the synthetic colours are bluer than the photometric ones by  $\lesssim$ 0.1 mag. 
To summarise, we find that the errors on the continuum in our data are of $\lesssim$ 20\% for $\lambda$ $<$ 4000 \AA, $\lesssim$ 15\% for $\lambda$ $>$ 6800 \AA,
and $\lesssim$ 10\% elsewhere. These uncertainties are
consistent with those previously 
reported in the literature (see Gavazzi et al. 2004; Moustakas \& Kennicutt 2006).

\section{Emission line flux measurements} 

\subsection{Contribution of the underlying Balmer absorption}

We measure the emission lines of each spectra by visually inspecting them using \textsc{splot}, and obtain a measurement of the relative flux and 
equivalent width for the detected emission lines. The typical signal-to-noise of the emission lines, measured as the ratio of the peak flux of the line and
the sigma measured for the continuum either sides of the line, is
S/N $\gtrsim$ 20 for \ha, S/N $\sim$ 3-12 for the other lines with exception of the [\oii] lines where S/N $\sim$ 2-8 due to the low sensitivity of
the CCD shortward of 4000 \AA. 
The \hb \ line often displays underlying stellar absorption which must be corrected for, 
otherwise the measured flux of the emission line is underestimated (see Fig. 8 of Gavazzi et al. 2004). The underlying absorption 
also affects the other main Balmer line, \ha.\\
Different techniques have been proposed in the literature for removing the contribution of the underlying Balmer absorption to the
emission lines. Moustakas \& Kennicutt (2006) and Moustakas et al. (2010) proposed to fit the observed 
spectra of galaxies with population synthesis models using different star
formation histories and than subtract them from the observed spectra to obtain pure line emission spectra.
Their analysis have shown that, because of quite different star formation histories, the equivalent
width of the \hb ~ underlying absorption of star forming galaxies 
can vary in between 3.9 and 5.9 \AA. Using this technique, they estimate that the mean underlying Balmer absorption in their sample of galaxies 
is E.W.H$\beta_{abs}$ = 4.4 $\pm$ 0.63 \AA ~ and E.W.H$\alpha_{abs}$ = 2.80 $\pm$ 0.38 \AA, respectively (Moustakas \& Kennicutt 2006). \\
Our spectra are unfortunately quite noisy in the blue range, shortward of $\sim$ 4000 \AA. 
Furthermore the HRS sample includes bright, massive galaxies chracterised by a high surface brightness as well as relatively low luminosity,
low surface brightness objects. The set of spectroscopic data in our hand is thus quite non homogeneous in terms of S/N.
Since the accuracy of the fit depends on the S/N (Oh et al. 2011),
these spectra can be hardly used without other photometric data to constrain the star formation 
history of the target galaxies and measure the underlying Balmer absorption using population synthesis models.\\
Moreover, the fitting procedures using population synthesis models have also their own limits. Groves et al. (2012) have indeed shown that, for a fixed 
age of the underlying stellar population, different population synthesis models give
differences in the H$\beta$ and H$\alpha$ Balmer absorption lines up to $\sim$ 2-3 \AA. There are also
indications that the use of different fitting procedures can lead to different measurements of line emission, in particular 
whenever the emission is weak (Oh et al. 2011).  For these reasons we prefer to adopt a simpler approach
by directly measuring the underlying absorption in the H$\beta$ line from the spectra, and use a constant 
correction for the H$\alpha$ line.

Indeed the spectral resolution and the sensitivity of our data allow us to directly measure the underlying absorption under \hb \ in most of the 
target galaxies (181/238). The comparison of direct measurements on the spectra with those determined using population sysnthesis models 
on SDSS data done by Cid Fernandes et al (2005) produced very consistent results.  
The \hb \ absorption feature is deblended from the emission line using 
\textsc{splot}. To be consistent with Gavazzi et al. (2004), a mean additive correction of 1.8 in flux and -1.4 \AA \ in EW is 
applied to those H$\beta$ lines where underlying absorption is not detected. 
These values correspond to the fraction of the (broader) absorption feature that lies under the emission feature.\\
The spectral resolution, combined with the contamination of the two [\nii] lines, however, prevent the measurement 
of the underlying H$\alpha$ absorption. Their measurement is thus quite tricky since the three lines tend to overlap at the continuum. 
Hence, the \ha \ and [\nii] lines are all measured by simultaneously fitting gaussian profiles to each line in the triplet using the 
same baseline fit to the continuum. For the underlying absorption we adopt a fixed correction
of E.W.H$\alpha_{abs}$ = 2.80 \AA ~ for all galaxies, a value consistent with the mean value of Moustakas \& Kennicutt (2006). 
This correction can be simply added to the observed emission line as:

\begin{equation}
{E.W.H\alpha_{corr} {\rm(\AA)} = E.W.H\alpha_{obs} + 2.8}
\end{equation}

\noindent
for the equivalent width, where E.W.H$\alpha_{corr}$ is the corrected equivalent width and E.W.H$\alpha_{obs}$ 
the observed one, and 

\begin{equation}
{f(H\alpha)_{corr} = f(H\alpha)_{obs} \times (1 + \frac{2.8}{E.W.H\alpha_{obs}})}
\end{equation}
  
\noindent
for the flux. We remark that this correction is significantly larger than the one determined by Gavazzi et al. (2011) using the SDSS
spectra of 881 passive galaxies in the Coma supercluster (E.W.H$\alpha_{abs}$ = 1.3 \AA). For consistency, the same correction is applied 
to the Gavazzi et al. (2004) data.\\

To quantify any possible systematic effect due to this assumption for the determination of the \ha \ line emission, as well as
in the direct measurement of \hb \ , we have run the MILES population synthesis models (Vazdekis et al. 2010) assuming two different star formation histories,
the Sandage law, whose analytical form is presented in Gavazzi et al. (2001) and Boselli (2011), and that determined by the chemo-spectrophotometric models 
of Boissier \& Prantzos (1999), presented in Buat et al. (2008). Although not tuned to reproduce any possible effect due to the 
interaction with the environment, these star formation laws are well adapted for representing galaxies with a smooth secular evolution 
as expected for our K-band selected sample (Boissier \& Prantzos 2000; Boissier et al. 2001; 2003; Gavazzi et al. 2001; Boselli et al. 2001).
The MILES population synthesis models have been chosen because they have the sufficient spectral resolution (2.51 \AA) necessary for this analysis.
The Sandage law is indicised on a delayed exponential time scale $\tau$, while the Boissier \& Prantzos (1999) star formation history on the 
stellar mass. We have thus run the MILES population synthesis models for 5 different $\tau$ (2, 5, 8, 10, 20 Gyr) \footnote{The value of $\tau$ gives the typical 
age of the peak of star formation in this delayed exponentially declining star formation history.}
and stellar masses (log $M_{star}$ = 8.89, 9.92, 10.52, 10.94, 11.25 M$_{\odot}$), and assuming 
two different metallicities ($Z=Z_{\odot}$; $Z=1/4 Z_{\odot}$). These parameters are representative of the dynamic range observed in our sample (Boselli et al. 2012; Hughes et al. 2012).
We have then measured the equivalent width of the \ha \ and \hb \ underlying Balmer absorption lines of the different model galaxies 
consistently with what done on the real data. We have also checked that the results do not depend on the resolution, as claimed by
Moustakas \& Kennicutt (2006), by degrading the resolution from the nominal value of 2.51 \AA, down to 3 \AA, 4 \AA ~ and 6 \AA \footnote{The
following analysis is based on the 3 \AA ~ resolution data.}. The results of this exercise are shown in Fig. \ref{sandage}.  
Figure \ref{sandage} shows the expected variation of the equivalent width of the \ha \ (upper panel) and \hb \ (middle panel) lines as a 
function of $\tau$ (left) and $M_{star}$ (right). The lower histograms show the expected distribution for the whole galaxy sample,
and for galaxies separated in two bins of metallicity (12+log(O/H) $\leq$ 8.5, blue; 12+log(O/H) $>$ 8.5, red, 
with metallicities taken from Hughes et al. 2012). The threshold in metallicity (12+log(O/H) $=$ 8.5) has been chosen to include $\simeq$ 
the same number of objects in the 2 defined bins. For a solar neighborhood metallicity of Z$_{\odot}$ = 8.69 (Asplund et al. 2009), 
the $Z=1/4$ Z$_{\odot}$ metallicity of the models shown in the lower and middle panels 
corresponds to 12+log(O/H) $=$ 8.09, a value below the lower limit in the range of metallicities
observed in our sample.

  \begin{figure*}
   \centering
   \includegraphics[width=15cm]{sandagebuat.epsi}
   \caption{Variation of the equivalent width of the \ha \ (upper panel) and \hb \ (middle panel) Balmer absorption lines, in \AA, predicted by the MILES population
   synthesis models, as a function of the delayed exponentially declining time scale $\tau$ for a Sandage law (left) and 
   of the stellar mass for the Boissier \& Prantzos (2000) chemo-spectrophotometric models of galaxy secular evolution (right).
   The red line is for solar metallicities, the blue one for $Z=1/4$ Z$_{\odot}$. The Balmer absorption lines
   are measured for five different $\tau$ (2, 5, 8, 10, 20 Gyr) and stellar masses (log $M_{star}$ = 8.89, 9.92, 10.52, 10.94, 11.25 M$_{\odot}$).
   The horizontal solid and dotted lines give the mean value $\pm$ the standard deviation of E.W.H$\alpha_{abs}$ (2.8 $\pm$ 0.38 \AA) and E.W.H$\beta_{abs}$ (4.4 $\pm$ 0.63 \AA)
   of Moustakas \& Kennicutt (2006) for their sample of galaxies.
   The lower panel gives the expected distribution of $\tau$ and $M_{star}$ for the late-type HRS galaxies analysed in this work. The values of $\tau$ 
   are estimated using the $\tau$ - $L_H$ luminosity relation given in Gavazzi et al. (2002): log $\tau$ = -0.149 $\times$ log $L_H$ + 2.221.
   The blue and red histograms give the observed distribution of HRS galaxies with metallicity 12+log(O/H) $\leq$ 8.5 and 12+log(O/H) $>$ 8.5, respectively.
   The right histogram in the middle row shows the distribution of the observed underlying Balmer absorption at \hb \ for the whole sample (black)
   and for metal poor (blue) and metal rich (red) galaxies.
   }
   \label{sandage}%
   \end{figure*}

\noindent
Figure \ref{sandage} shows that overall the observed values of E.W.H$\beta_{abs}$ are consistent with those predicted by 
population synthesis models for the ranges of $\tau$ and/or $M_{star}$ covered by the HRS galaxies (3 $\lesssim$ E.W.H$\beta_{abs}$ $\lesssim$ 6.5). 
There is, however, a systematic difference in the observed distribution of 
metal poor (blue) and metal rich (red) galaxies, the former with higher values of E.W.H$\beta_{abs}$
than the latter (histogram in the right panel of the middle raw). Overall this systematic difference is opposite to that 
predicted by population synthesis models. However, this difference between models and observations might partly be due to the fact
that the E.W.H$\beta_{abs}$ decreases with increasing mass, making metal rich, massive galaxies with E.W.H$\beta_{abs}$
comparable to those of metal poor, low mass systems.
The values of E.W.H$\beta_{abs}$ determined by fitting Bruzual \& Charlot (2003) 
population synthesis models to the observed spectra by Moustakas \& Kennicutt (2006) are slightly lower than those measured 
in our data (see sect. 7.5) or those predicted by the MILES models for solar metallicities, but perfectly match
the dynamic range covered by the MILES population synthesis models for $Z=1/4$ Z$_{\odot}$ metallicities.
Figure \ref{sandage} thus indicates that the difference in the estimate of the equivalent width of the Balmer absorption under H$\beta$ 
determined fitting different population synthesis models (Bruzual \& Charlot 2003 vs. Vazdekis et al. 2010; $\simeq$ 2 \AA) is expected to be comparable to the
width of the distribution of the observed values ($\sim$ 1.5 \AA) shown in sect. 7.5. \\
In the expected range of $\tau$ and $M_{star}$ covered by the HRS galaxies, the MILES population synthesis models 
indicate that E.W.H$\alpha_{abs}$ varies between 2.3 and 3.5 \AA, consistently with the values determined by 
Moustakas \& Kennicutt (2006) (E.W.H$\alpha_{abs}$ 2.8 $\pm$ 0.38 \AA). Variations with the metallicity are at 
most of 0.3 \AA, thus $\lesssim$ 10 \% ~ of the E.W.H$\alpha_{abs}$. Sandage model star formation histories give values of E.W.H$\alpha_{abs}$
in the range 2.3-3 \AA, slightly lower than those predicted by the Boissier \& Prantzos models (2.6 $\lesssim$ E.W.H$\alpha_{abs}$ $\lesssim$ 3.5 \AA)
in the range of $\tau$ and $M_{star}$ expected for the HRS galaxies.
With the data in our hand we are at present in the impossibility of identifying which between the two models (Sandage vs. Boissier \& Prantzos) 
better reproduces the observations, making the choice of a constant value of E.W.H$\alpha_{abs}$ reasonable.
Figure \ref{sandage} shows that the adoption of a constant value of E.W.H$\alpha_{abs}$ might induce systematic effects in the dataset 
presented and analysed in this work. These effects, however, should be very minor. The mean equivalent width of the \ha \ line 
measured from our spectra is of the order of 25 \AA ~ (see sect. 7.2), thus significantly larger than any possible systematic variation in
E.W.H$\alpha_{abs}$ ($\lesssim$ 0.4 \AA). The other emission lines might be affected through the extinction correction based 
on the Balmer decrement. The \ha \ over \hb \ ratio, however, can be measured only whenever both the \ha \ and \hb \ lines are detected. Figure
\ref{CHbEW} shows that this is the case only whenever E.W.H$\alpha_{emi}$ $\gtrsim$ 10 \AA. Systematic errors of $\lesssim$ 0.4 \AA ~ on \ha \
are small compared to the mean uncertainty on the line emission measurements ($\simeq$ 10-20 \%) and could thus be neglected in the following analysis. 
Although possibly present (Rosa-Gonzalez et al. 2002), we do not apply any correction due to stellar absorption in other emission lines.

  \begin{figure}
   \centering
   \includegraphics[width=8cm]{CHbEW.epsi}
   \caption{Observed variation of the Balmer decrement $C(H\beta)$ as a function of the \ha \ (upper panel) and \hb \ (middle panel) equivalent widths,
   in \AA. Red symbols indicate HI-deficient cluster galaxies ($HI-def$ $>$ 0.4), blue ones HI-normal, field objects ($HI-def$ $\leq$ 0.4). Filled symbols
   indicate galaxies hosting an AGN, empty symbols normal galaxies.
   The typical uncertainty on $C(H\beta)$ ranges from $\sim$ 0.3 for E.W.H$\beta_{abs}$ $>$ 2 \AA ~ to 0.5 below.
   }
   \label{CHbEW}%
   \end{figure}

\subsection{Line uncertainties}

\begin{figure*}
   \centering
   \includegraphics[width=15cm]{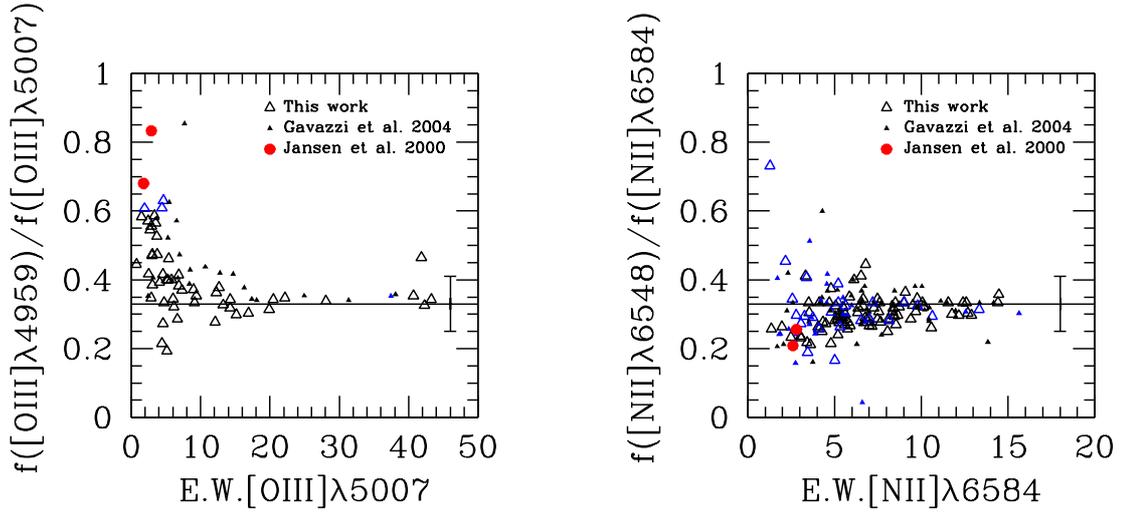}
   \caption{The relationship between the [\oiii]$\lambda$4959/[\oiii]$\lambda$5007 flux line ratio
   and the equivalent width of [\oiii]$\lambda$5007 (left) and of the [\nii]$\lambda$6548/[\nii]$\lambda$6584 flux line ratio
   and the equivalent width of [\nii]$\lambda$6584 (right). The solid line gives the theoretical value (1/3). Empty triangles
   are for galaxies observed in this work, filled triangles for those presented in Gavazzi et al. (2004), red dots for objects in Jansen et al.
   (2000). Galaxies hosting an AGN are marked with a blue symbol. The vertical line indicates a 18\% ~ error in the flux line emission line.}
   \label{sanity}%
   \end{figure*}

An empirical way of quantitatively checking the quality of the line emission measurements is that of comparing the observed distribution
of the [\oiii]$\lambda$4959/[\oiii]$\lambda$5007 and of the [\nii]$\lambda$6548/[\nii]$\lambda$6584 flux line ratios to the theoretical values (1/3;
Osterbrock \& Ferland 2005). 
This comparison is done in Figure \ref{sanity}, where the [\oiii]$\lambda$4959/[\oiii]$\lambda$5007 and the 
[\nii]$\lambda$6548/[\nii]$\lambda$6584 flux line ratios are plotted versus the equivalent width of the [\oiii] $\lambda$ 5007 and [\nii] $\lambda$ 6584
lines, respectively. 
Figure \ref{sanity} shows that the ratios of the two doublets are close to the theoretical values for large equivalent widths, 
with a scatter from the expected relations increasing with decreasing intensity of the emission lines.
We do not see any evident systematic difference in the ratios of galaxies hosting an AGN,
consistent with the idea that the nuclear contribution is generally minor in these integrated spectra (see Sect. 7.2).
For the [\nii] line doublet, the mean ratio for the whole sample (including the data of Gavazzi et al. 2004)
differs from the theoretical value by less then 10\%, with a mean scatter of $\sim$ 20\% (see Table \ref{Tabsanity}).
The difference with respect to the theoretical values for our new set of data significantly reduces for galaxies with E.W.[\nii]$\lambda$6548 $>$ 2 \AA ~
([\nii]$\lambda$6548/[\nii]$\lambda$6584 = 0.32 $\pm$ 0.04).

\begin{table*}
\caption{Line ratios: comparison with theoretical values}
\label{Tabsanity}
{
\[
\begin{tabular}{ccccc}
\hline
\noalign{\smallskip}
\hline
Line~ratio					& Sample	& Condition				& N.~object	& ratio	\\
\hline
[\nii]$\lambda$6548/[\nii]$\lambda$6584		& T.W. + G04	& -					& 191		& 0.31$\pm$0.07	\\
						& T.W.		& -					& 118		& 0.31$\pm$0.06	\\
						& T.W. + G04	& [\nii]$\lambda$6548 $>$ 2 \AA		& 86		& 0.33$\pm$0.06	\\
						& T.W. 		& [\nii]$\lambda$6548 $>$ 2 \AA		& 55		& 0.32$\pm$0.04	\\
						& T.W. + G04	& [\nii]$\lambda$6548 $>$ 3 \AA		& 39		& 0.33$\pm$0.06	\\
						& T.W. 		& [\nii]$\lambda$6548 $>$ 3 \AA		& 21		& 0.33$\pm$0.02	\\
						& T.W. + G04	& [\nii]$\lambda$6548 $\leq$ 2 \AA	& 105		& 0.30$\pm$0.08	\\
						& T.W. 		& [\nii]$\lambda$6548 $\leq$ 2 \AA	& 63		& 0.30$\pm$0.08	\\
\hline
[\oiii]$\lambda$4959/[\oiii]$\lambda$5007	& T.W. + G04	& -					& 85		& 0.42$\pm$0.11	\\
						& T.W.		& -					& 62		& 0.40$\pm$0.10	\\
						& T.W. + G04	& [\oiii]$\lambda$4959 $>$ 3 \AA	& 40		& 0.39$\pm$0.11	\\
						& T.W. 		& [\oiii]$\lambda$4959 $>$ 3 \AA	& 24		& 0.35$\pm$0.04	\\
						& T.W. + G04	& [\oiii]$\lambda$4959 $\leq$ 3 \AA	& 45		& 0.44$\pm$0.11	\\
						& T.W. 		& [\oiii]$\lambda$4959 $\leq$ 3 \AA	& 38		& 0.44$\pm$0.11	\\
\hline				
\noalign{\smallskip}
\hline
\end{tabular}
\]
}
\end{table*}

The ratio [\oiii]$\lambda$4959/[\oiii]$\lambda$5007 of our new observations is close to the theoretical value whenever 
E.W.[\oiii] $\lambda$4959 $\gtrsim$ 3 \AA ~ ([\oiii]$\lambda$4959/[\oiii]$\lambda$5007 = 0.35 $\pm$ 0.04), 
while is significantly larger below this limit, or whenever the data of Gavazzi et al. (2004) are included (see Table \ref{Tabsanity}). 
An accurate inspection of the spectra suggests that this discrepancy is due to a systematic overestimate of 
the [\oiii] $\lambda$4959 line for E.W.[\oiii] $\lambda$4959 $<$ 3 \AA ~ resulting from the contamination of an important 
NaI sky line ($\lambda$ 4983 \AA; Osterbrock \& Martel 1992) which falls close to this emission line at the typical
redshift of our targets \footnote{The comparison of observed flux ratios with the expected theoretical values allowed us
to identify a few spurious measurements in Gavazzi et al. (2004). For these galaxies, indicated with a note in the following Tables, we have remeasured fluxes and equivalent
widths.}. \\
The deviations of the [\oiii] and [\nii] line ratios from the expected values can be used to quantify the uncertainty in the line emission measurement.
Here we make the conservative assumption that the uncertainty in the line ratio is totally due to the measure 
of the weakest of the two lines of the doublet. This test indicates that the mean uncertainty in the measure of the [\nii] line intensity 
for the new observations presented here (T.W.) is $\lesssim$ 6\% 
whenever the equivalent width of the emission line is larger than 3 \AA, $\lesssim$ 12\% for 
equivalent widths larger than 2 \AA ~ and increases to $\lesssim$ 24 \% below. The uncertainty in the [\oiii] $\lambda$4959 is $\lesssim$ 12\% ~ 
whenever the equivalent width of the emission line is larger than 3 \AA, but it increases to $\sim$ 35 \% ~ with a systematic trend 
(the line emission is overestimated) for smaller equivalent widths.
The errors are slightly larger whenever the data of Gavazzi et al. (2004) are included (T.W.+G04).
These uncertainties are comparable to those obtained with the same test using the integrated spectra of Jansen et al. (2000) 
or those quoted by Moustakas \& Kennicutt (2006) who measured [\oiii]$\lambda$4959/[\oiii]$\lambda$5007 ratios within $\pm$ 4\% ~ from the 
theoretical value whenever E.W.[\oiii] $\lambda$4959 $\gtrsim$ 3 \AA. They are also consistent with the uncertainties derived by the comparison 
with narrow band imaging data (see Sect. 6.2).
As previously mentioned, the large uncertainty in the [\oiii] $\lambda$4959 line is related to 
the contamination of a strong NaI sky line at $\lambda$ 4983 \AA \footnote{Given that the sky emission strongly contaminates only the [\oiii] $\lambda$4959 line,
its emission can be deduced with a smaller uncertainty from the [\oiii] $\lambda$5007 \AA ~ line assuming a constant ratio.}. Given that the signal to noise in the
spectra around 4800-5000 \AA ~ is comparable to that at $\sim$ 6500 \AA, we can assume that the mean uncertainties in the measurement of
the [\oiii] $\lambda$5007 and H$\beta$ lines are comparable to those determined for the [\nii] lines. This is also valid for the [\sii] lines.
On the other hand, given the noise of the spectra below 4000 \AA ~ and the observed systematic difference at these short wavelengths
with respect to other sets of data available in the literature ($\lesssim$ 20\%), the uncertainty on the measurements of the [\oii] $\lambda$3727 \AA ~ line
is significantly larger (we indeed detected only emission lines with equivalent widths $\gtrsim$ 10 \AA), 
at best of 20\% but never exceding 50\%. 

\subsection{Balmer decrement}

The emission line intensities can be corrected for internal extinction using the Balmer decrement $C(H\beta)$. With the \hb \ line corrected 
for underlying absorption, the Balmer decrement is given by (Lequeux et al. 1981):  

\begin{eqnarray}\label{eq:c1}
C(H\beta) = \left[\log{ \left( \frac{H{\alpha}}{H{\beta}}\right)}_{theor} - \log{\left(\frac{H{\alpha}}{H{\beta}}\right)}_{obs} \right]/f(H{\alpha})
\end{eqnarray}

\noindent
where $\log{(H{\alpha}/H{\beta})}_{theor}$ is the theoretically expected ratio between \ha \ and \hb , $\log{(H{\alpha}/H{\beta})}_{obs}$ 
is the observed value and $f$(\ha) is the reddening function relative to \hb . 
The theoretical ratio depends on the electron density $n$ and the gas temperature $T$. Assuming that $T$ = 10000 K and 
$n$ = 100 e cm$^{-3}$ and that all the Lyman line photons are absorbed by the diffuse gas located around the emitting star  
(case B from Osterbrock \& Ferland 2006), which are values typical of HII regions, then $(H{\alpha}/H{\beta})_{theor}$ = 2.86. 
The measured line fluxes can be corrected for internal extinction using the relation:

\begin{eqnarray}\label{eq:fcorr}
\log{ \left( \frac{\lambda}{H{\beta}}\right)}_{corr} = \log{ \left( \frac{\lambda}{H{\beta}}\right)}_{obs} + {C(\lambda)} {\times f(\lambda)}
\end{eqnarray}

\begin{table}
\caption{Extinction coefficients for optical emission lines, from Fitzpatrick \& Massa (2007)}
\label{Fitz}
{\scriptsize
\[
\begin{tabular}{cccc}
\hline
\noalign{\smallskip}
line	& $\lambda$	& $k_{MW}(\lambda)^a$	& $f(\lambda)^{a,b}$	\\	
	& AA\		&			&		\\
\hline
$[OII]$	& 3727/29	& 4.751			& +0.324	\\
H$\beta$& 4861		& 3.588			&  0.000	\\
$[OIII]$& 4959		& 3.497			& -0.025	\\
$[OIII]$& 5007		& 3.452			& -0.038	\\
$[NII]$	& 6548		& 2.524			& -0.294	\\
H$\alpha$&6563		& 2.517			& -0.297	\\
$[NII]$	& 6584		& 2.507			& -0.301	\\
$[SII]$	& 6717		& 2.444			& -0.319	\\
$[SII]$	& 6731		& 2.437			& -0.321	\\
\noalign{\smallskip}
\hline
\end{tabular}
\]
}
Notes: adapted from Boselli (2011).\\
a) $k(\lambda)$ and $f(\lambda)$ are determined assuming $R_{MW}(V)$=3.08.\\
b) $f(\lambda)$ is relative to H$\beta$.
\end{table}

\noindent
and adopting the reddening function $f(\lambda$) of Fitzpatrick \& Massa (2007) (see Table \ref{Fitz}).
In those galaxies where \hb \ is undetected, we do not derive an upper limit to $C(H\beta)$ based on the \ha \ equivalent width, as in Gavazzi et al. (2004), 
since these estimates are very uncertain. We present in Table \ref{tab:measlinesflux} the observed line emission 
fluxes (normalised to H$\alpha$) of the main emission lines detected in our spectra, i.e. [\oii]3727 \AA, H$\beta$4861 \AA, [\oiii]4959 and [\oiii]5007 \AA, 
[\nii]6548 \AA, H$\alpha$6563 \AA, [\nii]6584 \AA, [\sii]6717 \AA ~ and [\sii]6731 \AA.
Table \ref{tab:measlinesflux} is arranged as follow:

\begin{itemize}
\item{Column 1: $Herschel$ Reference Sample (HRS) name.}
\item{Column 2: References for the data: 1: this work; 2: Gavazzi et al. (2004); 3: Moustakas et al. (2010); 4: Moustakas \& Kennicutt (2006); 
5: Jansen et al. (2000); 6: Kennicutt (1992a; 1992b) .}
\item{Column 3: $E(B-V)$, from Schlegel et al. (1998).}
\item{Column 4: Balmer decrement $C(H\beta)$. The contribution of the Milky Way is subtracted using the Galactic extinction map of 
Schlegel et al. (1998) combined with the Fitzpatrick \& Massa (2007) Galactic extinction law\footnote{The values of $C(H\beta)$ differ from those
of Gavazzi et al. (2004) because those published in that paper include the Galactic extinction and have been measured using a different Galactic
extinction law.}.}
\item{Columns 5-13: Observed line intensities normalized to \ha. We normalise to \ha ~ since this is the only line detected in all emission line 
galaxies.}
\end{itemize}

We also present in Table \ref{tab:measlinesew} the corresponding equivalent widths of the main emission lines detected in our spectra. Table \ref{tab:measlinesew} is arranged as follow:

\begin{itemize}
\item{Column 1: $Herschel$ Reference Sample (HRS) name.}
\item{Column 2: References for the data: 1: this work; 2: Gavazzi et al. (2004); 3: Moustakas et al. (2010); 4: Moustakas \& Kennicutt (2006); 
5: Jansen et al. (2000); 6: Kennicutt (1992a; 1992b) .}
\item{Column 3: $E(B-V)$, from Schlegel et al. (1998).}
\item{Columns 4-12: Equivalent widths (in \AA).}
\item{Column 13: Equivalent width of the underlying H$\beta$ absorption (in \AA).}
\end{itemize}

\section{Comparison with the literature}

\subsection{Comparison with integrated spectra}

\begin{figure*}
   \centering
   \includegraphics[width=15cm]{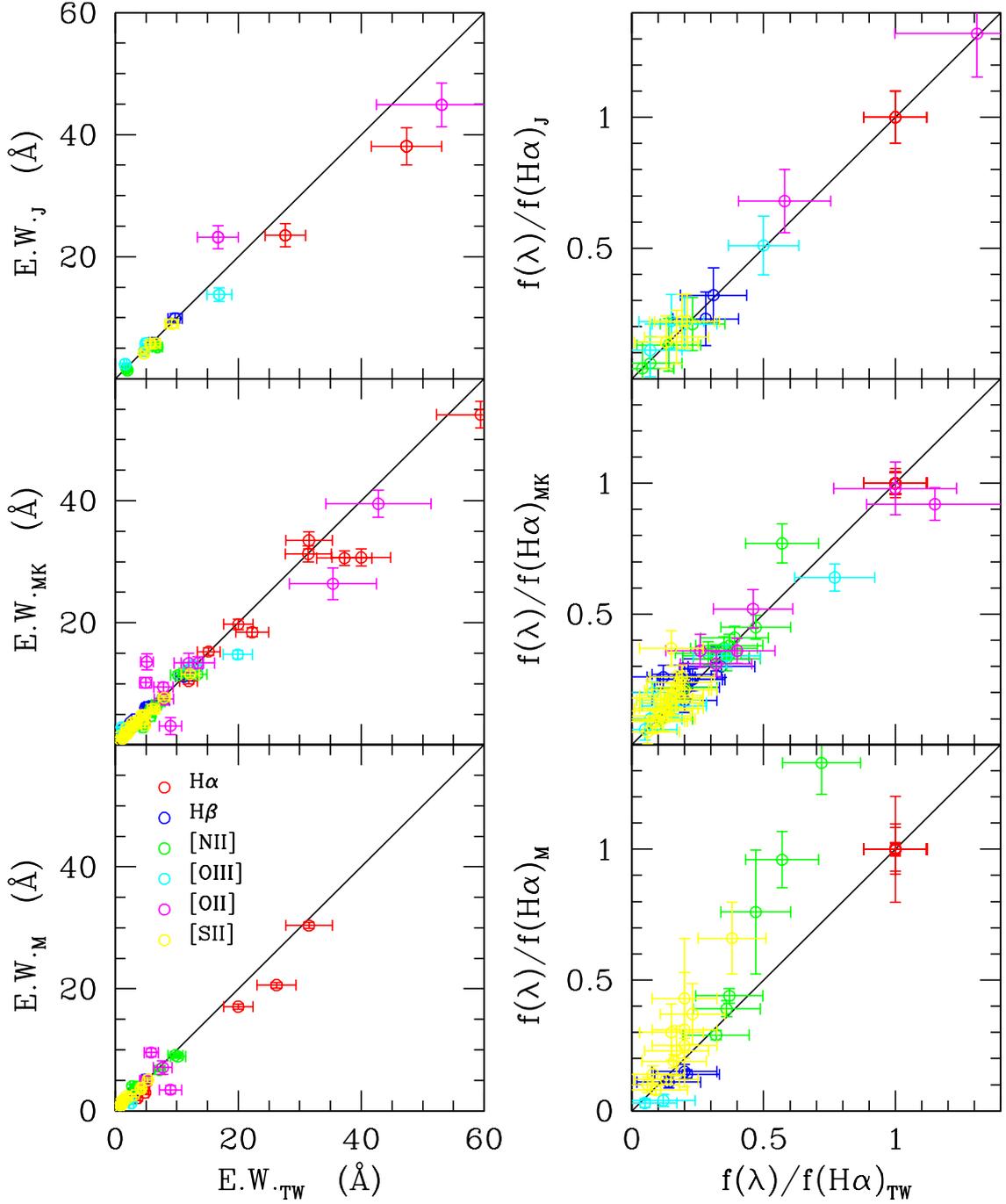}
   \caption{The comparison between the equivalent width (left) and the normalised flux (right) of the main emission lines measured in this work (TW)
   with those available in the literature from
   Jansen et al. (2000; J; upper panel), Moustakas \& Kennicutt (2006; MK; middle panel) and Moustakas et al. (2010; M;
   lower panel). Different colour codes are used for the different emission lines: red symbols for H$\alpha$; blue for
   H$\beta$, green for [\nii]6548 and [\nii]6584 \AA, cyan for [\oiii] and 5007 \AA, magenta for [\oii] 3727 \AA, and yellow for
   [\sii] 6717 and 6731 \AA ~(the two lines composing a doublet are plotted with the same symbol). The solid line shows the 1:1 relationship.
   }
   \label{letteraturaEWflux}%
   \end{figure*}

A few HRS galaxies have been observed using the same drift scan technique adopted in this work
by Kennicutt (1992a; 4 objects), Jansen et al. (2000; 5), Moustakas \& Kennicutt (2006; 16) and Moustakas et al. (2010; 8). 
Their integrated spectroscopic data are given in Tables \ref{tab:TablitEW} and \ref{tab:Tablitflux} and can be compared to those obtained in this 
work or in Gavazzi et al. (2004). Figure \ref{letteraturaEWflux} 
shows the relationship between the equivalent width (left) and the normalised flux (right) of the most important emission lines for galaxies in common.
We consider here the data published in Gavazzi et al. (2004) together with those collected in this more recent observational campaign, as both datasets have been
taken using the same telescope with the same instrumental configuration.

Figure \ref{letteraturaEWflux} shows that for the galaxies in common, the different sets of data are consistent within the quoted errors. 
There is, however, a systematic difference in the normalised fluxes of the [\nii] \ and [\sii] \ lines with Moustakas et al. (2010), who give 
values larger than those obtained in this work or in Gavazzi et al. (2004). The largest differences are observed in galaxies 
hosting an AGN (NGC 4579, NGC 4569 and NGC 4450). AGNs are characterised by larger [\nii]/\ha \ and [\sii]/\ha \ line ratios than star forming discs
(e.g. Kewley et al. 2006).
Given that the integrated spectra of Moustakas et al. (2010) are covering only a radial strip (including the nucleus) which is just a fraction 
of the optical disc covered by our observations, we expect that the contamination of the central AGN to the total emission
is more important in Moustakas et al. (2010) than in our data. The mean ratio of the fluxes normalised to H$\alpha$
of the galaxies in common with Jansen et al. (2000) and Moustakas \& Kennicutt (2006) is of 0.93 $\pm$ 0.18, while that of the 
equivalent width is of 1.01 $\pm$ 0.30, and drops to 1.00 $\pm$ 0.22 excluding the most uncertain [\oii] line. 
These differences are consistent with the errors on the different line emission estimated in the previous section.\\

   \begin{figure}
   \centering
   \includegraphics[width=9cm]{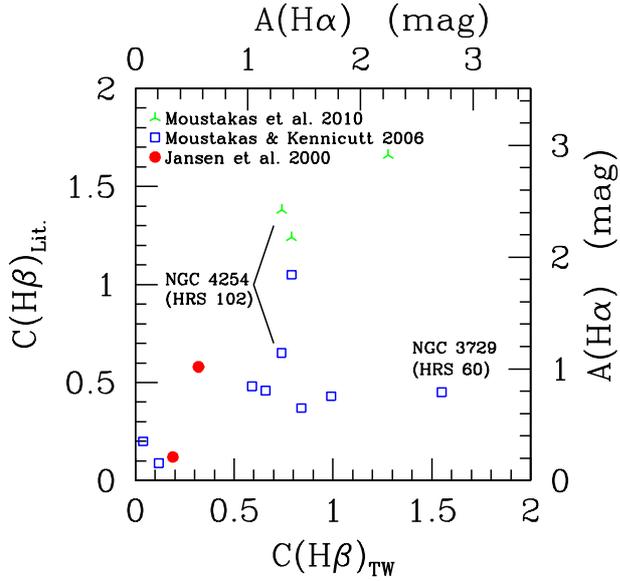}
   \caption{The comparison between the Balmer decrement $C(H\beta)$ (or equivalently $A(H\alpha)$) measured in this work
   (TW) with that obtained using data available in the literature (Lit). Green three points crosses are for data from
   Moustakas et al. (2010), blue open squares from Moustakas \& Kennicutt (2006) and red filled dots from Jansen et al.
   (2000). The typical uncertainty on $C(H\beta)$ from our measurements is 0.3-0.5.
   }
   \label{letteraturaCHb}%
   \end{figure}

\noindent
Figure \ref{letteraturaCHb} shows the relationship between $C(H\beta)$ Balmer decrement for 8 HRS galaxies 
in the literature with detected H$\alpha$ and H$\beta$ lines.
The poor statistics prevents us to see whether the agreement between the different sets of data is satisfactory.
The most discrepant values are those relative to NGC 4254 (HRS 102) and NGC 3729 (HRS 60). 
NGC 4254 has been observed by Moustakas et al. (2010) and Moustakas \& Kennicutt
(2006), yielding to $C(H\beta)$ = 1.38 and 0.65, respectively. These values can be compared to $C(H\beta)$ = 0.74 as determined 
from our own data. If we consider the difference between the two values obtained by the team of Moustakas as indicative of
the uncertainty on this value or of any systematic difference related to the observed region, we can 
conclude that our value is consistent with their measurements. NGC 3729 is the
HRS galaxy within the Moustakas \& Kennicutt (2006) sample with the lowest value of H$\beta$ E.W. (H$\beta$ E.W. = 1.2
\AA~ from this work, H$\beta$ E.W. = 2.56 \AA~ from Moustakas \& Kennicutt (2006)). The mean ratio of the $C(H\beta)$ determined from this work
to those given in the literature for galaxies in common is of $C(H\beta)_{T.W.}/C(H\beta)_{Lit.}$ = 1.13 $\pm$ 0.65. The scatter in the ratio
is slightly larger than an error of $\sim$ 0.3-0.4 determined from the uncertainty in the measurements of the H$\alpha$ and H$\beta$ lines.
We thus conclude that the total uncertainty on $C(H\beta)$ from our data is of 0.3-0.5, with the largest uncertainties 
for the highest values of the Balmer decrement.


\subsection{Comparison with H$\alpha$+[\nii] narrow band imaging}

\begin{figure}
   \centering
   \includegraphics[width=8cm]{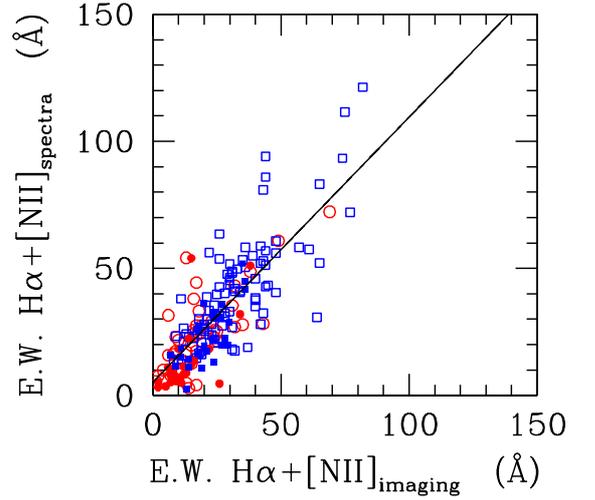}
   \caption{The comparison between the E.W.H$\alpha$+[\nii] obtained in our spectroscopic survey with those obtained in
   imaging mode by Boselli et al. (in preparation) for 198 galaxies in common. 
   Blue crosses are for star forming galaxies with a normal HI gas content ($HI-def$ $\leq$ 0.4), open red circles for HI-deficient
   objects ($HI-def$ $>$ 0.4). Filled symbols indicate galaxies hosting an AGN. The solid line shows the linear fit E.W.H$\alpha$+[\nii]$_{spectra}$ =
   1.040($\pm$0.058)E.W.H$\alpha$+[\nii]$_{imaging}$ + 5.370($\pm$1.766) ($r$=0.79). The typical uncertainty in the equivalent width 
   measured from spectroscopic data is 12-18\%, in imaging mode is $\sim$ 20\%.
   }
   \label{HaEW}%
   \end{figure}

Most of the late-type HRS galaxies have been recently observed in imaging mode using narrow band interferential 
filters using the San Pedro Martir 2.1m telescope (Boselli et al. in prep.). Those in the Virgo cluster have narrow band 
imaging data from Koopmann et al. (2001), Boselli et al. (2002a), Boselli \& Gavazzi (2002) and Gavazzi et al. (2002; 2006). 
H$\alpha$+[\nii] equivalent widths\footnote{The interferential filters have a FWHM of $\sim$ 80 \AA ~and thus encompass the nearby [\nii] lines at 
6548 and 6584 \AA.} obtained with integrated spectroscopy can be thus directly compared to those obtained in imaging mode
(Fig. \ref{HaEW}). To take into account that the imaging data do include the contribution of the underlying absorption,
the spectroscopic H$\alpha$+[\nii] equivalent width is defined as:

\begin{eqnarray}
{E.W.H\alpha +[NII]_{spectra} = E.W.H\alpha + E.W.[NII]6548 }\nonumber \\
{+ E.W.[NII]6564 - E.W.H\alpha_{abs}}
\end{eqnarray}

\noindent
where $E.W.H\alpha_{abs}$ = 2.8 \AA~ (see previous section).

\noindent
Figure \ref{HaEW} shows that the E.W.H$\alpha$+[\nii]$_{spectra}$ obtained in integrated spectroscopic mode 
are consistent with those obtained in imaging mode. The best fit to the data gives 
E.W.H$\alpha$+[\nii]$_{spectra}$ = 1.040($\pm$0.058)E.W.H$\alpha$+[\nii]$_{imaging}$ + 5.370($\pm$1.766) ($r$=0.79). It also indicates that the drift scan spectroscopy
is tracing the global properties of the observed galaxies.

\section{Analysis}

Because of its definition, the HRS sample can be used to trace the line emission statistical properties 
of a volume limited, K-band selected sample of nearby late-type galaxies spanning a large range in stellar mass, 
luminosity, morphological type and belonging to different environments (cluster vs. fields). In this section we study
the mean statistical properties of the HRS sample and the dependence of the derived spectral parameters (emission line 
equivalent widths, Balmer decrement, line diagnostics, underlying Balmer absorption)
as a function of different physical parameters (morphological type, stellar mass and surface density, birthrate parameter, 
metallicity, star formation rate and gas column density). 

\subsection{Ancillary data and the derived physical parameters} 

Different sets of multifrequency data are required to characterise the physical properties of the target galaxies. Near-IR and optical photometry, taken
from 2MASS (Jarrett et al. 2003), GOLDMine (Gavazzi et al. 2003) or NED, are used to estimate total stellar masses using the colour-dependent 
recipes given in Boselli et al. (2009). The near-IR photometry is also used to measure the H-band effective surface brightness $\mu_e(H)$ (in mag arcsec$^{-2}$), 
defined as the mean surface brightness within the effective radius (radius including half of the total stellar light; see Gavazzi et al. (2000)).
This entity gives a direct measure of the intensity of the general interstellar radiation field, i.e. the intensity of the radiation emitted by the bulk of
the stellar component. Stellar masses, combined with
UV GALEX (Boselli et al. 2011; Cortese \& Hughes 2009; Hughes \& Cortese 2009) and H$\alpha$+[\nii] imaging data (Boselli et al. in prep.), are also used to quantify different 
direct tracers of the star formation history of the galaxies. This is done by measuring the birthrate parameter $b$ (Kennicutt et al. 1994), that in a closed box model can be defined as in
Boselli et al. (2001):

\begin{equation}
{b = \frac{SFR}{<SFR>}=\frac{SFR t_0 (1-R)}{M_{star}}}
\end{equation}

\noindent
with $t_0$ the age of the galaxy (13 Gyr) and $R$ the returned gas fraction, here assumed to be $R$=0.3 (Boselli et al. 2001). 
As remarked in Boselli et al. (2012), the birthrate parameter is equivalent to the specific star formation rate $SSFR$.
The star formation rate is measured using extinction corrected H$\alpha$+[\nii] imaging data and FUV GALEX images using the prescription 
described in Boselli et al. (2009). The current set of spectroscopic data is used to correct H$\alpha$+[\nii] fluxes for the [\nii] contamination and
the Balmer decrement. 
UV data are corrected for dust attenuation using the prescriptions of Cortese et al. (2008) based on the far infrared to UV flux ratios. Star formations are 
then determined using the UV and H$\alpha$ calibrations of Kennicutt (1998). We assume an escape fraction of zero and a fraction of ionising photons absorbed by dust before ionising the gas of zero ($f$=1).
Although unphysical (see Boselli et al. 2009), this choice has been done to allow a direct comparison with the results obtained in the literature 
using other star formation rates determined using
H$\alpha$ data. These works generally assume $f$=1. Our most recent results indicate that $f$ $\sim$ 0.6 (Boselli et al. 2009).
When both H$\alpha$ and UV data are available, the $SFR$ is determined as the mean value.\\
As defined, $b$ measures the ratio of the ionising (photons with $\lambda$ $<$ 912 \AA) to non ionising ($\lambda$ = 1.65 $\mu$m) 
radiation and is thus a direct tracer of the hardness of the interstellar radiation field. 
Galaxies with a $b$ parameter $>$ 1 have 
a present day star formation activity more important than their mean star formation activity since their birth. They are 
characterised by very blue colours and thus have hard interstellar radiation fields.\\
The birthrate parameter is proportional to the specific star formation rate $SSFR$ defined as (Brinchmann et al. 2004):

\begin{equation}
{SSFR = \frac{SFR }{M_{star}} = \frac{b}{t_0 (1-R)}}
\end{equation}

\noindent
This parameter is also important since it is often used to discriminate the far infrared properties of galaxies in blind infrared cosmological surveys 
such as H-GOODS and H-ATLAS (e.g. Elbaz et al. 2011; Smith et al. 2012b). We thus use either of the two parameters in the following analysis.\\

Gas metallicities, 12+log(O/H), are derived using the set of data described in this paper as indicated in Paper II (Hughes et al. 2012).
We follow the conversions of Kewley \& Ellison (2008) for five different metallicity calibrations
from the literature and adopt the PP04 O3N2 calibration on the [\nii] and [\oiii] emission
lines (Pettini \& Pagel 2004) as the base metallicity. We then 
determine the average oxygen abundance in units of 12+log(O/H) for each galaxy.
Integrated HI gas data, available for almost the totality of the late-type galaxies of the sample, are used to detect those objects that might 
have suffered any kind of perturbation induced by the Virgo cluster environment. 
Interferometric data, mainly taken in order of preference from VIVA (Chung et al. 2009a),  Cayatte  et al. (1994) and Warmels (1986), 
are used to measure HI gas column densities (in M$\odot$ pc$^{-2}$). For the VIVA galaxies, column densities are determined using the isophotal HI
radius\footnote{For the galaxy HRS 102 (NGC 4254) we use the effective radius since the isophotal radius gives HI column densities a factor of
$\sim$ 1000 larger than those determined by Cayatte et al. (1994). The effective radius is indeed more representative of the HI distribution within the
disc of this galaxy.}. For the sake of homogeneity, we normalise the HI column densities 
given in Cayatte et al. (1994) and in Warmels (1986) by calculating the mean ratio of the column densities measured for galaxies in common with the VIVA
survey. This gives log $\Sigma HI(VIVA)$ = 0.92 ($\pm$ 0.58) log $\Sigma HI(Cayatte)$\footnote{We exclude in the measure of the mean gas column density ratio 
the galaxy HRS 257 (NGC 4698) for which the difference in the gas column density with Cayatte et al. (1994) is of $\sim$ a factor of 10.} 
and log $\Sigma HI(VIVA)$ = 1.04 ($\pm$ 0.45) log $\Sigma HI(Warmels)$.
$^{12}CO(1-0)$ imaging data are available for a small fraction of the HRS galaxies. We estimate H$_2$ gas column densities using
the set of data of Chung et al. (2009b). For consistency with the HI, column densities are measured using the isophotal radius. The total H$_2$ 
gas mass is estimated using the relation:

\begin{equation}
{M(H_2) \rm{(M\odot)} = 3.922 \times 10^{-17} X_{CO} D^2 S_{CO} \rm{(Jy km s^{-1})}}
\end{equation} 

\noindent
where $X_{CO}$ is the CO to H$_2$ conversion factor, in units of mol cm$^{-2}$ (K km s$^{-1}$)$^{-1}$. Here we use the
H-band luminosity dependent calibration of Boselli et al. (2002b).

To quantify the effects of the
environment on the statistical properties of the HRS galaxies we code them according to their HI gas content.
There are indeed strong indications that this gas component is removed during the interactions of galaxies 
with the hostile cluster environment (Boselli \& Gavazzi 2006). We assume as normal, unperturbed objects those with an HI-deficiency parameter $HI-def$ $\leq$ 0.4,
where $HI-def$ is defined as the difference in logarithmic scale between the expected and the observed HI mass of a galaxy of given angular size and morphological type 
(Haynes \& Giovanelli 1984). HI-deficiencies for all the target galaxies have been measured using the recent calibrations of Boselli \& Gavazzi (2009).

\subsection{Line diagnostic}

\begin{figure*}
   \centering
   \includegraphics[width=15cm]{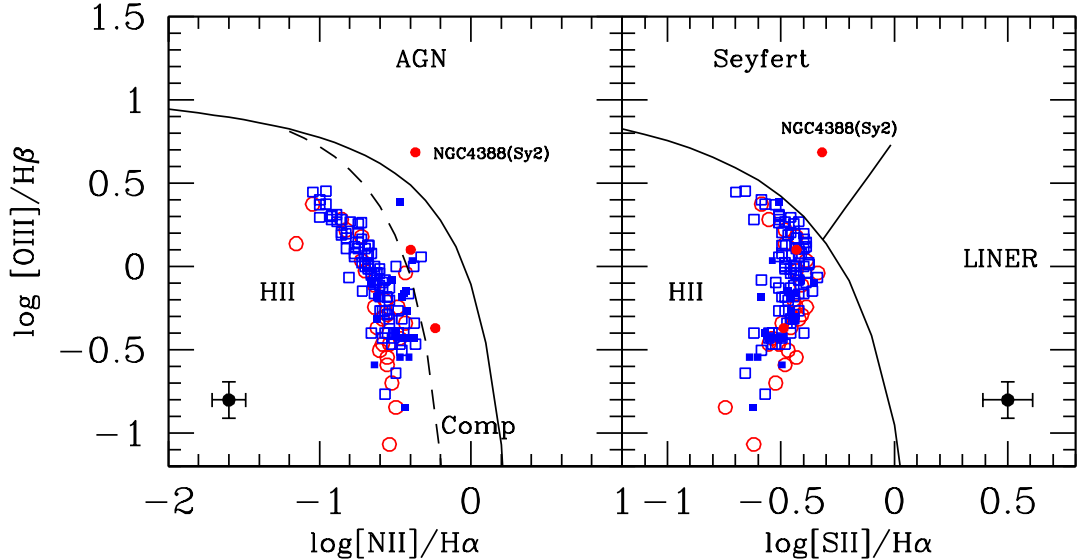}
   \caption{The emission line diagnostic diagrams [\oiii]$\lambda$5007/H$\beta$ vs. [\nii]$\lambda$6584/H$\alpha$ and 
   [\oiii]$\lambda$5007/H$\beta$ vs. [\sii]$\lambda$6717-6731/H$\alpha$ for all HRS galaxies. The solid and dashed lines gives the limits
   between star forming (HII), active (AGN, Seyfert, LINER) and composite galaxies of Kewley \& Ellison (2008).
   Blue symbols are for galaxies with a normal HI gas content ($HI-def$ $\leq$ 0.4), red symbols for HI-deficient
   objects ($HI-def$ $>$ 0.4). Filled symbols are for galaxies hosting an AGN.}
   \label{diagnostic}%
   \end{figure*}

Spectral line emission is often used to characterise, using different diagnostic diagrams, the nature of the
emitting source. The set of data in our hands allow us to construct two of the three mostly used diagnostic
diagrams, the [\oiii]$\lambda$5007/H$\beta$ vs. [\nii]$\lambda$6584/H$\alpha$ and the 
[\oiii]$\lambda$5007/H$\beta$ vs. [\sii]$\lambda$6717-6731/H$\alpha$ (Fig. \ref{diagnostic}).
To discriminate active galaxies from normal, star forming objects here we use the recent definitions of Kewley \& Ellision (2008).

\noindent
Despite the presence of a large fraction of active galaxies in the sample\footnote{The nuclear data necessary for this classification are
available for 233/260 of the late-type galaxies analysed in this work.}, the diagnostic diagrams shown in Fig.
\ref{diagnostic} indicate that the integrated spectra of the HRS galaxies are mainly those of star forming
objects (HII), with a few composite spectra (Comp). Only one galaxy, NGC 4388, has an integrated spectrum
typical of a Seyfert galaxy. This galaxy is a well known edge-on Seyfert 2 galaxy located close to M86.\\
The analysis of Fig. \ref{diagnostic} does not show any strong systematic difference between 
gas poor and gas rich late-type galaxies in the [\oiii]$\lambda$5007/H$\beta$ vs. [\nii]$\lambda$6584/H$\alpha$
diagram. There might be a slight (hardly quantifiable) shift towards high [\sii]$\lambda$6717-6731/H$\alpha$ ratios
in the [\oiii]$\lambda$5007/H$\beta$ vs. [\sii]$\lambda$6717-6731/H$\alpha$ diagram of the HI-deficient objects
(red circles). If real, we think that this mild difference between gas rich and gas poor objects comes from the
fact that in HI-deficient objects star formation is suppressed in the outer disc (Boselli et al. 2006; Boselli \&
Gavazzi 2006). In these objects HII regions, from where most of the line emission comes from, are located only in the
inner disc. 63 out of the 260 HRS late-type galaxies are 
known to host an AGN (Seyfert, LINER or retired AGN). The contribution of the nuclear emission to the integrated spectrum is thus more
important in HI-deficient objects than in gas rich systems. 
Despite the contribution of the AGN to the integrated spectra is minor, 
objects hosting an active nucleus should be removed in any study of the physical properties of the ISM 
since their line emission can be partly contaminated by the nuclear activity. The $Herschel$ Reference Survey, however, 
has been designed not only to gather a complete set of multifrequency data ideal for the study the physical properties of the ISM 
in nearby galaxies, but also for providing a well defined reference for high redshift studies.
In these studies of distant, unresolved galaxies the spectra cover a large fraction of the galaxy disc and are thus directly comparable to 
the integrated spectra obtained in this work (Kobulnicky et al. 1999). For this reason we keep in the following analysis 
the whole galaxy sample, but we clearly identify and remove active galaxies whenever necessary for the physical interpretation of the data.

\subsection{Emission lines equivalent widths}

   \begin{figure*}
   \centering
   \includegraphics[width=15cm]{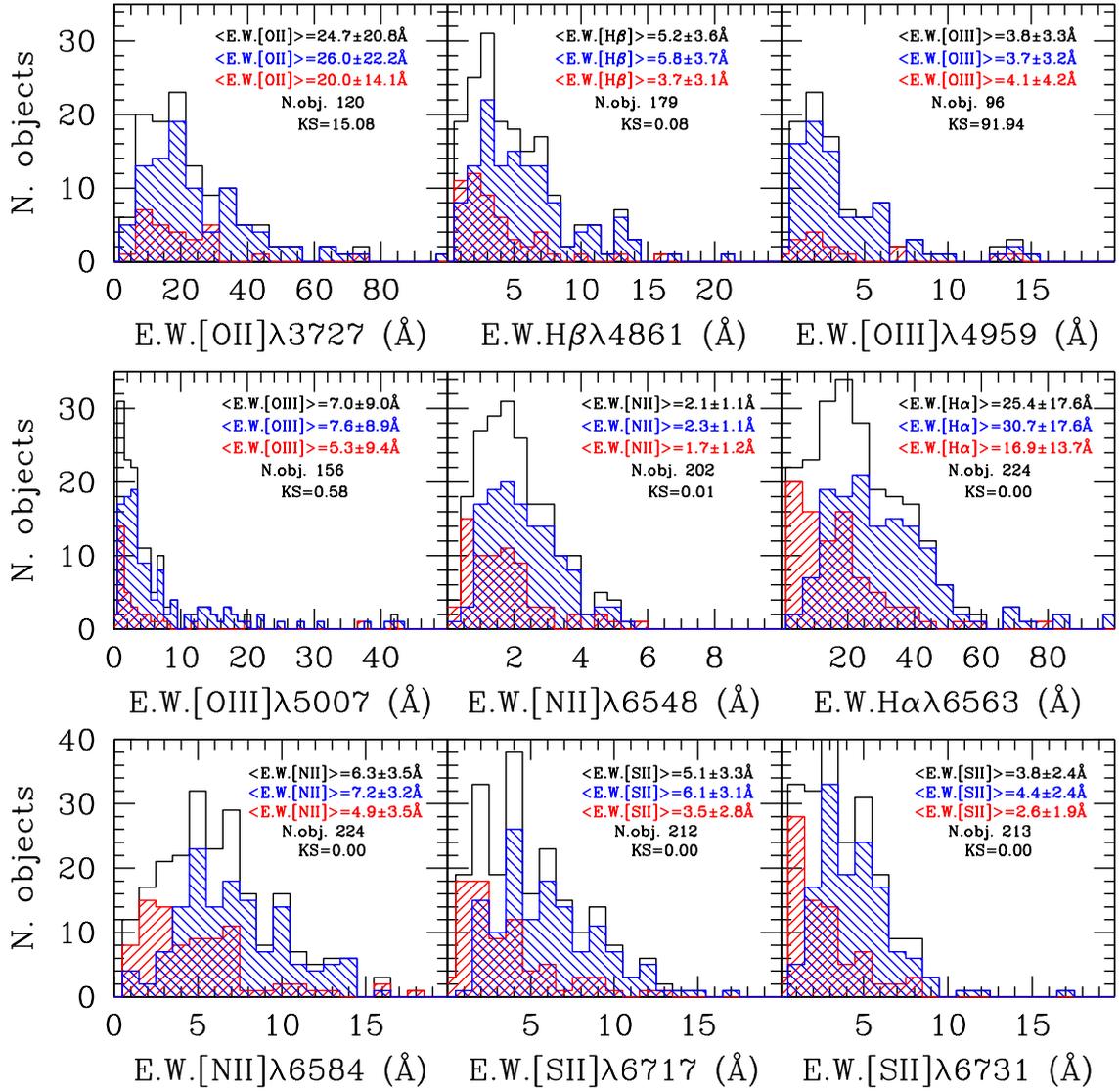}
   \caption{The distribution of the equivalent width of the different emission lines in the HRS sample. The black line
   indicates the whole sample of late-type galaxies, the blue line that of galaxies with a normal HI gas content ($HI-def$ $\leq$ 0.4) 
   and the red one that of HI-deficient Virgo cluster objects ($HI-def$ $>$ 0.4). Each panel gives the total number of available data,
   the mean value of the equivalent width for the whole distribution
   (black) and separately for galaxies with a normal HI gas content ($HI-def$ $\leq$ 0.4; blue) and for HI-deficient objects (red), and the probability 
   that the HI-normal and HI-deficient galaxy distributions are driven by the same parent distribution, as derived by the Kolmogorov-Smirnov (KS) test. 
   }
   \label{EWdist}%
   \end{figure*}

The equivalent widths are useful parameters in statistical analysis since they are normalised entities only moderately sensitive to internal attenuation 
(in the hypothesis that line and the continuum are equally attenuated). The H$\alpha$ equivalent width, historically used in the study of the star formation activity of
galaxies (e.g. Kennicutt 1983a), for instance, is tightly related to the birthrate parameter and the specific star formation rate. Figure \ref{EWdist}
shows the distribution of the equivalent width of all the emission lines detected in our survey. The black solid line gives the
distribution of the whole sample and can thus be considered as the typical distribution of a K-band selected sample of nearby late-type galaxies.
The number of objects (shown in each single panel in Fig. \ref{EWdist}) indicates whether the distribution can be considered as representative (the whole sample 
is composed of 260 objects).
The blue line gives the distribution of the galaxies with a normal HI content ($HI-def$ $\leq$ 0.4) while the red one
that of HI-deficient objects. The Kolmogorov-Smirnov test indicates that the probability that the HI-normal and HI-deficient distributions
are driven by the same parent distribution is not null only for [\oii]$\lambda$3727 \AA ~ and [\oiii]$\lambda$4959 \AA.
These two lines, however, are detected only in a few HI-deficient cluster galaxies. The observed distributions of these two lines, whose detection rate
is low compared to the other lines, might thus not be representative in particular for the quiescent, HI-deficient cluster objects. 
We also remind that the equivalent width of the [\oiii]$\lambda$4959 \AA ~ line can be significantly overestimated (by $\sim$
40 \%) whenever the E.W.[\oiii]$\lambda$4959 $\lesssim$ 3 \AA ~ (see Sect. 5.2).\\
A systematic difference in the mean H$\alpha$ equivalent width of HI-normal and HI-deficient galaxies has been already observed in nearby clusters by Kennicutt (1983b)
and Gavazzi et al. (1991; 2002; 2006). This is the first evidence for a systematic difference in the distribution of the equivalent widths of the 
other emission lines. The observed difference in the H$\beta$ line is
not surprising since, as H$\alpha$, H$\beta$ is a direct tracer of star formation. [\oiii], [\nii] and [\sii] are not directly related to star formation
since in galaxies they can be ionised by shocks (e.g. Dopita \& Sutherland 1996). Furthermore, their intensity also depends on metallicity (Kewley \& Ellison 2008).

\subsection{Balmer decrement}

The Balmer decrement gives a direct measure of the dust extinction of the hydrogen emission lines within HII regions.
The importance of this quantity resides in the fact that, under the assumption that all 
emission lines come from the same HII regions, it can be used to correct for dust extinction all UV, optical and near-IR
emission lines of galaxies. At the same time the Balmer decrement is related to the far-IR to UV flux 
ratio and is thus often used for correcting the UV to near-IR 
stellar continuum in star forming galaxies when far infrared data are not available (Calzetti 2001).
Figure \ref{CHbdist} traces the distribution of the Balmer decrement $C(H\beta)$ (see eq. \ref{eq:c1}), or equivalently
of $A(H\alpha)$ ($A(H\alpha)$ = 1.754 $C(H\beta)$), in the HRS sample. We include in the following analysis 
all galaxies with detected H$\alpha$ and H$\beta$ emission lines without any restriction on their equivalent width.
This choice is dictated by the fact that any cut in the H$\beta$ line to restrict the analysis to
high quality data would induce a systematic bias in the analysed sample. There is indeed a tight correlation between $C(H\beta)$
and E.W.H$\beta_{emi}$ (shown in Fig. \ref{CHbEW}) indicating that the most uncertain values are those relative to the high attenuations.
We remind that the mean uncertainty in the measure of $C(H\beta)$ is $\sim$ 0.3-0.5.

   \begin{figure}
   \centering
   \includegraphics[width=8cm]{CHbdist.epsi}
   \caption{The distribution of the Balmer decrement $C(H\beta)$ or $A(H\alpha)$ for the whole the HRS sample (above) and excluding AGNs (below). The black line
   indicates the whole sample of late-type galaxies, the blue line that of galaxies with a normal HI gas content ($HI-def$ $\leq$ 0.4) 
   and the red one that of HI-deficient Virgo cluster objects ($HI-def$ $>$ 0.4).
   }
   \label{CHbdist}%
   \end{figure}

\noindent
Figure \ref{CHbdist} shows an asymmetric distribution with values of $C(H\beta)$ from 0 to $\sim$ 2 (0 $\leq$ $A(H\alpha)$ $\leq$ 3.5 mag)
peaked at $C(H\beta)$ = 0.79 $\pm$ 0.49 ($A(H\alpha)$ = 1.38 $\pm$ 0.87 mag) when the whole sample is considered, 
$C(H\beta)$ = 0.74 $\pm$ 0.47 ($A(H\alpha)$ = 1.30 $\pm$ 0.82 mag) if AGN are excluded. 
This value is smaller than the values obtained by Moustakas et al. (2006) (0 $\lesssim$ $A(H\alpha)$
$\lesssim$ 2.5 mag, with a mean value of $A(H\alpha)$ 0.51 $\pm$ 0.50) but fairly consistent with
Kennicutt (1992b) ($A(H\alpha)$ $\simeq$ 1), or with Boselli et al. (2001) for a subsample of normal, late-type galaxies such as those
analysed in this work extracted from the spectroscopic atlas of Kennicutt (1992b) ($A(H\alpha)$ = 0.78 $\pm$ 0.47 \AA)\footnote{The values of Kennicutt (1992b)
and Boselli et al. (2001) have been measured without accounting for any H$\alpha$ underlying absorption and are thus systematically underestimated with
respect to the values given in this work.}. 
We see a systematic difference between HI-deficient cluster galaxies ($C(H\beta)$ = 1.00 $\pm$ 0.54; $A(H\alpha)$ = 1.76 $\pm$ 0.94 mag
for the whole sample, $C(H\beta)$ = 0.94 $\pm$ 0.51; $A(H\alpha)$ = 1.65 $\pm$ 0.89 mag excluding AGN)
and field, HI-normal objects ($C(H\beta)$ = 0.71 $\pm$ 0.45; $A(H\alpha)$ = 1.24 $\pm$ 0.80 mag for the whole sample,
$C(H\beta)$ = 0.66 $\pm$ 0.43; $A(H\alpha)$ = 1.16 $\pm$ 0.75 mag excluding AGN). A Kolmogorov-Smirnov test 
indicates that the probability that the two distributions are driven by the same parent distribution is only of 0.15-0.30\%.\\
The H$\alpha$ and H$\beta$ emission lines used to trace the distribution given in Fig. \ref{CHbdist} can be measured in 178 
out of the 238 HRS late-type galaxies with spectroscopic data. In a large fraction of our sample (60 over 238 observed galaxies, 25\%),
the H$\beta$ emission line is not detected. This can be explained by the fact that the H$\beta$ line is intrinsically 2.86 times 
less intense than the H$\alpha$ line, thus hardly detectable in low surface brightness objects with low signal to noise spectra. These galaxies
are generally low mass (Gavazzi et al. 1996), metal poor objects (Tremonti et al. 2004), where dust attenuation is very low (Boselli et al. 2009; 
see also Fig. \ref{CHbfisici}).
The presence of low surface brightness objects in our sample where $C(H\beta)$ cannot be measured 
might thus bias the observed $C(H\beta)$ distribution given in Fig. \ref{CHbdist}
towards highly attenuated objects. At the same time, however, the H$\beta$ line can be undetected in highly attenuated objects
where H$\alpha$/H$\beta$ $\gg$ 2.86. The two effects should compensate each other, and the resulting mean value of $C(H\beta)$ and $A(H\alpha)$
given above should thus be representative of the entire sample. At the same time we do not have any reason to believe that any systematic
bias can be at the origin of the observed distribution of HI-deficient and HI-normal objects. We have to remark, however, that
because of the metallicity gradient in late-type galaxies, the inner disc of cluster objects, whose ISM is unaffected
during the interaction with the hostile environment (Boselli et al. 2006; Cortese et al. 2011b), is characterised by a higher dust extinction than the metal poor outer disc. 
The contribution of this outer disc in the integrated value of $C(H\beta)$ only in unperturbed field objects might thus explain their systematic lower values
with respect to HI-deficient systems. \\

We can study how the Balmer decrement $C(H\beta)$ depends on different intrinsic quantities characterising the observed galaxies. For this
purpose we plot in Fig. \ref{CHbfisici} the relationship between the Balmer decrement $C(H\beta)$ and the morphological type, the stellar mass $M_{star}$, 
the H band effective surface brightness $\mu_e(H)$, the birthrate parameter $b$, the metallicity index 12+log(O/H), and the star formation rate.
Different symbols are used to indicate normal (empty) from active (filled) galaxies.
Table \ref{tab:Tabfisici} gives the Spearman correlation coefficients of the relations between the Balmer decrement and these physical parameters
for the whole sample, while Table \ref{tab:Tabfisicifit} the best fit for the different subsamples of galaxies with a normal HI gas content 
($HI-def$ $\leq$ 0.4) and HI-deficient objects ($HI-def$ $>$ 0.4) as well as for star forming galaxies and for AGNs.

   \begin{figure*}
   \centering
   \includegraphics[width=15cm]{CHbfisici.epsi}
   \caption{The relationship between the Balmer decrement $C(H\beta)$ and different parameters characterising the observed galaxies.
   From left to right, upper line: the morphological type, the logarithm of the total stellar mass (in M$\odot$), the H band
   effective surface brightness (in mag arcsec$^{-2}$); lower line: the logarithm of the birthrate parameter $b$ or equivalently
   the specific star formation rate $SSFR$, the metallicity index 12+log(O/H), and the 
   star formation rate SFR (in M$\odot$ yr$^{-1}$). Blue symbols are for galaxies with a normal HI gas content ($HI-def$ $\leq$ 0.4), red symbols for HI-deficient
   objects ($HI-def$ $>$ 0.4). Filled symbols are for galaxies hosting an AGN. The black cross shows the typical uncertainty on the data.
   The blue and red lines indicate the best fit to the data for HI-normal and HI-deficient galaxies respectively
   whenever evident correlations are present. Solid lines indicate the best fit obtained including all galaxies, the dotted lines excluding those objects hosting an AGN.
   The black dotted line indicates the relation obtained by Gilbank et al. (2010) using SDSS data (increased by 13 \% ~ to take into account the difference between
   the two extinction laws used in their (Seaton 1979) and our work (Fitzpatrick \& Massa 2007)). The black short-dashed line indicates the relationships
   obtained by Garn \& Best (2010) using SDSS, increased by 30 \% ~ to take into account the difference between our extinction law and that used in their work
   (Calzetti et al 2000). The orange dotted line is the best fit for 0.75 $\leq$ $z$ $\leq$ 1.5 galaxies given by Dominguez et al. (2012), 
   the yellow, long-dashed line is the relation obtained by Lee et al. (2009) once B band absolute magnitudes are transformed into
   stellar masses using the relations log$L_H$ = -0.455 $\times$ $M_B$+1.289 and $B-V$ = 0.711 $\times$ log$L_H$ - 4.439,
   combined with the relations given in Boselli et al. (2009) for measuring stellar masses. 
   The green dotted-dashed line is the best fit given in Boselli et al. (2009).
   }
   \label{CHbfisici}%
   \end{figure*}

\noindent
The analysis of Fig. \ref{CHbfisici} and of Table \ref{tab:Tabfisici} and \ref{tab:Tabfisicifit} indicates that: \\
1) Despite a huge dispersion in the distribution of $C(H\beta)$, the attenuation of the emission lines is more important in early-type 
spirals (Sa-Sc) than in late-type Im and BCD (Stasinska et al. 2004). This evidence is consistent with the decrease of the attenuation 
of the stellar continuum determined using the far-IR to UV flux ratio ($A(FUV)$) with increasing morphological 
type observed by Buat \& Xu (1996) and Cortese et al. (2008). In early-type spirals ($\leq$ Sbc) we do not observe
any strong systematic difference between cluster HI-deficient and field normal objects. Later morphological types, however, are mainly 
objects with a normal gas content. We can thus deduce that the observed systematic difference in the $C(H\beta)$ distribution shown in Fig. \ref{CHbdist}
can also be due to the lack of HI-deficient late-type spirals. The lack of this class of objects in the HRS can be easily explained 
considering that these low mass systems are rapidly transformed into quiescent ellipticals in the Virgo cluster after a ram pressure stripping event able to remove
their total gas content and quenching their star formation activity (Boselli et al. 2008a,b). \\ 
2) $C(H\beta)$ increases with the stellar mass. The polynomial fit to the data of HI-normal galaxies (blue solid line), 
given in Table \ref{tab:Tabfisicifit} and shown in Fig. \ref{CHbfisici}, is comparable with that obtained by Boselli et al. (2009) on a 
similar set of data or with the median values determined by Stasinska et al. (2004) on SDSS data\footnote{The $C(H\beta)$
values determined from Stasinska et al. (2004) assuming the same Galactic extinction law used in this work ranges from $\sim$ 0.77 at log$M_{star}$ $\sim$ 9.37
to $C(H\beta)$ $\sim$ 1.05 at log$M_{star}$ $\sim$ 10.57.}. It
is, however, significantly steeper than those obtained using SDSS data by Gilbank et al. (2010; black dotted line) and
Garn \& Best (2010). Part of this difference is due to the fact that the two works use different extinction laws, the first Seaton (1979) which underestimates
the extinction by $\sim$ 13 \%, the second (Calzetti et al.
2000) by 30 \%. The difference in stellar mass resulting from the adoption of different recipes is of the order of $\simeq$ 0.1 in dex. Despite possible 
uncertainties resulting from the transformation of $M_B$ absolute magnitudes into stellar masses, the difference between our data for isolated, unperturbed galaxies
with the fit of Lee et al. (2009) is mainly due to selection effects. Indeed, to avoid objects with very uncertain data 
Lee et al. (2009) use only galaxies not hosting an AGN and with an equivalent width
of the \hb \ emission line larger than 5 \AA. Applying the same selection criteria, our data roughly match the fit given by
Lee et al. (2009). Our sample, however, is dominated by galaxies with E.W.H$\beta$ $\leq$ 5 \AA, and the same selection would thus drop the sample with useful data
to $\sim$ 1/3 of the total sample (see for example Fig. \ref{CHbEW} and \ref{EWdist}). Furthermore, this selection criterium might introduce a strong selection bias since low
\hb \ emission lines might come from galaxies with a strong attenuation. To check whether the high $C(H\beta)$ values determined in our sample are 
driven mainly by uncertainties in the measurement of the \hb \ line or are rather indicative of high attenuations, we have inspected the spectra and the images of
those objects with $C(H\beta)$ $>$ 1.5. Ten out of the 17 galaxies with $C(H\beta)$ $>$ 1.5 are edge-on galaxies with prominent dust lanes 
(HRS 21, 23, 149, 197, 233, 264, 273, 284, 308, 323), where the attenuation is expected to be very
important (Xiao et al. 2012). Furthermore the quality of the spectra of these 17 galaxies is sufficiently good to get fairly accurate measurements. We also checked whether the inclusion of
active galaxies (AGN, Seyferts, LINERS and retired following the classification of Stasinska et al. (2008)) might change the results. The fits given in Table
\ref{tab:Tabfisicifit} do not change significantly including or not these active galaxies, consistent with the fact that our integrated spectra are representative of entire galaxies
and thus typical of star forming regions (see Sect. 7.2).\\
Variations of the H$\alpha$ over H$\beta$ ratio with the total B band luminosity has been reported by Moustakas et al. (2006)
on both their sample of nearby galaxies with integrated spectra (Moustakas \& Kennicutt 2006) and on a much larger SDSS sample of star forming objects.
Variations with the $r$ band luminosity has been also reported by Stasinska et al. (2004). 
This trend is similar to the increase of $A(FUV)$ with the H band luminosity, 
proxy for the total stellar mass, observed by Cortese et al. (2006) or that reported by Iglesias-Paramo et al. (2007) or Salim et al. (2007)
in the local universe, and by Ly et al. (2012) at $z$ $\sim$ 0.4. \\
3) Our Balmer decrement vs. stellar mass relation for isolated galaxies is significantly steeper than that of Dominguez et al. (2012) for
objects in the redshift range 0.75 $\leq$ $z$ $\leq$ 1.5, indicative of a significant evolution.\\
4) The Balmer decrement increases with the H band effective surface brightness of the target galaxies. 
Once again, these results agrees with the variation of the dust attenuation of the stellar continuum $A(FUV)$ with $\mu_e(H)$ observed by Cortese et al. (2006)
and Johnson et al. (2007) or with the log(H$\alpha$/H$\beta$) vs. $r$-band surface brightness relation found on SDSS galaxies by Stasinska et al. (2004).\\
5) Although we do not observe any obvious relation between the Balmer decrement and the birthrate parameter $b$, the most active galaxies 
in star formation ($b$ $\gtrsim$ 4) all have $C(H\beta)$ $\lesssim$ 0.5. We should remark, however, that the dynamic range sampled in $b$ 
is quite small because relatively quiescent late-type galaxies are generally undetected in \hb. A qualitatively similar behavior has been previously observed between 
the far infrared to UV flux ratio and $b$ by Martin et al. (2007), and between log(H$\alpha$/H$\beta$) and the H$\alpha$ equivalent width, proxy of the birthrate parameter,
by Stasinska et al. (2004).\\
6) As the stellar attenuation $A(FUV)$, $C(H\beta)$ increases with the gas metallicity (Stasinska et al. 2004; Cortese et al. 2006; Johnson et al. 2007; Garn \& Best 2010; Xiao et al. 2012). 
This relationship is expected, given that metals are the major constituents of dust grains.\\
7) There is an evident but dispersed trend between $C(H\beta)$ and the star formation rate of galaxies. This trend has been already observed by 
Wang \& Heckman (1996), Sullivan et al. (2001), Perez-Gonzalez et al. (2003), Stasinska et al. (2004), Garn \& Best (2010) and Xiao et al. (2012).
Variations of the dust attenuation measured using
the far infrared to UV flux ratio with different tracers of the star formation activity have been also reported by Wang \& Heckman (1996), Buat et al. (2005), 
Iglesias-Paramo et al. (2006).\\
8) The observed trends are shared by galaxies with a normal HI gas content (blue squares) and HI-deficient objects (red circles). Consistently with 
the distributions shown in Fig. \ref{CHbdist}, however, HI-deficient galaxies tend to have slightly higher values of $C(H\beta)$ than normal objects.
We also observe the well know decrease of the star formation activity of cluster galaxies with respect to that of unperturbed field objects
due to the removal of the atomic gas reservoir (Gavazzi et al. 2002; Boselli \& Gavazzi 2006; Boselli et al. 2006, 2008a). \\
9) We do not see in any of these relations any strong systematic difference between galaxies hosting an AGN and normal, star forming objects.\\

\noindent
All these results seem robust vs. the uncertainty in the determination of the Balmer decrement, which is of the order of 0.3-0.5.

\begin{table*}
\caption{Spearman correlation coefficients $\rho$ of the relations between the Balmer decrement and the underlying \hb \ 
absorption and the physical parameters for all galaxies (Fig. \ref{CHbfisici} and \ref{UlHbfisici})}
\label{tab:Tabfisici}
{
\[
\begin{tabular}{cccccc}
\hline
\noalign{\smallskip}
Variable		& log $M_{star}$	& $\mu_e(H)$ 	   	&  log b	& 12+log(O/H)	& log $SFR$		  	\\
Units			& M$_{\odot}$		& AB mag arcsec$^{-2}$  &  		&		& M$_{\odot}$ yr$^{-1}$ 	\\
\hline
All~ galaxies		&			&			&		&		&		\\
\hline
$C(H\beta)$     	& 0.39		     	& -0.38	     		& -0.16		& 0.49		& 0.33	       \\
$E.W.H\beta_{abs}$     	&-0.16 		     	&  0.08	     		&  0.04		&-0.31 		&-0.07	       \\
\hline
Excluding ~ AGN		&			&			&		&		&		\\
\hline
$C(H\beta)$     	& 0.35		     	& -0.36	     		& -0.22		& 0.50		& 0.25	       \\
\noalign{\smallskip}
\hline
\end{tabular}
\]
}
Notes: determined using both HI-normal and HI-deficient galaxies.
\end{table*}

\begin{table*}
\caption{Relationships between $C(H\beta)$ and the different physical parameters (Fig. \ref{CHbfisici})}
\label{tab:Tabfisicifit}
{
\[
\begin{tabular}{c|cc|cc}
\hline
\noalign{\smallskip}
Sample				& All ~ galaxies							&		 & Excluding ~ AGN 							&		 \\	 
\hline
				& Regression								& 	$\rho$	 & Regression	 							&	$\rho$	\\
\hline
Non-deficient			& $C(H\beta)$ = 0.096$\times$log $M_{star}$ $^2$ - 1.344log $M_{star}$ + 4.804	& 0.41 &$C(H\beta)$ = 0.100$\times$log $M_{star}$ $^2$ - 1.400log $M_{star}$ + 5.000  & 0.39	\\       
Deficient			& $C(H\beta)$ = 0.151$\times$log $M_{star}$ $^2$ - 2.114log $M_{star}$ + 7.499	& 0.40 &$C(H\beta)$ = 0.151$\times$log $M_{star}$ $^2$ - 2.110log $M_{star}$ + 7.484  & 0.33	\\
Non-deficient			& $C(H\beta)$ = -0.503$\times$ $\mu_e(H)$ + 10.500	&	-0.43	 &		$C(H\beta)$ = -0.510$\times$ $\mu_e(H)$ + 10.631      &       -0.41          	    	\\       
Deficient			& $C(H\beta)$ = -0.450$\times$ $\mu_e(H)$ + 9.806	&	-0.23	 &		$C(H\beta)$ = -0.554$\times$ $\mu_e(H)$ + 11.893      &       -0.30          	    	\\
Non-deficient			& $C(H\beta)$ = 1.882$\times$[12+log O/H] - 15.413	&	0.48	 &		$C(H\beta)$ = 1.929$\times$[12+log O/H] - 15.798      &       0.48          	    	\\       
Deficient			& $C(H\beta)$ = 1.681$\times$[12+log O/H] - 13.482	&	0.36 	 &		$C(H\beta)$ = 1.352$\times$[12+log O/H] - 10.674      &       0.35          	    	\\
Non-deficient			& $C(H\beta)$ = 0.692$\times$log $SFR$ + 0.797		&	0.39	 &		$C(H\beta)$ = 0.541$\times$log $SFR$ + 0.772	      &       0.27           	    	\\       
Deficient			& $C(H\beta)$ = 0.756$\times$log $SFR$ + 1.308		&	0.47 	 &		$C(H\beta)$ = 0.722$\times$log $SFR$ + 1.307	      &       0.48          	    	\\       
\noalign{\smallskip}
\hline
\end{tabular}
\]
}
Notes: linear relations are determined using a bisector fit.
\end{table*}

\begin{figure*}
   \centering
   \includegraphics[width=15cm]{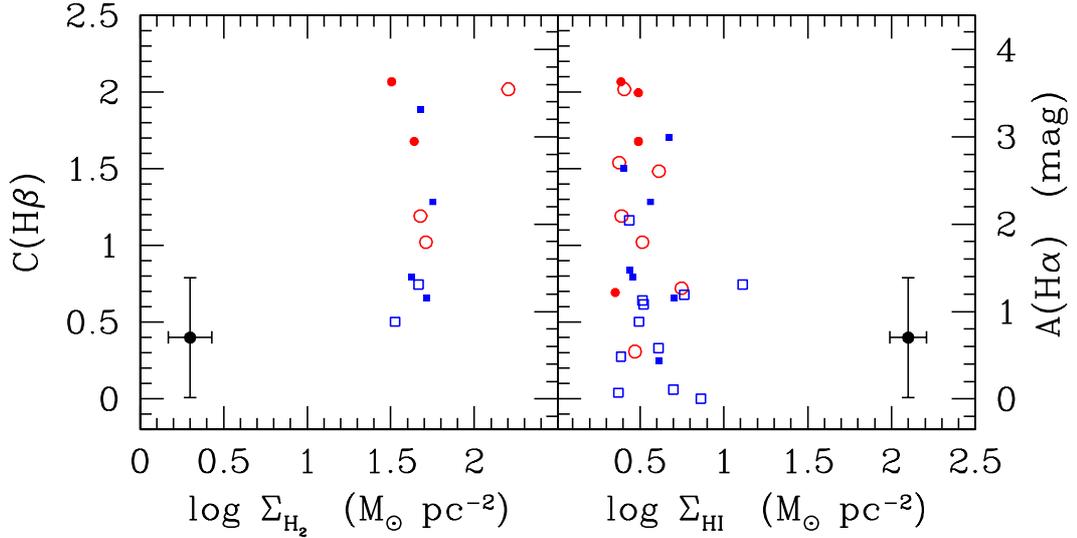}
   \caption{The relationship between the Balmer decrement $C(H\beta)$ and the atomic (right) and molecular (left) hydrogen column density (in M$\odot$ pc$^{-2}$). 
   Blue symbols are for galaxies with a normal HI gas content ($HI-def$ $\leq$ 0.4), red symbols for HI-deficient
   objects ($HI-def$ $>$ 0.4). Filled symbols are for galaxies hosting an AGN. The black cross shows the typical uncertainty on the data.
   }
   \label{CHbgas}%
   \end{figure*}

Semi-analytical models of galaxy evolution often quantify the dust attenuation within galaxies by means of their total gas column density (e.g. Guiderdoni et al.
1998).
Although limited to a very small subsample of objects having high resolution HI and CO data, we can test whether this assumption is valid
in normal, star forming galaxies such as those belonging to the HRS. Figure \ref{CHbgas} shows the relationships between
the Balmer decrement $C(H\beta)$ and the molecular and atomic gas column densities. We do not consider the total gas column density 
which is obviously dominated by the molecular gas phase (see Figure \ref{CHbgas}).

\noindent
The Balmer decrement does not depend on the HI and H$_2$ gas column density. Given the small number of galaxies with HI and CO imaging data, however, and the
small dynamic range sampled in $C(H\beta)$, $\Sigma_{HI}$ and $\Sigma_{H_2}$, we cannot exclude that we do not see any relation just because of the poor statistics. 
Indeed quite dispersed variations have been already reported in the literature between $A(FUV)$ and the total gas column density (Xu et al. 1997; Boquien et al., in prep.),
although based on data of much lower quality.\\

\begin{figure}
   \centering
   \includegraphics[width=9cm]{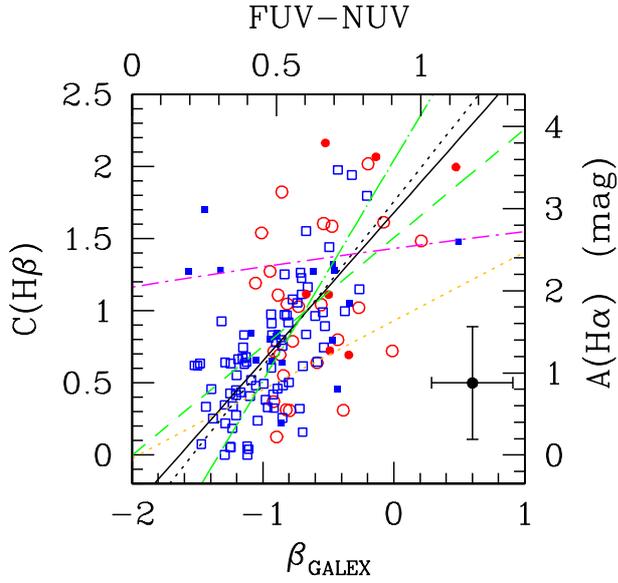}
   \caption{The relationship between the Balmer decrement $C(H\beta)$ and the UV slope $\beta_{GALEX}$ determined using GALEX data.  
   Blue symbols are for galaxies with a normal HI gas content ($HI-def$ $\leq$ 0.4), red symbols for HI-deficient
   objects ($HI-def$ $>$ 0.4). Filled symbols are for galaxies hosting an AGN. The balck cross shows the typical uncertainty on the data.
   The black solid line indicates the best bisector fit to our data
   when all galaxies are included, the dotted line excluding AGNs. The orange dotted line shows the fit obtained by Hao et al. (2011), 
   the green dotted-long dashed line and the dashed line the best fit for normal and starburst galaxies by Cortese et al. (2006), and the magenta
   dotted short-dashed line that of GAMA/H-ATLAS galaxies of Wijesinghe et al. (2011).}
   \label{CHbbeta}%
   \end{figure}

The Balmer decrement can be compared to other standard tracers of the dust attenuation within galaxies. Among these, the most widely used are the far-IR to UV
flux ratio (Buat \& Xu 1996, Gordon et al. 2000, Witt \& Gordon 2000), and the slope of the UV spectrum $\beta$ (Meurer et al. 1999, Calzetti et al. 2000,
Calzetti 2001). The far-IR to UV flux ratio has been generally measured using FUV GALEX and
either IRAS or Spitzer IR data for different samples of galaxies. Owing to the advent of the $Herschel$ space mission, which is producing imaging data in the
70 to 500 $\mu$m spectral domain, we can significantly increase the quality of the measure of the total far-IR luminosity of nearby galaxies
using spectral energy distribution fitting codes. The SPIRE observations of the HRS galaxies have been recently completed (Ciesla et al. 2012.). PACS observations, more critical for the determination of the total far-IR luminosity since they cover the 70-170 $\mu$m spectral domain,
just started. For this reason we will postpone the investigation of the relationship of the $C(H\beta)$ vs. total far-IR to UV flux ratio to a future communication. 
Here we focus on the comparison of the Balmer decrement with the slope of the UV spectrum $\beta$, defined as in Overzier et al. (2011):

\begin{equation}
{\beta_{GALEX} = 2.22[FUV-NUV]-2.0}
\end{equation} 

\noindent
where $FUV$ and $NUV$ are the GALEX data in the far ($\lambda$ = 1539 \AA) and near ($\lambda$ = 2316 \AA) UV bands, corrected for galactic extinction using
the Schlegel et al. (1998) Milky Way extinction map and the Fitzpatrick \& Massa (2007) extinction law. 
Figure \ref{CHbbeta} shows the relationship between the Balmer decrement $C(H\beta)$ and the $\beta_{GALEX}$ parameter for the HRS galaxies.

\noindent

Figure \ref{CHbbeta} shows a very dispersed correlation between the two variables (bisector fit: $C(H\beta)$ = 1.03 $\times$ $\beta_{GALEX}$ +1.68; 
Spearman coefficient $\rho$ = 0.56) shared by HI-deficient and HI-normal galaxies (solid line). A very similar relation is obtained excluding those galaxies 
hosting an AGN (bisector fit: $C(H\beta)$ = 1.14 $\times$ $\beta_{GALEX}$ +1.75; Spearman coefficient $\rho$ = 0.61).
Given the large dispersion of the observed relation,
our data are $\sim$ consistent with the best fit of Cortese et al. (2006) for normal galaxies (dotted-long dashed line),
and with that of starbursts (dashed line). Although roughly matching their extremely dispersed data, our fit is completely inconsistent with the best fit given by 
the GAMA/H-ATLAS collaboration by Wijesinghe et al. (2011; dotted short-dashed line). This significant difference in the fit (but not necessarily in the data)
might be partly due to the quite uncertain aperture corrections necessary to transform nuclear emission line data form GAMA/H-ATLAS to integrated values.
The best fit given by Hao et al. (2011) is significantly flatter than the relation observed in our data. It is worth noticing that the 
Hao et al. sample, however, spans a smaller range in Balmer decrement ($A(H\alpha)$ $\lesssim$ 2) than ours, and at the same time includes 
quite a few objects with $\beta_{GALEX}$ $\gtrsim$ 0.

   \begin{figure}
   \centering
   \includegraphics[width=8cm]{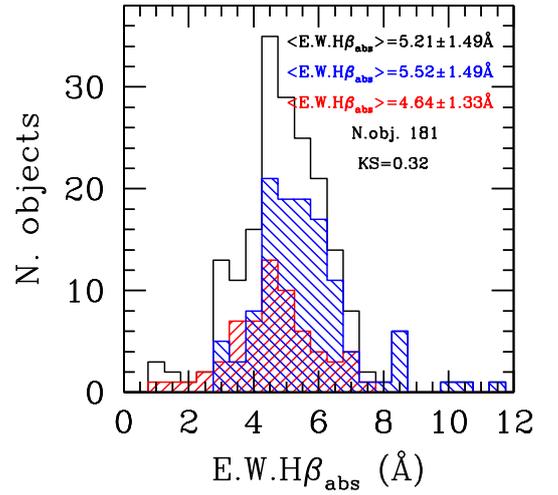}
   \caption{The distribution of the equivalent width of the H$\beta$ underlying absorption line (in \AA) in the HRS sample. The black line
   indicates the whole sample of late-type galaxies, the blue line that of galaxies with a normal HI gas content ($HI-def$ $\leq$ 0.4) 
   and the red one that of HI-deficient Virgo cluster objects ($HI-def$ $>$ 0.4).
   }
   \label{UlHbdist}%
   \end{figure}

\subsection{Underlying Balmer absorption}

The underlying Balmer absorption features observed in the spectra of galaxies are often used as direct tracers of the mean age of the 
underlying stellar populations since they are due to the absorption of the stellar continuum in the atmosphere of warm A-type stars 
(Poggianti \& Barbaro 1997; Thomas et al. 2004; Le Borgne et al. 2004). 
Thanks to the spectral resolution of CARELEC, the H$\beta$ line is easily observable in the spectra of the HRS galaxies. 
We can thus trace the distribution of the equivalent width of the H$\beta$ underlying absorption in a complete, volume limited 
sample of nearby galaxies. Figure \ref{UlHbdist} shows the distribution of the E.W.H$\beta_{abs}$ for the 181/260 late-type 
galaxies where this feature has been observed and measured.

\begin{figure*}
   \centering
   \includegraphics[width=15cm]{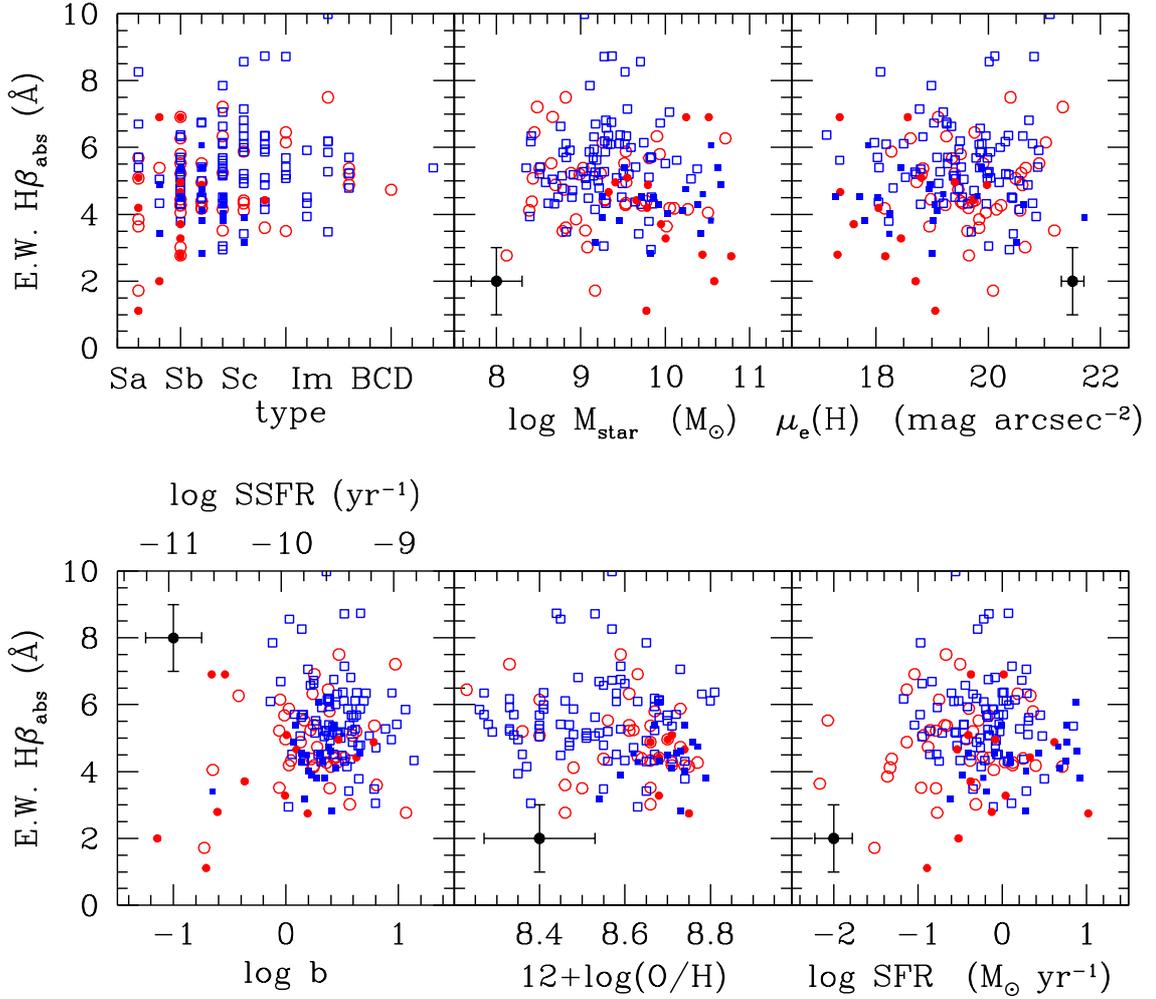}
   \caption{The relationship between the underlying Balmer absorption E.W.H$\beta_{abs}$ and different parameters characterising the observed galaxies.
   From left to right, upper line: the morphological type, the logarithm of the total stellar mass (in M$\odot$), the H band
   effective surface brightness (in mag arcsec$^{-2}$); lower line: the logarithm of the birthrate parameter $b$, or equivalently the specific star formation rate $SSFR$,
    the metallicity index 12+log(O/H), and the star formation rate (in M$\odot$ yr$^{-1}$). Blue symbols are for galaxies with a normal 
    HI gas content ($HI-def$ $\leq$ 0.4), red symbols for HI-deficient
   objects ($HI-def$ $>$ 0.4). Filled symbols indicate galaxies hosting an AGN. The black cross shows the typical uncertainty on the data for a galaxy of equivalent width of 5
   \AA.
   }
   \label{UlHbfisici}%
   \end{figure*}

\begin{figure*}
\centering
\includegraphics[width=0.33\textwidth]{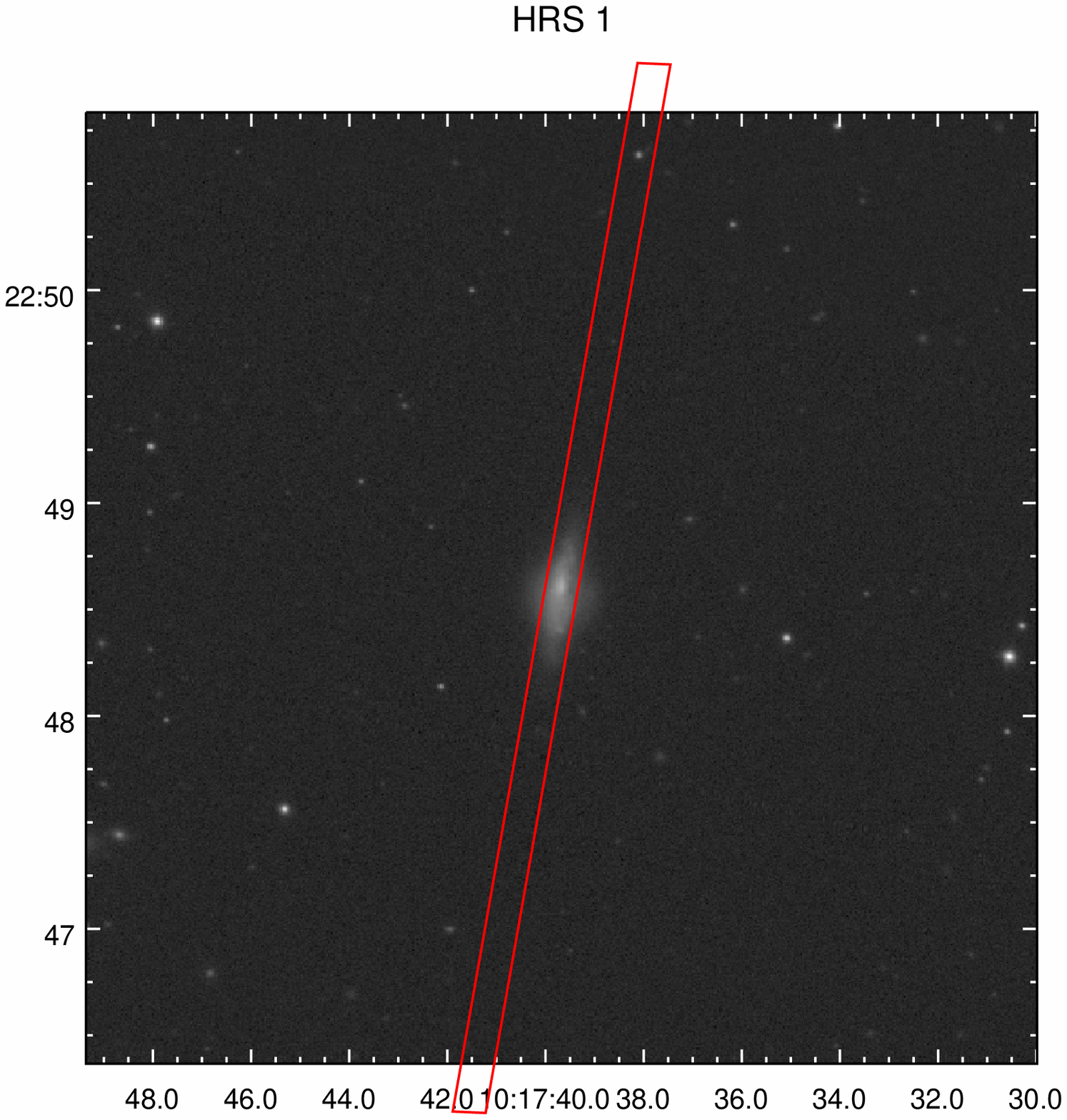}
\includegraphics[width=0.66\textwidth]{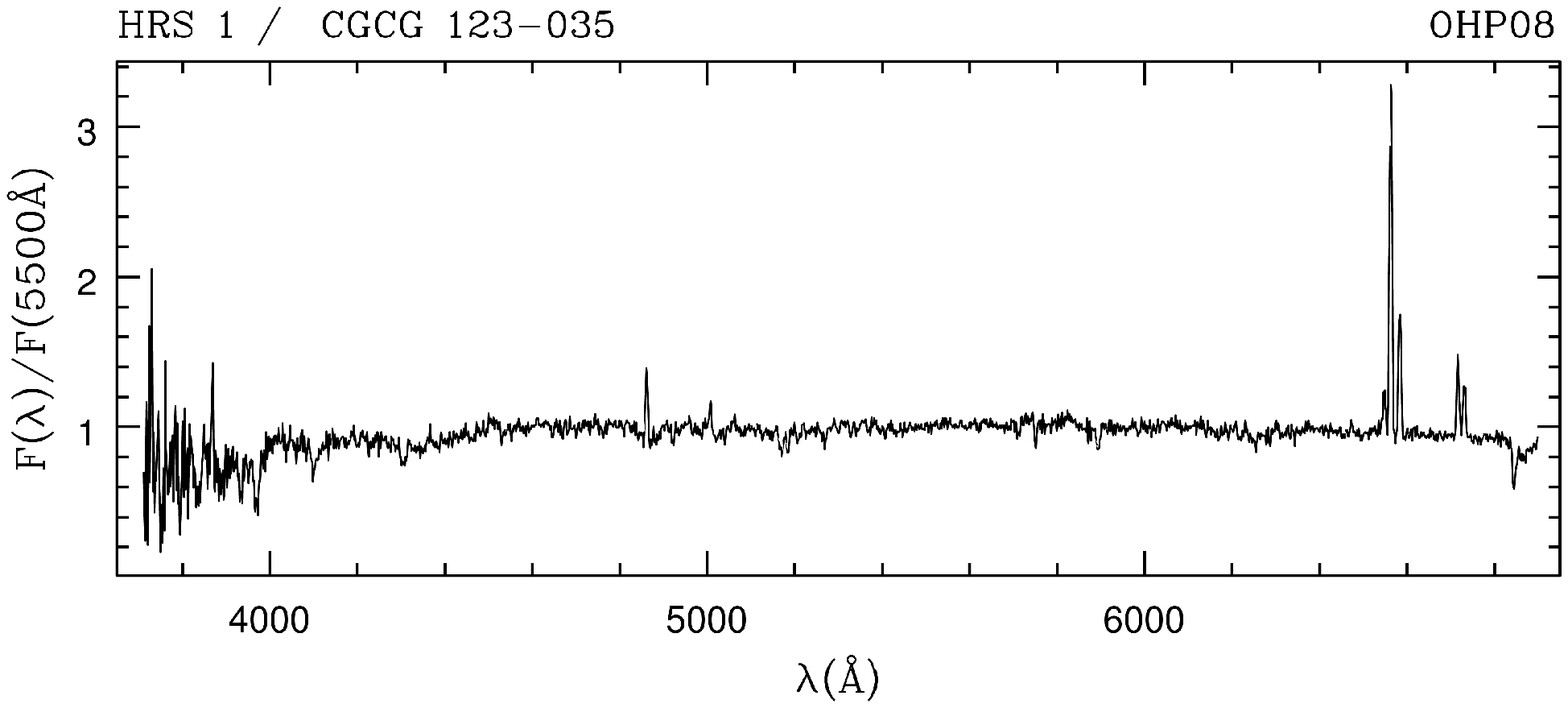}\\
\includegraphics[width=0.33\textwidth]{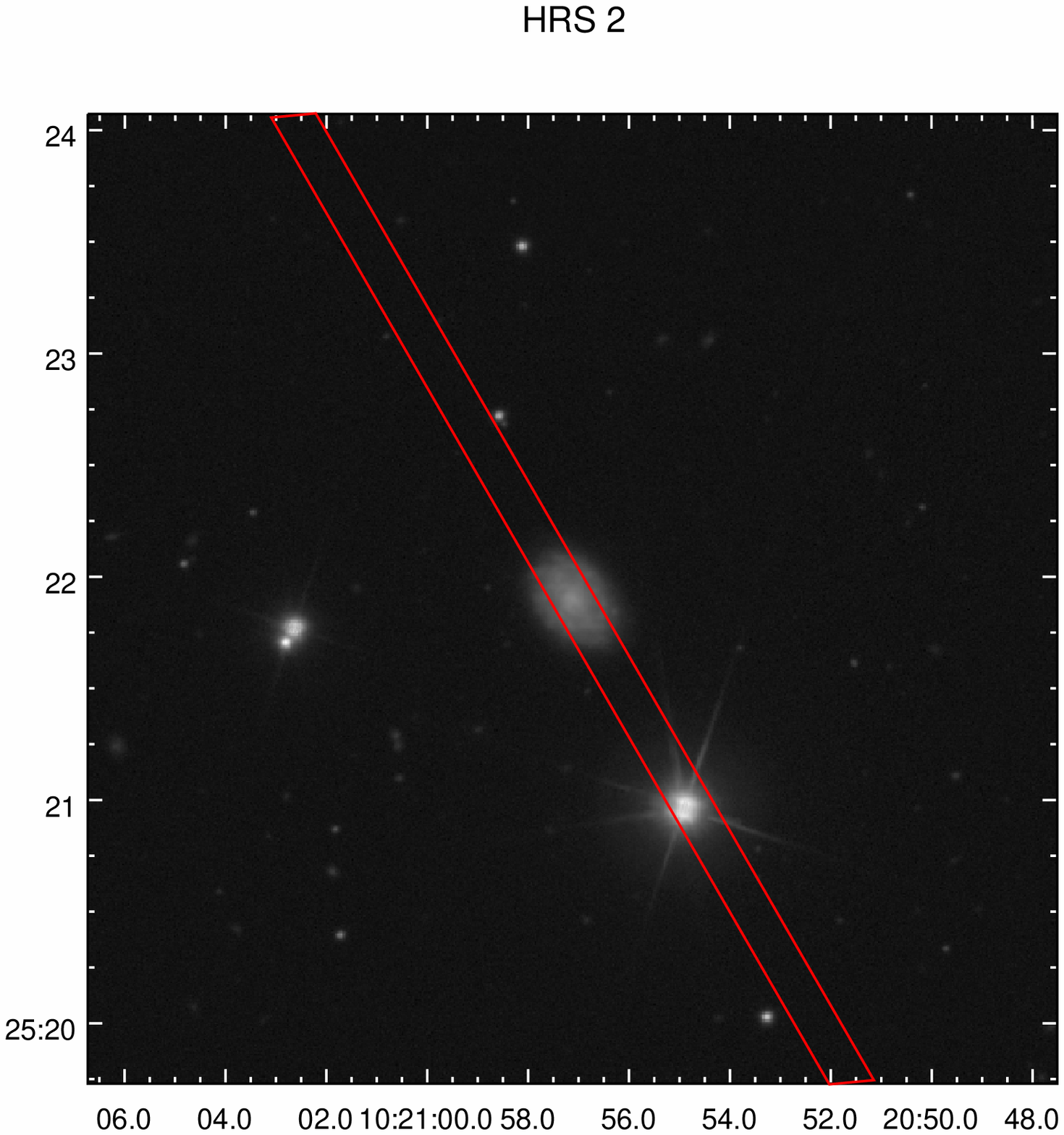}
\includegraphics[width=0.66\textwidth]{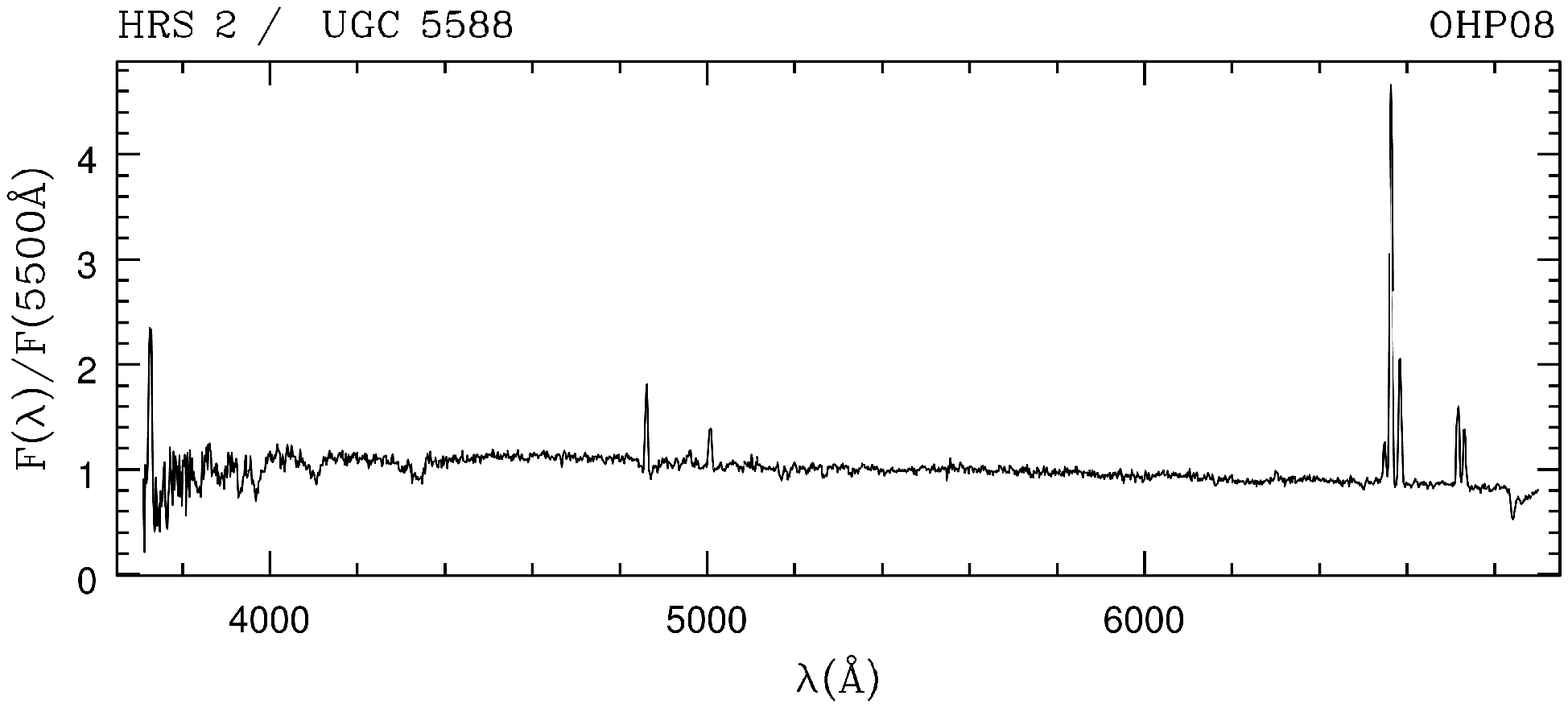}\\
\includegraphics[width=0.33\textwidth]{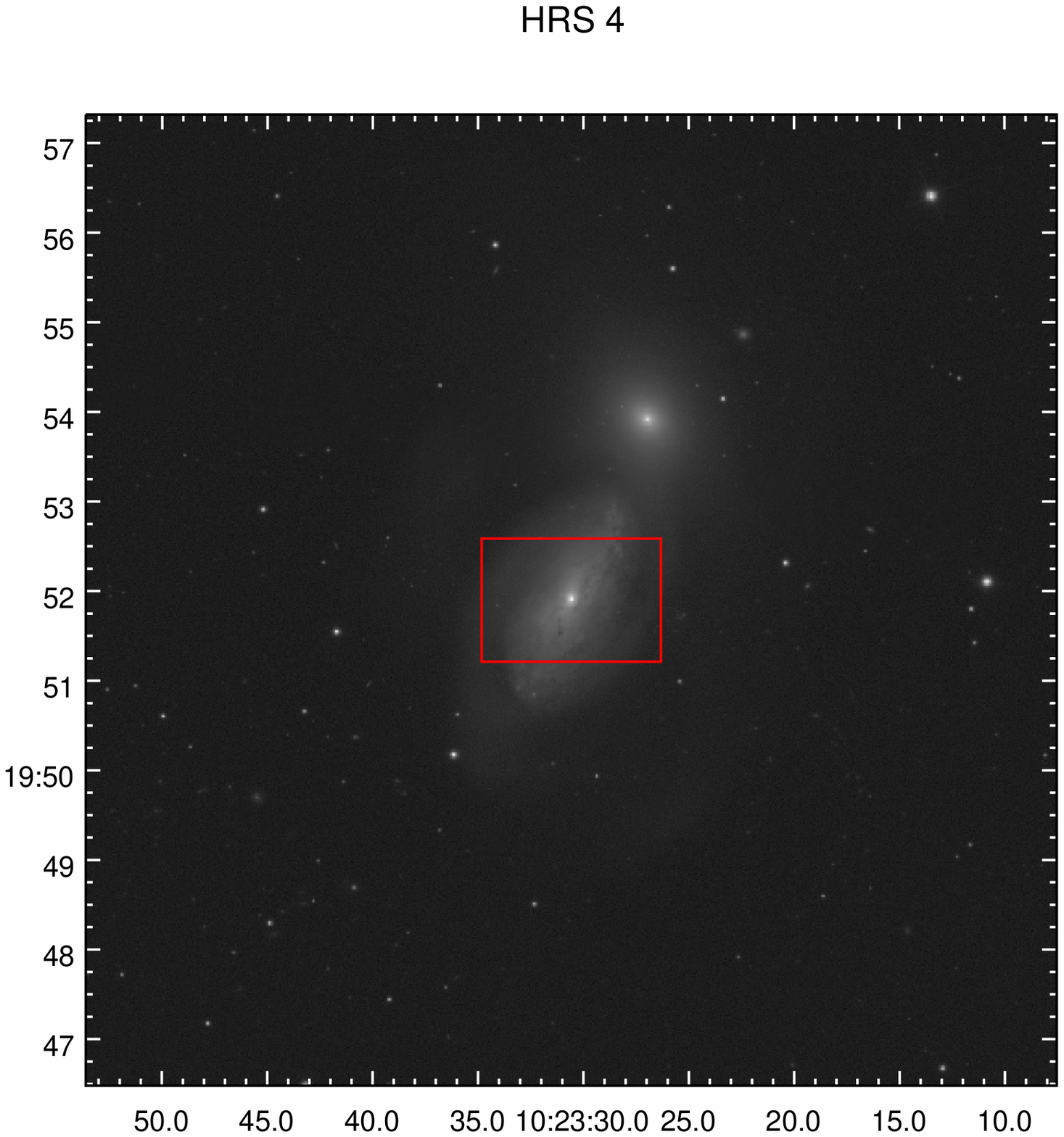}
\includegraphics[width=0.66\textwidth]{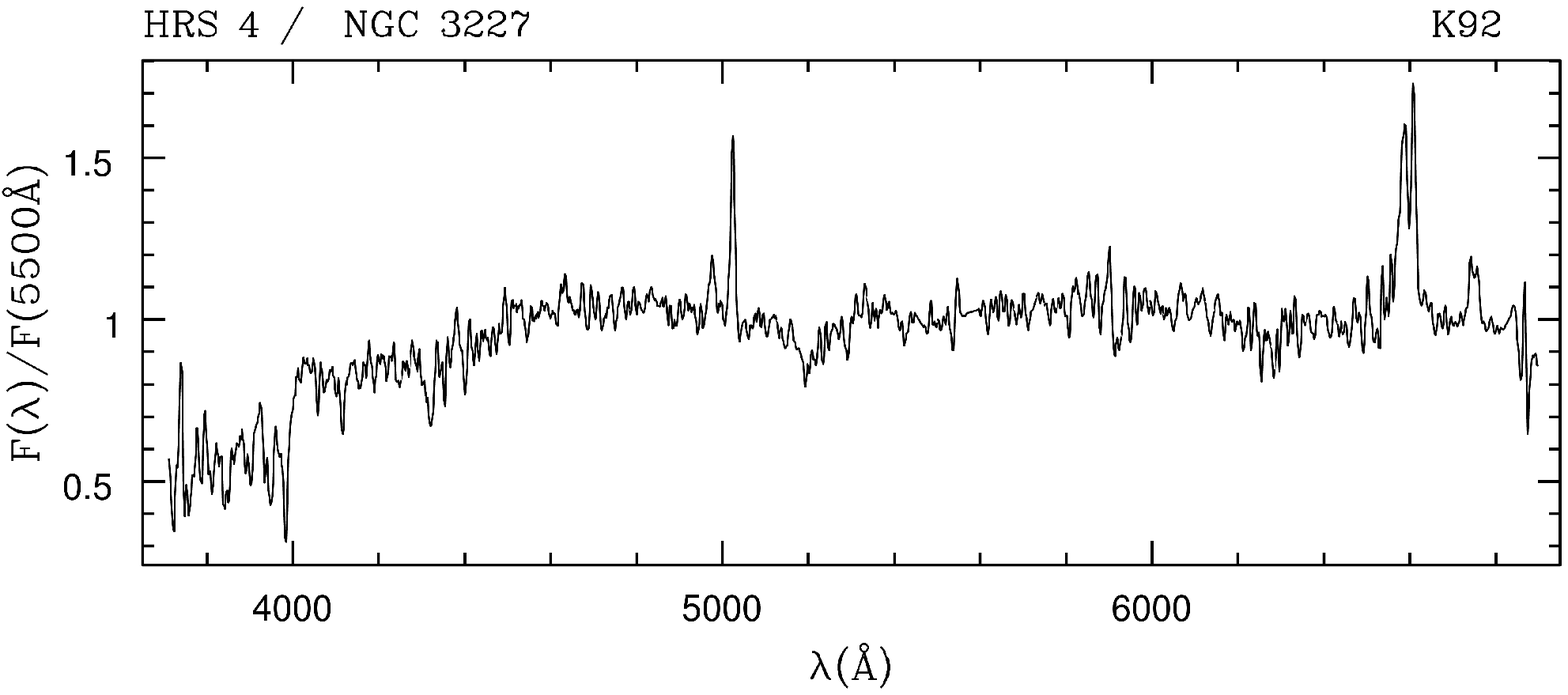}\\
\includegraphics[width=0.33\textwidth]{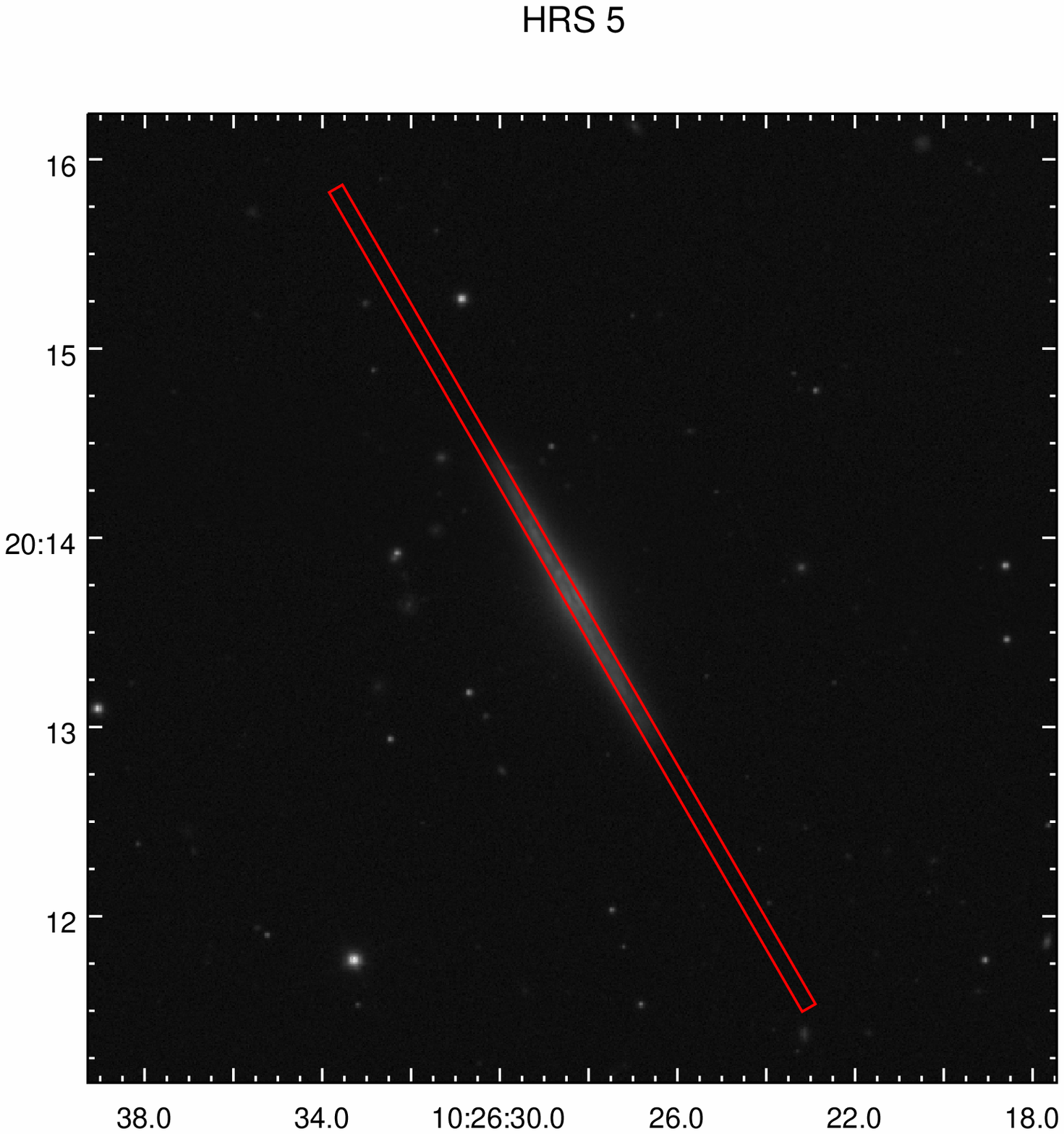}
\includegraphics[width=0.66\textwidth]{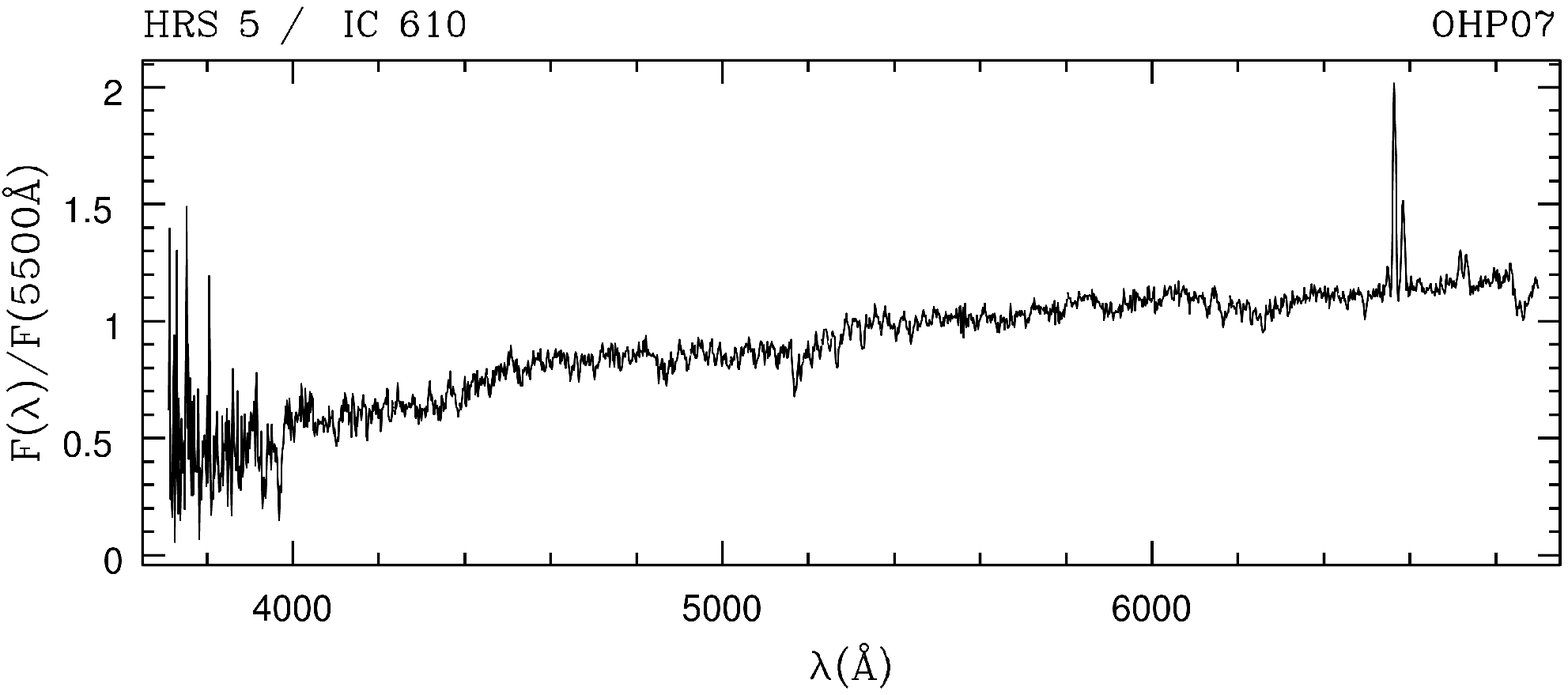}\\
\caption{Integrated spectral atlas of the HRS galaxies, in order of increasing HRS name. Left: the $r$ band SDSS image of the observed galaxies with overimposed the 
position covered by the slit during the observations (red solid line). Right: observed spectra normalised to their mean intensity between $\lambda$ = 5400 - 5600 \AA.
The full atlas is available in electronic format on the HeDaM database. }
   \label{spectra}%
   \end{figure*}

\noindent
The mean value of the distribution is E.W.H$\beta_{abs}$ = 5.21 $\pm$ 1.49 \AA, a value consistent with that measured by Gavazzi et al. (2004) (E.W.H$\beta_{abs}$ = 5.7 $\pm$ 1.9 \AA).
By fitting population synthesis models to the observed spectra of their galaxies, Moustakas \& Kennicutt (2006) found a mean E.W.H$\beta_{abs}$ = 4.40 $\pm$ 0.63 \AA,
while Moustakas et al. (2010) give E.W.H$\beta_{abs}$ = 2.5 $\pm$ 1.0 \AA.
Combining the integrated spectra of Kennicutt (1992b) with their own measurements of dwarf irregular galaxies, Kobulnicky et al. (1999) found E.W.H$\beta_{abs}$ 
ranging from 1 and 6 \AA, with a mean value of E.W.H$\beta_{abs}$ = 3 $\pm$ 2 \AA. Smaller values have been obtained for BCD galaxies by Izotov et al. (1994) 
(0.0 $\leq$ E.W.H$\beta_{abs}$
$\leq$ 3.5), or for extragalactic HII regions by McCall et al. (1985) (E.W.H$\beta_{abs}$ = 1.4 $\pm$ 0.2 \AA). Figure \ref{UlHbdist} also shows a systematic 
significant difference between the distribution of the underlying Balmer absorption at \hb \ of galaxies with a normal HI gas content 
(E.W.H$\beta_{abs}$ = 5.52 $\pm$ 1.49 \AA) and HI-deficient objects (E.W.H$\beta_{abs}$ = 4.64 $\pm$ 1.33 \AA). A Kolmogorov-Smirnov test
indicates that the probability that the two distributions are driven by the same parent distribution is only of 0.32\%.
We are not aware of any other systematic study of any possible dependence of the E.W.H$\beta_{abs}$ index observed in 
late-type galaxies belonging to different environments. Variations of E.W.H$\beta_{abs}$ as a function of the environment have been reported in Smith et al. (2009) 
for dwarf early-type galaxies. This work has shown that dwarf ellipticals belonging to the core of the Coma cluster have, on average, stronger 
H$\beta$ absorption lines than similar objects at the cluster periphery. All these evidences are consistent with the predictions of our chemo-spectrophotometric 
models of galaxy evolution expecially tailored to take into account the effects of 
the interaction with the cluster environment (Boselli et al. 2006; 2008a, 2008b). These models predict that ram pressure stripping is able to remove
a fraction of the cold gas component quanching, on relatively short time scales ($\lesssim$ 1 Gyr), the activity of star formation. The quenching of the star formation activity
induces a decrease of the equivalent width of the H$\beta$ absorption line on time scales of the order of 0.5-0.8 Gyrs, as depicted in Fig. 20 of Boselli et al. (2008a). 
Late-type HI-deficient galaxies are thus
expected to have a lower equivalent width of the H$\beta$ absorption line than gas-rich star forming systems of similar luminosity, as indeed shown in Fig. \ref{UlHbdist}.
\\
Figure \ref{UlHbfisici} shows how the underlying absorption of the H$\beta$ line is related to other physical parameters characterising the sample galaxies.
The Spearman regression coefficients of the different relations are given in Table \ref{tab:Tabfisici}. 
Figure \ref{UlHbfisici} and Table \ref{tab:Tabfisici} do not show any evident strong trend between E.W.H$\beta_{abs}$ and any of the parameters, not even the birthrate parameter $b$, which is 
tightly related to star formation history of galaxies. There are however some indications that the \hb \ underlying absorption is barely anticorrelated to the
metallicity of the parent galaxy ($\rho$ = -0.31).
Variations of the equivalent width of the H$\beta$ absorption line with the $r$ band absolute magnitude have been reported by Smith et al. (2009) 
for the early-type galaxy population in the Coma cluster. A direct comparison of our results with those of Smith et al. (2009) 
is, however, difficult because our sample is limited to the late-type galaxy population while their includes only early-types. 
In unperturbed late-type galaxies the underlying absorption comes mainly from A-type stars formed through a rather constant star formation activity, while in 
elliptical galaxies these stars have been produced by the last, and generally short-lived, episode of star formation. 
Despite the fact that the H$\beta$ equivalent width of the underlying absorption is tightly connected to the mean age
of the underlying stellar population, a direct interpretation of this index and of its variations in terms of age is quite difficult
in star forming systems such as those analysed in this work.
As depicted in Fig. \ref{sandage}, in late-type galaxies E.W.H$\beta_{abs}$ is expected to be fairly constant for objects with stellar masses 
10$^{8.8}$ $\lesssim$ $M_{star}$ $\lesssim$ 9$^{9.8}$ M$\odot$, which is the stellar mass range mainly covered by our sample. A small
decrease of the E.W.H$\beta_{abs}$ is expected only for higher stellar masses, where the decrease of the star formation over cosmic time 
has been sufficiently rapid to produce any effect. For comparison, Le Borgne et al. (2004) give the variation of the H$\beta$ equivalent 
width as a function of the stellar age for an instantaneous burst, 
representaive of the star formation history of massive ellipticals or globular clusters.
As previously mentioned, however, the equivalent width of the
H$\beta$ absorption line also depends on metallicity (higher values of E.W.H$\beta_{abs}$
are expected in metal-rich systems, see Fig. \ref{sandage}). Given the mass-metallicity relation observed also in our sample (Hughes et al. 2012),
any variation of E.W.H$\beta_{abs}$ with stellar age is neutralized by the dependence of E.W.H$\beta_{abs}$ on the metallicity. For the same reason 
we expect that any possible dependence of the observed E.W.H$\beta_{abs}$ with any tracer of the star formation activity of the parent galaxies
($SFR$, $b$ and indirectly 12+log(O/H)) is removed by the combined dependence of this parameter on age and metallicity.

\section{Conclusions}

We present new long-slit integrated spectroscopy in the spectral range 3600-6900 \AA ~ at a resolution $R$ $\simeq$ 1000 
of 138 late-type galaxies belonging to the $Herschel$ Reference Survey. Combined with those already collected by our team
(Gavazzi et al. 2004) or those available in the literature, these observations provide us with spectroscopic data for 238 out of the 260 late-type galaxies of the sample.
The completeness of the observations allow us to make a statistical analysis of the spectroscopic properties of a complete, K-band selected, volume limited sample of nearby galaxies
(15 $\leq$ $D$ $\leq$ 25 Mpc) spanning a large range in morphology and luminosity, and belonging to different environments (Virgo cluster vs. field).\\
This analysis has shown systematic differences in the spectroscopic properties of normal, gas rich field late-type galaxies and cluster HI-deficient objects. 
HI-deficient ($HI-def$ $>$ 0.4) Virgo cluster galaxies have, on average, equivalent widths of all the sampled emission lines significantly smaller than those of 
similar objects with a normal HI content. The observed difference in those lines sensitive to the present day star formation activity (H$\alpha$, H$\beta$) 
is a further confirmation that the star formation activity of cluster galaxies is quenched whenever the gas reservoir is removed during any interaction with the hostile environment (e.g. Boselli
\& Gavazzi 2006). The observed difference in the other lines ([\oiii], [\nii] and [\sii]) is probably a geometrical effect due to the smaller extension of the emitting disc
of HI-deficient galaxies linked to the suppression of star formation in the outer disc (e.g. Boselli et al. 2006).\\
This analysis also shows that the Balmer decrement is systematically higher in HI-deficient cluster galaxies with respect to field objects with a normal gas content.
This difference might be due to two different effects: 1) a lack of HI-deficient low luminosity, late-type spirals and 2) the mean extinction of gas and dust truncated discs 
observed in HI-deficient objects is higher than in normal, field galaxies just because measured only in the metal rich inner disc. We also observe that the underlying
Balmer absorption is systematically lower in cluster, HI-deficient objects than in field galaxies.\\
The Balmer decrement is correlated with the stellar mass, the stellar surface density, the metallicity and the star formation rate of the observed galaxies.
The correlation with stellar mass is significantly steeper than that determined in other works based either on integrated spectroscopy or SDSS data.
Although dust is tightly associated to the gaseous phase of the interstellar medium, we do not observe any relation between the Balmer decrement and  
the column density of the atomic and molecular gas. On the contrary, the underlying Balmer absorption looks almost independent from
stellar mass, stellar surface density, birthrate parameter, star formation rate and gas metallicity. 
We also observe a relation, although scattered, between the Balmer decrement and the slope of the UV spectrum, an independent tracer of the UV attenuation in galaxies.
The relation is consistent with that found in different samples of resolved, nearby galaxies, but is significantly different from that determined 
for the GAMA/H-ATLAS collaboration. The huge dispersion observed in all these relations clearly indicates that none of the sampled physical parameters can be used without introducing a large
uncertainty as a proxy of the dust attenuation in galaxies. This result might thus have major implications in the interpretation of spectro-photometric data gathered in cosmological 
surveys and in the calibration of semi-analytical models of galaxy evolution.

\begin{acknowledgements}

We are grateful to the anonymous referee for the accurate and detailed reviewing and for the constructive comments that
significantly helped to increase the quality of the manuscript.
We want to thank the OHP telescope operators J. Taupenas et J.P. Troncin 
for their invaluable support during the observations and the PNCG for their generous allocation
of time at the OHP telescope. A.B thanks the ESO visiting program committee for inviting him
at the Garching headquarters for a two months staying. We are also grateful to P. Anders, S. Boissier, M. Boquien, L. Ciesla and M. Fossati
for their help during the preparation of the figures. We thank R. Jansen for providing us 
with the fits files of HRS galaxies in common. TMH acknowledges the support of a Kavli Research Fellowship. 
The research leading to these results has received funding from the European Community's Seventh 
Framework Programme (/FP7/2007-2013/) under grant agreement No 229517. This research has made use of the 
NASA/IPAC Extragalactic Database (NED) 
which is operated by the Jet Propulsion Laboratory, California Institute of 
Technology, under contract with the National Aeronautics and Space Administration
and of the GOLDMine database (http://goldmine.mib.infn.it/).
IRAF is distributed by the National Optical Astronomy Observatory, 
which is operated by the Association of Universities for Research in Astronomy 
(AURA) under cooperative agreement with the National Science Foundation.

\end{acknowledgements}

\onecolumn
\input{hrsobsparam.tex}
\twocolumn
\onecolumn
\input{tabella_noncor_flux.tex}

\twocolumn
\onecolumn
\input{tabella_noncor_ew.tex}

\twocolumn
\onecolumn
\input{EWletteratura.tex}
\twocolumn
\onecolumn
\input{letteratura_flux.tex}
\twocolumn

\clearpage

\end{document}

%% file: hrsobsparam.tex
\begin{landscape}
\begin{center} \tiny 
\begin{longtable}{l c c c c c c c c l c c c c c c c}
\caption[The general properties of the observed galaxies.]{The general properties of observed late-type galaxies of the HRS sample.\label{tab:hrsgalaxies}}\\
\hline \hline \\[-2ex]
\multicolumn{1}{c}{HRS} &
\multicolumn{1}{c}{CGCG} &
\multicolumn{1}{c}{VCC} &
\multicolumn{1}{c}{UGC} &
\multicolumn{1}{c}{NGC} &
\multicolumn{1}{c}{IC} &
\multicolumn{1}{c}{R.A.(h m s)} &
\multicolumn{1}{c}{Dec.($^{\circ \, \prime \, \prime \prime}$)} &
\multicolumn{1}{c}{D (Mpc)} &
\multicolumn{1}{c}{Type} &
\multicolumn{1}{c}{K$_{Stot}$} &
\multicolumn{1}{c}{D$_{25}$ ($\prime$)} &
\multicolumn{1}{c}{V (km s$^{-1}$)} &
\multicolumn{1}{c}{Member} &
\multicolumn{1}{c}{Run} &
\multicolumn{1}{c}{Phot.} &
\multicolumn{1}{c}{T$_{exp} / s$} \\
\multicolumn{1}{c}{(1)} &
\multicolumn{1}{c}{(2)} &
\multicolumn{1}{c}{(3)} &
\multicolumn{1}{c}{(4)} &
\multicolumn{1}{c}{(5)} &
\multicolumn{1}{c}{(6)} &
\multicolumn{1}{c}{(7)} &
\multicolumn{1}{c}{(8)} &
\multicolumn{1}{c}{(9)} &
\multicolumn{1}{c}{(10)} &
\multicolumn{1}{c}{(11)} &
\multicolumn{1}{c}{(12)} &
\multicolumn{1}{c}{(13)} &
\multicolumn{1}{c}{(14)} &
\multicolumn{1}{c}{(15)} &
\multicolumn{1}{c}{(16)} &
\multicolumn{1}{c}{(17)} \\ [0.5ex] \hline
   \\[-1.8ex]
\endfirsthead
\multicolumn{17}{c}{{\tablename} \thetable{} -- Continued} \\[0.5ex]
  \hline \hline \\[-2ex]
\multicolumn{1}{c}{HRS} &
\multicolumn{1}{c}{CGCG} &
\multicolumn{1}{c}{VCC} &
\multicolumn{1}{c}{UGC} &
\multicolumn{1}{c}{NGC} &
\multicolumn{1}{c}{IC} &
\multicolumn{1}{c}{R.A.(h m s)} &
\multicolumn{1}{c}{Dec.($^{\circ \, \prime \, \prime \prime}$)} &
\multicolumn{1}{c}{D (Mpc)} &
\multicolumn{1}{c}{Type} &
\multicolumn{1}{c}{K$_{Stot}$} &
\multicolumn{1}{c}{D$_{25}$ ($\prime$)} &
\multicolumn{1}{c}{V (km s$^{-1}$)} &
\multicolumn{1}{c}{Member} &
\multicolumn{1}{c}{Run} &
\multicolumn{1}{c}{Phot.} &
\multicolumn{1}{c}{T$_{exp} / s$} \\
\multicolumn{1}{c}{(1)} &
\multicolumn{1}{c}{(2)} &
\multicolumn{1}{c}{(3)} &
\multicolumn{1}{c}{(4)} &
\multicolumn{1}{c}{(5)} &
\multicolumn{1}{c}{(6)} &
\multicolumn{1}{c}{(7)} &
\multicolumn{1}{c}{(8)} &
\multicolumn{1}{c}{(9)} &
\multicolumn{1}{c}{(10)} &
\multicolumn{1}{c}{(11)} &
\multicolumn{1}{c}{(12)} &
\multicolumn{1}{c}{(13)} &
\multicolumn{1}{c}{(14)} &
\multicolumn{1}{c}{(15)} &
\multicolumn{1}{c}{(16)} &
\multicolumn{1}{c}{(17)} \\ [0.5ex] \hline
  \\[-1.8ex]
\endhead 
 \\[-1.8ex] \hline
  \multicolumn{17}{c}{{Continued on next page\ldots}} \\
\endfoot

  \\[-1.8ex] \hline \hline
\endlastfoot
  1     &   123035   & -        &  -	    &  -        &  -		  & 	     10 17 39.6      &  +22 48 36.1     & 16.79        & Pec		    & 11.59	   & 1.0       & 1175	    &  Leo Cl.  		  &   O08      &   C	   &   1  x 1800   \\ 
  2     &   124004   & -        &  5588     &  -        &  -		  & 	     10 20 57.2      &  +25 21 54.0     & 18.44        & S?		    & 11.03	   & 0.52      & 1291	    &  Leo Cl.  		  &   O08      &   P	   &   1  x 1800   \\ 
  5     &   94052    & -        &  -	    &  -        &  610  	  & 	     10 26 28.4      &  +20 13 39.8     & 16.71        & Sc		    & 9.94	   & 1.86      & 1170	    &  Leo Cl.			  &   O07      &   P	   &   1  x 1200   \\ 
  10    &   183028   & -        &  5738     &  -        &  -		  & 	     10 34 29.8      &  +35 15 24.0     & 21.66        & S?		    & 11.31	   & 0.91      & 1516	    &  Leo Cl.			  &   O09      &   C	   &   1  x 1800   \\ 
  11    &   124038   & -        &  5742     &  3287     &  -		  & 	     10 34 47.3      &  +21 38 52.0     & 18.93        & SB(s)d 	    & 9.78	   & 2.09      & 1325	    &  Leo Cl.			  &   O07      &   P	   &   1  x 1800   \\ 
  12    &   124041   & -        &  -	    &  -        &  -		  & 	     10 35 42.0      &  +26 07 33.0     & 19.89        & cI		    & 11.98	   & 0.59      & 1392	    &  Leo Cl.  		  &   O08      &   C	   &   1  x 1800   \\ 
  13    &   183030   & -        &  5753     &  3294     &  -		  & 	     10 36 16.2      &  +37 19 28.8     & 22.47        & SA(s)c 	    & 8.38	   & 3.55      & 1573	    &  Leo Cl.  		  &   O07      &   P	   &   1  x 1200   \\ 
  16    &   94116    & -        &  5842     &  3346     &  -		  & 	     10 43 38.9      &  +14 52 18.6     & 18.00        & SB(rs)cd	    & 9.59	   & 2.69      & 1260	    &  Leo Cl.  		  &   O07      &   P	   &   1  x 1800   \\
\end{longtable}
\end{center}											                  
\end{landscape}

%% file: tabella_noncor_flux.tex
\begin{center} \tiny \begin{longtable}{c c c c c c c c c c c c c }
\caption[Normalised emission line properties of the HRS galaxies.]{Fluxes of the emission lines of the HRS galaxies.\label{tab:measlinesflux}}\\
\hline \hline \\[-2ex]
\multicolumn{1}{c}{ } &
\multicolumn{1}{c}{ } &
\multicolumn{1}{c}{ } &
\multicolumn{1}{c}{ } &
\multicolumn{1}{c}{Flux f(\ha)=1} &
\multicolumn{1}{c}{ } &
\multicolumn{1}{c}{ } &
\multicolumn{1}{c}{ } &
\multicolumn{1}{c}{ } &
\multicolumn{1}{c}{ } &
\multicolumn{1}{c}{ } &
\multicolumn{1}{c}{ } &
\multicolumn{1}{c}{ } \\
\multicolumn{1}{c}{HRS} &
\multicolumn{1}{c}{ref} &
\multicolumn{1}{c}{E(B-V)} &
\multicolumn{1}{c}{C(\hb)} &
\multicolumn{1}{c}{[\oii]} &
\multicolumn{1}{c}{\hb} &
\multicolumn{1}{c}{[\oiii]} &
\multicolumn{1}{c}{[\oiii]} &
\multicolumn{1}{c}{[\nii]} &
\multicolumn{1}{c}{\ha} &
\multicolumn{1}{c}{[\nii]} &
\multicolumn{1}{c}{[\sii]} &
\multicolumn{1}{c}{[\sii]}  \\
\multicolumn{1}{c}{ } &
\multicolumn{1}{c}{ } &
\multicolumn{1}{c}{mag} &
\multicolumn{1}{c}{ } &
\multicolumn{1}{c}{$\lambda$3727} &
\multicolumn{1}{c}{$\lambda$4861} &
\multicolumn{1}{c}{$\lambda$4958} &
\multicolumn{1}{c}{$\lambda$5007} &
\multicolumn{1}{c}{$\lambda$6548} &
\multicolumn{1}{c}{$\lambda$6563} &
\multicolumn{1}{c}{$\lambda$6584} &
\multicolumn{1}{c}{$\lambda$6717} &
\multicolumn{1}{c}{$\lambda$6731} \\
\multicolumn{1}{c}{(1)} &
\multicolumn{1}{c}{(2)} &
\multicolumn{1}{c}{(3)} &
\multicolumn{1}{c}{(4)} &
\multicolumn{1}{c}{(5)} &
\multicolumn{1}{c}{(6)} &
\multicolumn{1}{c}{(7)} &
\multicolumn{1}{c}{(8)} &
\multicolumn{1}{c}{(9)} &
\multicolumn{1}{c}{(10)} &
\multicolumn{1}{c}{(11)} &
\multicolumn{1}{c}{(12)} &
\multicolumn{1}{c}{(13)} \\ [0.5ex] \hline
   \\[-1.8ex]
\endfirsthead
\multicolumn{13}{c}{{\tablename} \thetable{} -- Continued} \\[0.5ex]
\hline \hline \\[-2ex]
\multicolumn{1}{c}{ } &
\multicolumn{1}{c}{ } &
\multicolumn{1}{c}{ } &
\multicolumn{1}{c}{ } &
\multicolumn{1}{c}{Flux f(\ha)=1} &
\multicolumn{1}{c}{ } &
\multicolumn{1}{c}{ } &
\multicolumn{1}{c}{ } &
\multicolumn{1}{c}{ } &
\multicolumn{1}{c}{ } &
\multicolumn{1}{c}{ } &
\multicolumn{1}{c}{ } &
\multicolumn{1}{c}{ } \\
\multicolumn{1}{c}{HRS} &
\multicolumn{1}{c}{ref} &
\multicolumn{1}{c}{E(B-V)} &
\multicolumn{1}{c}{C(\hb)} &
\multicolumn{1}{c}{[\oii]} &
\multicolumn{1}{c}{\hb} &
\multicolumn{1}{c}{[\oiii]} &
\multicolumn{1}{c}{[\oiii]} &
\multicolumn{1}{c}{[\nii]} &
\multicolumn{1}{c}{\ha} &
\multicolumn{1}{c}{[\nii]} &
\multicolumn{1}{c}{[\sii]} &
\multicolumn{1}{c}{[\sii]}  \\
\multicolumn{1}{c}{ } &
\multicolumn{1}{c}{ } &
\multicolumn{1}{c}{mag} &
\multicolumn{1}{c}{ } &
\multicolumn{1}{c}{$\lambda$3727} &
\multicolumn{1}{c}{$\lambda$4861} &
\multicolumn{1}{c}{$\lambda$4958} &
\multicolumn{1}{c}{$\lambda$5007} &
\multicolumn{1}{c}{$\lambda$6548} &
\multicolumn{1}{c}{$\lambda$6563} &
\multicolumn{1}{c}{$\lambda$6584} &
\multicolumn{1}{c}{$\lambda$6717} &
\multicolumn{1}{c}{$\lambda$6731} \\
\multicolumn{1}{c}{(1)} &
\multicolumn{1}{c}{(2)} &
\multicolumn{1}{c}{(3)} &
\multicolumn{1}{c}{(4)} &
\multicolumn{1}{c}{(5)} &
\multicolumn{1}{c}{(6)} &
\multicolumn{1}{c}{(7)} &
\multicolumn{1}{c}{(8)} &
\multicolumn{1}{c}{(9)} &
\multicolumn{1}{c}{(10)} &
\multicolumn{1}{c}{(11)} &
\multicolumn{1}{c}{(12)} &
\multicolumn{1}{c}{(13)}  \\ [0.5ex] \hline
  \\[-1.8ex]
\endhead 
 \\[-1.8ex] \hline
  \multicolumn{13}{c}{{Continued on next page\ldots}} \\
\endfoot

  \\[-1.8ex] \hline \hline
\endlastfoot
%
%
           1 &  1  & 	 0.031 &     1.20 &	  0.22 &      0.15 &	   -   &      0.06 &	  0.10 &      1.00 &	  0.29 &      0.18 &	  0.12     \\ 
           2 &  1  & 	 0.023 &     0.80 &	  0.51 &      0.20 &	  0.04 &      0.09 &	  0.10 &      1.00 &	  0.30 &      0.19 &	  0.14     \\ 
           4 &  6  & 	 0.023 &      -   &	   -   &       -   &	   -   &       -   &	   -   &       -   &	   -   &       -   &	   -	   \\ 
           5 &  1  & 	 0.020 &      -   &	   -   &       -   &	   -   &       -   &	  0.08 &      1.00 &	  0.34 &      0.14 &	  0.11     \\ 
          10 &  1  & 	 0.027 &     0.47 &	  0.75 &      0.25 &	  0.08 &      0.24 &	  0.06 &      1.00 &	  0.24 &      0.21 &	  0.15     \\ 
          11 &  1  & 	 0.023 &     0.76 &	  0.45 &      0.21 &	  0.06 &      0.12 &	  0.08 &      1.00 &	  0.27 &      0.23 &	  0.17     \\ 
          12 &  1  & 	 0.023 &     0.43 &	  0.91 &      0.26 &	  0.11 &      0.33 &	  0.07 &      1.00 &	  0.17 &      0.22 &	  0.17     \\ 
          13 &  1  & 	 0.019 &     1.28 &	   -   &      0.15 &	   -   &      0.06 &	  0.09 &      1.00 &	  0.32 &      0.17 &	  0.12     \\
\end{longtable}
\end{center}											                  
Notes:\\ 
a) spurious detections in the [\oiii]$\lambda$4959 line in Gavazzi et al. (2004).\\ 
b) new measurements on the original Gavazzi et al. (2004) data. \\

%% file: tabella_noncor_ew.tex
\begin{center} \tiny \begin{longtable}{c c c c c c c c c c c c c }
\caption[Spectral properties of galaxies comprising the HRS sample.]{Equivalent widths of the main emission lines of the HRS galaxies.\label{tab:measlinesew}}\\
\hline \hline \\[-2ex]
\multicolumn{1}{c}{ } &
\multicolumn{1}{c}{ } &
\multicolumn{1}{c}{ } &
\multicolumn{1}{c}{EW(\AA)  } &
\multicolumn{1}{c}{ } &
\multicolumn{1}{c}{ } &
\multicolumn{1}{c}{ } &
\multicolumn{1}{c}{ } &
\multicolumn{1}{c}{ } &
\multicolumn{1}{c}{ } &
\multicolumn{1}{c}{ } &
\multicolumn{1}{c}{ } &
\multicolumn{1}{c}{ } \\
\multicolumn{1}{c}{HRS} &
\multicolumn{1}{c}{ref} &
\multicolumn{1}{c}{E(B-V)} &
\multicolumn{1}{c}{[\oii]} &
\multicolumn{1}{c}{\hb} &
\multicolumn{1}{c}{[\oiii]} &
\multicolumn{1}{c}{[\oiii]} &
\multicolumn{1}{c}{[\nii]} &
\multicolumn{1}{c}{\ha} &
\multicolumn{1}{c}{[\nii]} &
\multicolumn{1}{c}{[\sii]} &
\multicolumn{1}{c}{[\sii]} &
\multicolumn{1}{c}{H$\beta_{abs}$} \\
\multicolumn{1}{c}{ } &
\multicolumn{1}{c}{ } &
\multicolumn{1}{c}{mag} &
\multicolumn{1}{c}{$\lambda$3727} &
\multicolumn{1}{c}{$\lambda$4861} &
\multicolumn{1}{c}{$\lambda$4958} &
\multicolumn{1}{c}{$\lambda$5007} &
\multicolumn{1}{c}{$\lambda$6548} &
\multicolumn{1}{c}{$\lambda$6563} &
\multicolumn{1}{c}{$\lambda$6584} &
\multicolumn{1}{c}{$\lambda$6717} &
\multicolumn{1}{c}{$\lambda$6731} &
\multicolumn{1}{c}{$\lambda$4861} \\
\multicolumn{1}{c}{(1)} &
\multicolumn{1}{c}{(2)} &
\multicolumn{1}{c}{(3)} &
\multicolumn{1}{c}{(4)} &
\multicolumn{1}{c}{(5)} &
\multicolumn{1}{c}{(6)} &
\multicolumn{1}{c}{(7)} &
\multicolumn{1}{c}{(8)} &
\multicolumn{1}{c}{(9)} &
\multicolumn{1}{c}{(10)} &
\multicolumn{1}{c}{(11)} &
\multicolumn{1}{c}{(12)} &
\multicolumn{1}{c}{(13)} \\ [0.5ex] \hline
   \\[-1.8ex]
\endfirsthead
\multicolumn{13}{c}{{\tablename} \thetable{} -- Continued} \\[0.5ex]
  \hline \hline \\[-2ex]
\multicolumn{1}{c}{ } &
\multicolumn{1}{c}{ } &
\multicolumn{1}{c}{ } &
\multicolumn{1}{c}{EW(\AA)  } &
\multicolumn{1}{c}{ } &
\multicolumn{1}{c}{ } &
\multicolumn{1}{c}{ } &
\multicolumn{1}{c}{ } &
\multicolumn{1}{c}{ } &
\multicolumn{1}{c}{ } &
\multicolumn{1}{c}{ } &
\multicolumn{1}{c}{ } &
\multicolumn{1}{c}{ } \\
\multicolumn{1}{c}{HRS} &
\multicolumn{1}{c}{ref} &
\multicolumn{1}{c}{E(B-V)} &
\multicolumn{1}{c}{[\oii]} &
\multicolumn{1}{c}{\hb} &
\multicolumn{1}{c}{[\oiii]} &
\multicolumn{1}{c}{[\oiii]} &
\multicolumn{1}{c}{[\nii]} &
\multicolumn{1}{c}{\ha} &
\multicolumn{1}{c}{[\nii]} &
\multicolumn{1}{c}{[\sii]} &
\multicolumn{1}{c}{[\sii]} &
\multicolumn{1}{c}{H$\beta_{abs}$} \\
\multicolumn{1}{c}{ } &
\multicolumn{1}{c}{ } &
\multicolumn{1}{c}{mag} &
\multicolumn{1}{c}{$\lambda$3727} &
\multicolumn{1}{c}{$\lambda$4861} &
\multicolumn{1}{c}{$\lambda$4958} &
\multicolumn{1}{c}{$\lambda$5007} &
\multicolumn{1}{c}{$\lambda$6548} &
\multicolumn{1}{c}{$\lambda$6563} &
\multicolumn{1}{c}{$\lambda$6584} &
\multicolumn{1}{c}{$\lambda$6717} &
\multicolumn{1}{c}{$\lambda$6731} &
\multicolumn{1}{c}{$\lambda$4861} \\
\multicolumn{1}{c}{(1)} &
\multicolumn{1}{c}{(2)} &
\multicolumn{1}{c}{(3)} &
\multicolumn{1}{c}{(4)} &
\multicolumn{1}{c}{(5)} &
\multicolumn{1}{c}{(6)} &
\multicolumn{1}{c}{(7)} &
\multicolumn{1}{c}{(8)} &
\multicolumn{1}{c}{(9)} &
\multicolumn{1}{c}{(10)} &
\multicolumn{1}{c}{(11)} &
\multicolumn{1}{c}{(12)} &
\multicolumn{1}{c}{(13)} \\ [0.5ex] \hline
  \\[-1.8ex]
\endhead 
 \\[-1.8ex] \hline
  \multicolumn{13}{c}{{Continued on next page\ldots}} \\
\endfoot

  \\[-1.8ex] \hline \hline
\endlastfoot
%
%
           1 &  1  & 	 0.031 &          6.53  &     3.11  &	   -	&     1.19  &	  2.28  &    22.12  &	  6.65  &     4.20  &	  2.73  &    -5.37\\ 
           2 &  1  & 	 0.023 &         21.36  &     5.66  &	  1.18  &     2.85  &	  3.81  &    36.76  &	 11.31  &     7.20  &	  5.18  &    -5.87\\ 
           4 &  6  & 	 0.023 &         10.00  &     4.00  &	  2.33  &     7.00  &	  0.00  &    34.80  &	     0  &     0.00  &	  5.00  &      -  \\ 
           5 &  1  & 	 0.020 &           -	&      -    &	   -	&      -    &	  0.68  &     9.03  &	  3.04  &     1.21  &	  1.00  &      -  \\ 
          10 &  1  & 	 0.027 &         22.40  &     6.11  &	  2.12  &     6.04  &	  2.13  &    33.64  &	  8.05  &     7.57  &	  5.38  &    -5.16\\ 
          11 &  1  & 	 0.023 &         18.10  &     4.90  &	  1.34  &     3.04  &	  2.42  &    30.78  &	  8.38  &     7.16  &	  5.30  &    -6.09\\ 
          12 &  1  & 	 0.023 &         34.24  &     7.02  &	  3.06  &     9.13  &	  2.44  &    37.57  &	  6.53  &     8.76  &	  6.60  &    -5.39\\ 
          13 &  1  & 	 0.019 &           -	&     3.09  &	   -	&     1.46  &	  2.26  &    24.69  &	  8.16  &     4.20  &	  2.96  &    -4.30\\
\end{longtable}
\end{center}			
Notes:\\ 
a) spurious detections in the [\oiii]$\lambda$4959 line in Gavazzi et al. (2004).\\ 
b) new measurements on the original Gavazzi et al. (2004) data. \\

%% file: EWletteratura.tex
\begin{landscape}
\begin{center} \tiny \begin{longtable}{c c c c c c c c c c c c c c c c c c c c}
\caption{HRS galaxies with data in the literature: equivalent widths (in \AA).\label{tab:TablitEW}}\\
\hline \hline \\[-2ex]
\multicolumn{1}{c}{K92} &
\multicolumn{1}{c}{ } &
\multicolumn{1}{c}{ } &
\multicolumn{1}{c}{ } &
\multicolumn{1}{c}{ } &
\multicolumn{1}{c}{ } &
\multicolumn{1}{c}{ } &
\multicolumn{1}{c}{ } &
\multicolumn{1}{c}{ } &
\multicolumn{1}{c}{ } &
\multicolumn{1}{c}{This} &
\multicolumn{1}{c}{ } &
\multicolumn{1}{c}{ } &
\multicolumn{1}{c}{ } &
\multicolumn{1}{c}{ } &
\multicolumn{1}{c}{ } &
\multicolumn{1}{c}{ } &
\multicolumn{1}{c}{ } &
\multicolumn{1}{c}{ } &
\multicolumn{1}{c}{ } \\
\multicolumn{1}{c}{ } &
\multicolumn{1}{c}{ } &
\multicolumn{1}{c}{ } &
\multicolumn{1}{c}{ } &
\multicolumn{1}{c}{ } &
\multicolumn{1}{c}{ } &
\multicolumn{1}{c}{ } &
\multicolumn{1}{c}{ } &
\multicolumn{1}{c}{ } &
\multicolumn{1}{c}{ } &
\multicolumn{1}{c}{work} &
\multicolumn{1}{c}{ } &
\multicolumn{1}{c}{ } &
\multicolumn{1}{c}{ } &
\multicolumn{1}{c}{ } &
\multicolumn{1}{c}{ } &
\multicolumn{1}{c}{ } &
\multicolumn{1}{c}{ } &
\multicolumn{1}{c}{ } &
\multicolumn{1}{c}{ } \\
\hline
\multicolumn{1}{c}{HRS} &
\multicolumn{1}{c}{[\oii]} &
\multicolumn{1}{c}{\hb} &
\multicolumn{1}{c}{[\oiii]} &
\multicolumn{1}{c}{[\oiii]} &
\multicolumn{1}{c}{[\nii]} &
\multicolumn{1}{c}{\ha} &
\multicolumn{1}{c}{[\nii]} &
\multicolumn{1}{c}{[\sii]} &
\multicolumn{1}{c}{[\sii]} &
\multicolumn{1}{c}{run} &
\multicolumn{1}{c}{[\oii]} &
\multicolumn{1}{c}{\hb} &
\multicolumn{1}{c}{[\oiii]} &
\multicolumn{1}{c}{[\oiii]} &
\multicolumn{1}{c}{[\nii]} &
\multicolumn{1}{c}{\ha} &
\multicolumn{1}{c}{[\nii]} &
\multicolumn{1}{c}{[\sii]} &
\multicolumn{1}{c}{[\sii]}\\
\hline
\end{longtable}
\end{center}
\end{landscape}

\noindent
Notes to Table :\\ 
a: References are K92 - Kennicutt (1992b); J00 - Jansen et al. (2000); MK06 - Moustakas \& Kennicutt (2006); M10 - Moustakas et al. (2010).\\
b: for consistency with our work, the original data of Kennicutt (1992b) are corrected for the underlying Balmer absorption in \hb ~(5 \AA) and \ha ~(2.8 \AA). 
The [\oiii]$\lambda$4959 and the [\nii]$\lambda$6548 equivalent widths are inferred from the [\oiii]$\lambda$5007 and the [\nii]$\lambda$6584 measurements assuming a standard ratio [\oiii]$\lambda$5007/[\oiii]$\lambda$4959=3 and
[\nii]$\lambda$6584/[\nii]$\lambda$6548=3 (Osterbrock \& Ferland 2005). The [\sii]$\lambda$6731 equivalent width includes the emission of both [\sii]$\lambda$6717 and [\sii]$\lambda$6731 lines.\\
c: the original data of Jansen et al. (2000) are corrected only for the underlying Balmer absorption in \ha ~(2.8 \AA), but not for that in \hb ~given that the spectral resolution of these data, comparable to our, allows the direct measurement of the underlying absorption. \\
d: the [\oiii]$\lambda$4959 and the [\nii]$\lambda$6548 equivalent widths of Moustakas et al. (2010) and Moustakas \& Kennicutt (2006) 
are inferred from the [\oiii]$\lambda$5007 and the [\nii]$\lambda$6584 measurements assuming a standard ratio [\oiii]$\lambda$5007/[\oiii]$\lambda$4959=3 and
[\nii]$\lambda$6584/[\nii]$\lambda$6548=3 (Osterbrock \& Ferland 2005).\\

%% file: letteratura_flux.tex
\begin{landscape}
\begin{center} \tiny \begin{longtable}{c c c c c c c c c c c c c c c c c c c c c c}
\caption{HRS galaxies with data in the literature: fluxes normalised to H$\alpha$.\label{tab:Tablitflux}} \\
\hline \hline \\[-2ex]
\multicolumn{1}{c}{K92} &
\multicolumn{1}{c}{ } &
\multicolumn{1}{c}{ } &
\multicolumn{1}{c}{ } &
\multicolumn{1}{c}{ } &
\multicolumn{1}{c}{ } &
\multicolumn{1}{c}{ } &
\multicolumn{1}{c}{ } &
\multicolumn{1}{c}{ } &
\multicolumn{1}{c}{ } &
\multicolumn{1}{c}{ } &
\multicolumn{1}{c}{This} &
\multicolumn{1}{c}{ } &
\multicolumn{1}{c}{ } &
\multicolumn{1}{c}{ } &
\multicolumn{1}{c}{ } &
\multicolumn{1}{c}{ } &
\multicolumn{1}{c}{ } &
\multicolumn{1}{c}{ } &
\multicolumn{1}{c}{ } &
\multicolumn{1}{c}{ } &
\multicolumn{1}{c}{ } \\
\multicolumn{1}{c}{ } &
\multicolumn{1}{c}{ } &
\multicolumn{1}{c}{ } &
\multicolumn{1}{c}{ } &
\multicolumn{1}{c}{ } &
\multicolumn{1}{c}{ } &
\multicolumn{1}{c}{ } &
\multicolumn{1}{c}{ } &
\multicolumn{1}{c}{ } &
\multicolumn{1}{c}{ } &
\multicolumn{1}{c}{ } &
\multicolumn{1}{c}{work} &
\multicolumn{1}{c}{ } &
\multicolumn{1}{c}{ } &
\multicolumn{1}{c}{ } &
\multicolumn{1}{c}{ } &
\multicolumn{1}{c}{ } &
\multicolumn{1}{c}{ } &
\multicolumn{1}{c}{ } &
\multicolumn{1}{c}{ } &
\multicolumn{1}{c}{ } &
\multicolumn{1}{c}{ } \\
\hline
\multicolumn{1}{c}{HRS} &
\multicolumn{1}{c}{C(\hb)} &
\multicolumn{1}{c}{[\oii]} &
\multicolumn{1}{c}{\hb} &
\multicolumn{1}{c}{[\oiii]} &
\multicolumn{1}{c}{[\oiii]} &
\multicolumn{1}{c}{[\nii]} &
\multicolumn{1}{c}{\ha} &
\multicolumn{1}{c}{[\nii]} &
\multicolumn{1}{c}{[\sii]} &
\multicolumn{1}{c}{[\sii]} &
\multicolumn{1}{c}{run} &
\multicolumn{1}{c}{C(\hb)} &
\multicolumn{1}{c}{[\oii]} &
\multicolumn{1}{c}{\hb} &
\multicolumn{1}{c}{[\oiii]} &
\multicolumn{1}{c}{[\oiii]} &
\multicolumn{1}{c}{[\nii]} &
\multicolumn{1}{c}{\ha} &
\multicolumn{1}{c}{[\nii]} &
\multicolumn{1}{c}{[\sii]} &
\multicolumn{1}{c}{[\sii]}\\
\multicolumn{1}{c}{ } &
\multicolumn{1}{c}{ } &
\multicolumn{1}{c}{$\lambda$3727} &
\multicolumn{1}{c}{$\lambda$4861} &
\multicolumn{1}{c}{$\lambda$4959} &
\multicolumn{1}{c}{$\lambda$5007} &
\multicolumn{1}{c}{$\lambda$6548} &
\multicolumn{1}{c}{$\lambda$6563} &
\multicolumn{1}{c}{$\lambda$6584} &
\multicolumn{1}{c}{$\lambda$6717} &
\multicolumn{1}{c}{$\lambda$6731} &
\multicolumn{1}{c}{} &
\multicolumn{1}{c}{} &
\multicolumn{1}{c}{$\lambda$3727} &
\multicolumn{1}{c}{$\lambda$4861} &
\multicolumn{1}{c}{$\lambda$4959} &
\multicolumn{1}{c}{$\lambda$5007} &
\multicolumn{1}{c}{$\lambda$6548} &
\multicolumn{1}{c}{$\lambda$6563} &
\multicolumn{1}{c}{$\lambda$6584} &
\multicolumn{1}{c}{$\lambda$6717} &
\multicolumn{1}{c}{$\lambda$6731} \\
  \\[-1.8ex]
 \\[-1.8ex] \hline 
  4 &	    -  &    0.13  &      -  &   0.07 &    0.20 & 	   - &	 1 &	  -  &	    -  &    0.16  &   - &    -   &   -   &      -   &       -   &       -   &       -   &	 -   &       -   &	  -   &       -   \\
 20 &	 0.30  &    0.80  &    0.28 &   0.13 &    0.39 & 	0.06 &	 1 &	0.17 &	    -  &    0.39  &   - &    -   &   -   &      -   &       -   &       -   &       -   &	 -   &       -   &	  -   &       -   \\
275 &    0.00  &    1.15  &    0.37 &   0.14 &    0.43 &        0.07 &	 1 &	0.21 &	    -  &    0.35  &   - &    -   &   -   &      -   &       -   &       -   &       -   &	 -   &       -   &	  -   &       -   \\
295 &	 0.46  &    0.25  &    0.25 &   0.01 &    0.04 & 	0.12 &	 1 &	0.37 &	    -  &    0.25  &   - &    -   &   -   &      -   &       -   &       -   &       -   &	 -   &       -   &	  -   &       -   \\
\hline \hline \\[-2ex]
\multicolumn{1}{c}{J00} &
\multicolumn{1}{c}{ } &
\multicolumn{1}{c}{ } &
\multicolumn{1}{c}{ } &
\multicolumn{1}{c}{ } &
\multicolumn{1}{c}{ } &
\multicolumn{1}{c}{ } &
\multicolumn{1}{c}{ } &
\multicolumn{1}{c}{ } &
\multicolumn{1}{c}{ } &
\multicolumn{1}{c}{ } &
\multicolumn{1}{c}{This} &
\multicolumn{1}{c}{ } &
\multicolumn{1}{c}{ } &
\multicolumn{1}{c}{ } &
\multicolumn{1}{c}{ } &
\multicolumn{1}{c}{ } &
\multicolumn{1}{c}{ } &
\multicolumn{1}{c}{ } &
\multicolumn{1}{c}{ } &
\multicolumn{1}{c}{ } &
\multicolumn{1}{c}{ } \\
\multicolumn{1}{c}{ } &
\multicolumn{1}{c}{ } &
\multicolumn{1}{c}{ } &
\multicolumn{1}{c}{ } &
\multicolumn{1}{c}{ } &
\multicolumn{1}{c}{ } &
\multicolumn{1}{c}{ } &
\multicolumn{1}{c}{ } &
\multicolumn{1}{c}{ } &
\multicolumn{1}{c}{ } &
\multicolumn{1}{c}{ } &
\multicolumn{1}{c}{work} &
\multicolumn{1}{c}{ } &
\multicolumn{1}{c}{ } &
\multicolumn{1}{c}{ } &
\multicolumn{1}{c}{ } &
\multicolumn{1}{c}{ } &
\multicolumn{1}{c}{ } &
\multicolumn{1}{c}{ } &
\multicolumn{1}{c}{ } &
\multicolumn{1}{c}{ } &
\multicolumn{1}{c}{ } \\
\hline
\multicolumn{1}{c}{HRS} &
\multicolumn{1}{c}{C(\hb)} &
\multicolumn{1}{c}{[\oii]} &
\multicolumn{1}{c}{\hb} &
\multicolumn{1}{c}{[\oiii]} &
\multicolumn{1}{c}{[\oiii]} &
\multicolumn{1}{c}{[\nii]} &
\multicolumn{1}{c}{\ha} &
\multicolumn{1}{c}{[\nii]} &
\multicolumn{1}{c}{[\sii]} &
\multicolumn{1}{c}{[\sii]} &
\multicolumn{1}{c}{run} &
\multicolumn{1}{c}{C(\hb)} &
\multicolumn{1}{c}{[\oii]} &
\multicolumn{1}{c}{\hb} &
\multicolumn{1}{c}{[\oiii]} &
\multicolumn{1}{c}{[\oiii]} &
\multicolumn{1}{c}{[\nii]} &
\multicolumn{1}{c}{\ha} &
\multicolumn{1}{c}{[\nii]} &
\multicolumn{1}{c}{[\sii]} &
\multicolumn{1}{c}{[\sii]}\\
\multicolumn{1}{c}{ } &
\multicolumn{1}{c}{ } &
\multicolumn{1}{c}{$\lambda$3727} &
\multicolumn{1}{c}{$\lambda$4861} &
\multicolumn{1}{c}{$\lambda$4959} &
\multicolumn{1}{c}{$\lambda$5007} &
\multicolumn{1}{c}{$\lambda$6548} &
\multicolumn{1}{c}{$\lambda$6563} &
\multicolumn{1}{c}{$\lambda$6584} &
\multicolumn{1}{c}{$\lambda$6717} &
\multicolumn{1}{c}{$\lambda$6731} &
\multicolumn{1}{c}{} &
\multicolumn{1}{c}{} &
\multicolumn{1}{c}{$\lambda$3727} &
\multicolumn{1}{c}{$\lambda$4861} &
\multicolumn{1}{c}{$\lambda$4959} &
\multicolumn{1}{c}{$\lambda$5007} &
\multicolumn{1}{c}{$\lambda$6548} &
\multicolumn{1}{c}{$\lambda$6563} &
\multicolumn{1}{c}{$\lambda$6584} &
\multicolumn{1}{c}{$\lambda$6717} &
\multicolumn{1}{c}{$\lambda$6731} \\
  \\[-1.8ex]
 \\[-1.8ex] \hline	
 29 &	 0.58  &    0.68  &    0.23 &    0.11 &   0.21 &       0.06  &	 1 &	0.21 &    0.22 &    0.16  &   1 &   0.32 &  0.58 &     0.28 &	   0.07 &      0.20 &	   0.07 &	  1  &	   0.23  &	0.21  &      0.17 \\
 35 &	  -    &     -    &    0.56 &     -   &     -  &       -     &	 1 &      -  &    -    &     -    &   - &    -   &   -   &       -  &	    -   &      -    &	    -   &        -   &	     -   &       -    &       -   \\
 64 &	 0.92  &    0.70  &    0.19 &    0.10 &   0.15 &       0.05  &	 1 &	0.19 &    0.27 &    0.19  &   - &    -   &   -   &       -  &	    -   &      -    &	    -   &	 -   &	     -   &       -    &       -   \\
 67 &	 0.12  &    1.32  &    0.32 &    0.22 &   0.51 &       0.04  &	 1 &	0.13 &    0.22 &    0.14  &   1 &   0.19 &  1.31 &     0.31 &	   0.15 &      0.50 &	   0.04 &	  1  &	   0.14  &	0.19  &      0.13 \\
271 &	 0.97  &    0.71  &    0.18 &    0.14 &   0.17 &       0.05  &	 1 &	0.16 &    0.28 &    0.20  &   - &    -   &   -   &      -   &	    -	&	-   &	    -	&	 -   &	     -	 &	 -    &       -   \\
\hline \hline \\[-2ex]
\multicolumn{1}{c}{MK06} &
\multicolumn{1}{c}{ } &
\multicolumn{1}{c}{ } &
\multicolumn{1}{c}{ } &
\multicolumn{1}{c}{ } &
\multicolumn{1}{c}{ } &
\multicolumn{1}{c}{ } &
\multicolumn{1}{c}{ } &
\multicolumn{1}{c}{ } &
\multicolumn{1}{c}{ } &
\multicolumn{1}{c}{ } &
\multicolumn{1}{c}{This} &
\multicolumn{1}{c}{ } &
\multicolumn{1}{c}{ } &
\multicolumn{1}{c}{ } &
\multicolumn{1}{c}{ } &
\multicolumn{1}{c}{ } &
\multicolumn{1}{c}{ } &
\multicolumn{1}{c}{ } &
\multicolumn{1}{c}{ } &
\multicolumn{1}{c}{ } &
\multicolumn{1}{c}{ } \\
\multicolumn{1}{c}{ } &
\multicolumn{1}{c}{ } &
\multicolumn{1}{c}{ } &
\multicolumn{1}{c}{ } &
\multicolumn{1}{c}{ } &
\multicolumn{1}{c}{ } &
\multicolumn{1}{c}{ } &
\multicolumn{1}{c}{ } &
\multicolumn{1}{c}{ } &
\multicolumn{1}{c}{ } &
\multicolumn{1}{c}{ } &
\multicolumn{1}{c}{work} &
\multicolumn{1}{c}{ } &
\multicolumn{1}{c}{ } &
\multicolumn{1}{c}{ } &
\multicolumn{1}{c}{ } &
\multicolumn{1}{c}{ } &
\multicolumn{1}{c}{ } &
\multicolumn{1}{c}{ } &
\multicolumn{1}{c}{ } &
\multicolumn{1}{c}{ } &
\multicolumn{1}{c}{ } \\
\hline
\multicolumn{1}{c}{HRS} &
\multicolumn{1}{c}{C(\hb)} &
\multicolumn{1}{c}{[\oii]} &
\multicolumn{1}{c}{\hb} &
\multicolumn{1}{c}{[\oiii]} &
\multicolumn{1}{c}{[\oiii]} &
\multicolumn{1}{c}{[\nii]} &
\multicolumn{1}{c}{\ha} &
\multicolumn{1}{c}{[\nii]} &
\multicolumn{1}{c}{[\sii]} &
\multicolumn{1}{c}{[\sii]} &
\multicolumn{1}{c}{run} &
\multicolumn{1}{c}{C(\hb)} &
\multicolumn{1}{c}{[\oii]} &
\multicolumn{1}{c}{\hb} &
\multicolumn{1}{c}{[\oiii]} &
\multicolumn{1}{c}{[\oiii]} &
\multicolumn{1}{c}{[\nii]} &
\multicolumn{1}{c}{\ha} &
\multicolumn{1}{c}{[\nii]} &
\multicolumn{1}{c}{[\sii]} &
\multicolumn{1}{c}{[\sii]}\\
\multicolumn{1}{c}{ } &
\multicolumn{1}{c}{ } &
\multicolumn{1}{c}{$\lambda$3727} &
\multicolumn{1}{c}{$\lambda$4861} &
\multicolumn{1}{c}{$\lambda$4959} &
\multicolumn{1}{c}{$\lambda$5007} &
\multicolumn{1}{c}{$\lambda$6548} &
\multicolumn{1}{c}{$\lambda$6563} &
\multicolumn{1}{c}{$\lambda$6584} &
\multicolumn{1}{c}{$\lambda$6717} &
\multicolumn{1}{c}{$\lambda$6731} &
\multicolumn{1}{c}{} &
\multicolumn{1}{c}{} &
\multicolumn{1}{c}{$\lambda$3727} &
\multicolumn{1}{c}{$\lambda$4861} &
\multicolumn{1}{c}{$\lambda$4959} &
\multicolumn{1}{c}{$\lambda$5007} &
\multicolumn{1}{c}{$\lambda$6548} &
\multicolumn{1}{c}{$\lambda$6563} &
\multicolumn{1}{c}{$\lambda$6584} &
\multicolumn{1}{c}{$\lambda$6717} &
\multicolumn{1}{c}{$\lambda$6731} \\
  \\[-1.8ex]
 \\[-1.8ex] \hline
 20 &	 0.26  &    0.86  &    0.29 &    0.13 &   0.40 &        0.07 &	 1 &	0.21 &    0.20 &    0.13  &   - &    -   &   -   &      -   &       -   &	-   &       -   &	 -   &	     -   &	 -    &	      -   \\
 27 &	 0.33  &    0.56  &    0.28 &    0.06 &   0.17 & 	0.10 &	 1 &	0.31 &    0.19 &    0.13  &   - &    -   &   -   &      -   &       -   &	-   &       -   &	 -   &	     -   &	 -    &	      -   \\
 36 &	 0.73  &    0.31  &    0.21 &    0.03 &   0.10 & 	0.19 &	 1 &	0.59 &    0.19 &    0.15  &   - &    -   &   -   &      -   &       -   &	-   &       -   &	 -   &	     -   &	 -    &	      -   \\
 60 &	 0.45  &    0.36  &    0.26 &    0.05 &   0.15 & 	0.15 &	 1 &	0.45 &    0.21 &    0.16  &   1 &   1.55 &  0.26 &     0.12 &       -   &      0.16 &      0.13 &          1 &      0.47 &       0.18 &	     0.15 \\
 73 &	 0.25  &    0.35  &    0.29 &    0.00 &   0.01 & 	0.13 &	 1 &	0.39 &    0.17 &    0.09  &   - &     -  &    -  &      -   &       -   &	 -  &       -   &	 -   &       -   &	  -   &       -   \\
 74 &	 0.48  &    0.36  &    0.25 &    0.05 &   0.16 & 	0.12 &	 1 &	0.37 &    0.18 &    0.13  &   1 &   0.59 &  0.40 &     0.23 &      0.10 &      0.16 &      0.11 &	   1 &      0.35 &	 0.15 &      0.11 \\
 76 &	 0.09  &    0.98  &    0.33 &    0.21 &   0.64 &        0.03 &	 1 &    0.10 &    0.19 &    0.10  &   1 &   0.12 &  1.00 &     0.32 &      0.24 &      0.77 &      0.03 &	   1 &      0.11 &	 0.15 &      0.12 \\
 85 &	 0.65  &    0.39  &    0.22 &    0.04 &   0.11 &        0.14 &	 1 &	0.43 &    0.21 &    0.14  &   - &     -	 &   -   &  	-   &       -   &	 -  &       -   &	 -   &       -   &	  -   &       -   \\
102 &	 0.65  &    0.27  &    0.22 &    0.02 &   0.06 & 	0.11 &	 1 &	0.34 &    0.14 &    0.09  &   2 &   0.74 &   -   &     0.21 &       -   &      0.05 &      0.10 &	   1 &      0.32 &	 0.13 &      0.09 \\
114 &	 0.46  &    0.31  &    0.25 &    0.03 &   0.10 & 	0.14 &	 1 &	0.41 &    0.14 &    0.11  &   2 &   0.66 &  0.32 &     0.22 &      0.13 &      0.07 &      0.12 &	   1 &      0.39 &	 0.15 &      0.10 \\
122 &	 1.05  &    0.10  &    0.17 &    0.02 &   0.07 & 	0.11 &	 1 &	0.33 &    0.09 &    0.05  &   1 &   0.79 &   -   &     0.20 &       -   &	-   &      0.12 &	   1 &      0.36 &	 0.08 &      0.06 \\
217 &	 2.00  &    0.43  &    0.09 &    0.10 &   0.31 & 	0.26 &	 1 &	0.77 &    0.37 &    0.21  &   2 &     -  &   -   & 	 -  &       -   &	 -  &      0.09 &	   1 &      0.57 &	 0.15 &      0.17 \\
246 &	 0.37  &    0.55  &    0.27 &    0.07 &   0.20 & 	0.13 &	 1 &	0.38 &    0.24 &    0.18  &   2 &   0.84 &   -   &     0.20 &      0.05 &      0.08 &      0.11 &	   1 &      0.37 &	 0.18 &      0.11 \\
250 &	 0.85  &    0.31  &    0.19 &    0.04 &   0.12 & 	0.13 &	 1 &	0.38 &    0.20 &    0.16  &   - &     -  &   -   &  	 -  &       -   &	 -  &        -  &	 -   &       -   &	  -   &       -   \\
262 &	 0.20  &    0.92  &    0.30 &    0.11 &   0.34 &        0.07 &	 1 &	0.22 &    0.21 &    0.14  &   1 &   0.04 &  1.15 &     0.34 &      0.14 &      0.36 &      0.06 &	   1 &      0.20 &	 0.20 &      0.13 \\
290 &	 0.43  &    0.52  &    0.26 &    0.06 &   0.17 & 	0.13 &	 1 &	0.35 &    0.26 &    0.17  &   1 &   0.99 &  0.46 &     0.18 &      0.12 &      0.14 &      0.08 &	   1 &      0.29 &	 0.19 &      0.14 \\
\hline \hline \\[-2ex]
\multicolumn{1}{c}{M10} &
\multicolumn{1}{c}{ } &
\multicolumn{1}{c}{ } &
\multicolumn{1}{c}{ } &
\multicolumn{1}{c}{ } &
\multicolumn{1}{c}{ } &
\multicolumn{1}{c}{ } &
\multicolumn{1}{c}{ } &
\multicolumn{1}{c}{ } &
\multicolumn{1}{c}{ } &
\multicolumn{1}{c}{ } &
\multicolumn{1}{c}{This} &
\multicolumn{1}{c}{ } &
\multicolumn{1}{c}{ } &
\multicolumn{1}{c}{ } &
\multicolumn{1}{c}{ } &
\multicolumn{1}{c}{ } &
\multicolumn{1}{c}{ } &
\multicolumn{1}{c}{ } &
\multicolumn{1}{c}{ } &
\multicolumn{1}{c}{ } &
\multicolumn{1}{c}{ } \\
\multicolumn{1}{c}{ } &
\multicolumn{1}{c}{ } &
\multicolumn{1}{c}{ } &
\multicolumn{1}{c}{ } &
\multicolumn{1}{c}{ } &
\multicolumn{1}{c}{ } &
\multicolumn{1}{c}{ } &
\multicolumn{1}{c}{ } &
\multicolumn{1}{c}{ } &
\multicolumn{1}{c}{ } &
\multicolumn{1}{c}{ } &
\multicolumn{1}{c}{work} &
\multicolumn{1}{c}{ } &
\multicolumn{1}{c}{ } &
\multicolumn{1}{c}{ } &
\multicolumn{1}{c}{ } &
\multicolumn{1}{c}{ } &
\multicolumn{1}{c}{ } &
\multicolumn{1}{c}{ } &
\multicolumn{1}{c}{ } &
\multicolumn{1}{c}{ } &
\multicolumn{1}{c}{ } \\
\hline
\multicolumn{1}{c}{HRS} &
\multicolumn{1}{c}{C(\hb)} &
\multicolumn{1}{c}{[\oii]} &
\multicolumn{1}{c}{\hb} &
\multicolumn{1}{c}{[\oiii]} &
\multicolumn{1}{c}{[\oiii]} &
\multicolumn{1}{c}{[\nii]} &
\multicolumn{1}{c}{\ha} &
\multicolumn{1}{c}{[\nii]} &
\multicolumn{1}{c}{[\sii]} &
\multicolumn{1}{c}{[\sii]} &
\multicolumn{1}{c}{run} &
\multicolumn{1}{c}{C(\hb)} &
\multicolumn{1}{c}{[\oii]} &
\multicolumn{1}{c}{\hb} &
\multicolumn{1}{c}{[\oiii]} &
\multicolumn{1}{c}{[\oiii]} &
\multicolumn{1}{c}{[\nii]} &
\multicolumn{1}{c}{\ha} &
\multicolumn{1}{c}{[\nii]} &
\multicolumn{1}{c}{[\sii]} &
\multicolumn{1}{c}{[\sii]}\\
\multicolumn{1}{c}{ } &
\multicolumn{1}{c}{ } &
\multicolumn{1}{c}{$\lambda$3727} &
\multicolumn{1}{c}{$\lambda$4861} &
\multicolumn{1}{c}{$\lambda$4959} &
\multicolumn{1}{c}{$\lambda$5007} &
\multicolumn{1}{c}{$\lambda$6548} &
\multicolumn{1}{c}{$\lambda$6563} &
\multicolumn{1}{c}{$\lambda$6584} &
\multicolumn{1}{c}{$\lambda$6717} &
\multicolumn{1}{c}{$\lambda$6731} &
\multicolumn{1}{c}{} &
\multicolumn{1}{c}{} &
\multicolumn{1}{c}{$\lambda$3727} &
\multicolumn{1}{c}{$\lambda$4861} &
\multicolumn{1}{c}{$\lambda$4959} &
\multicolumn{1}{c}{$\lambda$5007} &
\multicolumn{1}{c}{$\lambda$6548} &
\multicolumn{1}{c}{$\lambda$6563} &
\multicolumn{1}{c}{$\lambda$6584} &
\multicolumn{1}{c}{$\lambda$6717} &
\multicolumn{1}{c}{$\lambda$6731} \\
  \\[-1.8ex]
 \\[-1.8ex] \hline
102 &	  1.38  &   0.07  &    0.14 &   0.01 &    0.03 &        0.10 &	 1 &	0.29 &    0.12  &   0.08  &   2 &   0.74 &   -   &     0.21 &       -   &      0.05 &      0.10 &	   1 &      0.32 &	 0.13 &      0.09 \\
122 &	  1.24  &   0.10  &    0.15 &   0.01 &    0.04 & 	0.13 &	 1 &	0.39 &    0.14  &   0.10  &   1 &   0.79 &   -   &     0.20 &       -   &	 -  &      0.12 &	   1 &      0.36 &	 0.08 &      0.06 \\
170 &	  1.23  &   0.74  &    0.15 &   0.12 &    0.37 & 	0.25 &	 1 &	0.76 &    0.43  &   0.31  &   2 &     -  &   -   &	 -  &       -   &	 -  &      0.19 &	   1 &	    0.47 &	 0.20 &	     0.20 \\
205 &	  1.66  &   0.19  &    0.11 &   0.01 &    0.04 & 	0.15 &	 1 &	0.44 &    0.25  &   0.19  &   1 &   1.28 &   -   &     0.14 &       -   &      0.12 &      0.12 &          1 &      0.37 &       0.20 &	     0.16 \\
211 &	    -   &    -    &    1.00 &     -  & 	   -   & 	 -   &	 1 &     -   &	   -    &      -  &   2 &     -  &   -   &       -  &       -   &        -  &       -   &        -   &       -   &        -   &	      -   \\
217 &	  2.25  &   0.22  &    0.07 &   0.05 &    0.16 & 	0.32 &	 1 &	0.96 &    0.30  &   0.23  &   2 &     -  &   -   &	 -  &	    -	&	 -  &	   0.09 &	   1 &      0.57 &	 0.15 &      0.17 \\
220 &	  0.93  &   0.74  &    0.18 &   0.13 &    0.38 & 	0.44 &	 1 &	1.33 &    0.66  &   0.37  &   2 &     -  &   -   &       -  &       -   &        -  &	   0.20 &          1 &      0.72 &       0.38 &	     0.23 \\
263 &	    -   &    -	  &       - &    -   & 	   -   & 	   - &	 1 &	  -  &	   -    &     -   &   - &     -  &   -   &       -  &       -   &        -  &       -   &        -   &       -   &        -   &	      -   \\
\hline
\end{longtable}
\end{center}
\end{landscape}

\noindent
Notes to Table :\\ 
a: References are K92 - Kennicutt (1992b); J00 - Jansen et al. (2000); MK06 - Moustakas \& Kennicutt (2006); M10 - Moustakas et al. (2010).\\
b: for consistency with our work, the original data of Kennicutt (1992b) are corrected for the underlying Balmer absorption in \hb ~(5 \AA) and \ha ~(2.8 \AA). 
The [\oiii]$\lambda$4959 and the [\nii]$\lambda$6548 fluxes are inferred from the [\oiii]$\lambda$5007 and the [\nii]$\lambda$6584 measurements assuming a standard ratio [\oiii]$\lambda$5007/[\oiii]$\lambda$4959=3 and
[\nii]$\lambda$6584/[\nii]$\lambda$6548=3 (Osterbrock \& Ferland 2005). The [\sii]$\lambda$6731 flux includes the emission of both [\sii]$\lambda$6717 and [\sii]$\lambda$6731 lines.\\
c: the original data of Jansen et al. (2000) are corrected only for the underlying Balmer absorption in \ha ~(2.8 \AA), but not for that in \hb ~given that the spectral resolution 
of these data, comparable to our, allows the direct measurement of the underlying absorption. \\
d: the [\oiii]$\lambda$4959 and the [\nii]$\lambda$6548 fluxes of Moustakas et al. (2010) and Moustakas \& Kennicutt (2006) 
are inferred from the [\oiii]$\lambda$5007 and the [\nii]$\lambda$6584 measurements assuming a standard ratio [\oiii]$\lambda$5007/[\oiii]$\lambda$4959=3 and
[\nii]$\lambda$6584/[\nii]$\lambda$6548=3 (Osterbrock \& Ferland 2005). In the original works of Moustakas et al. (2010) and Moustakas \& Kennicutt (2006) 
fluxes are absolute values (not normalised) representative of the whole galaxies. The galaxy HRS 211 in Moustakas et al. (2010) is detected only in the \hb \ line.\\